\documentclass[useAMS,usenatbib]{mn2e}
\usepackage{graphicx}
\usepackage{txfonts}

\def\spose#1{\hbox to 0pt{#1\hss}}
\def\simlt{\mathrel{\spose{\lower 3pt\hbox{$\mathchar"218$}}
     \raise 2.0pt\hbox{$\mathchar"13C$}}}
\def\simgt{\mathrel{\spose{\lower 3pt\hbox{$\mathchar"218$}}
     \raise 2.0pt\hbox{$\mathchar"13E$}}}

\newcommand{\reduceme}{\mbox{R\raisebox{-0.35ex}{E}D\hspace{-0.05em}\raisebox{0.85ex}{uc}\hspace{-0.90em}\raisebox{-.35ex}{{m}}\hspace{0.05em}E}}

\title[Are Dry Mergers Dry, Moist, or Wet?]{Are Dry Mergers Dry, Moist, or Wet?}
\author[S\'anchez-Bl\'azquez et~al.]{P. S\'anchez-Bl\'azquez$^{1,2}$\thanks{E-mail: psanchez-blazquez@uclan.ac.uk}, B. K. Gibson$^{1}$, D. Kawata$^{3}$, N. Cardiel$^{4}$, and M. Balcells$^{2}$\\
$^{1}$Jeremiah Horrocks Insitute for Astrophysics \& Supercomputing, University of Central Lancashire, Preston, PR1~2HE, UK\\
$^{2}$Instituto de Astrof\'{\i}sica de Canarias, c/Via Lactea s/n, 38205, La Laguna, Tenerife, Spain\\
$^{3}$Mullard Space Science Laboratory, Holmbury St. Mary, RH5~6NT, UK\\
$^{4}$Departamento de Astrof\'{\i}sica y CC de la Atm\'osfera, Universidad Complutense de Madrid, Av Complutense s/n, 28040, Madrid, Spain}

\begin{document}

\date{Accepted}

\pagerange{\pageref{firstpage}--\pageref{lastpage}} \pubyear{2009}

\maketitle

\label{firstpage}

\begin{abstract}
We present a spectral analysis of a sample of red-sequence galaxies 
identified by van~Dokkum (2005) as dry merger remnants and ongoing dry 
mergers. Kinematics, stellar population absorption features, and 
ionisation from emission lines, are derived.  We find that  approximately half of the 
sample showing strong tidal features have younger 
stellar populations than a control sample at a given velocity 
dispersion.  Conversely, galaxies with weak tidal tails and/or ongoing 
mergers -- with the exception of one galaxy -- do not show this young 
component. This seems to indicate that the young stellar populations 
observed in a significant fraction of ellipticals is the consequence of 
star formation triggered by mergers.  This young component is consistent 
with a light ``frosting'' of young stars ($<$2\% by mass) superimposed 
upon a dominant, old ($\sim$11~Gyr), stellar population. In terms of 
stellar populations, these mergers are, in fact, fairly dry. We found, 
however, that merger remnants with young stellar populations are 
supported by rotation, contrary to the expectations of a major dry 
merger.  This suggests that the small amount of gas involved has been 
sufficient to produce a dynamically cold stellar component. Half of the galaxies
with strong tidal distortion, however, are slow rotating and have stellar populations
compatible with the control sample at a given velocity dispersion. Remarkably, 
none of the galaxies with velocity dispersions in excess of 
250~km~s$^{-1}$ have a young stellar component, independent of the 
merger stage.  
\end{abstract}


\section{Introduction}

An accurate understanding of the physics underlying the 
formation and evolution of massive early-type galaxies remains
a challenge.  Within the $\Lambda$-dominated Cold Dark Matter
hierarchical assembly framework, such galaxies are predicted
to form over time through the progressive mergers of smaller
galaxies (White \& Frenk 1991\nocite{WF91}; Cole et~al. 
2000\nocite{Cole00}), with the most massive systems assembling at
relatively late epochs ($z$$\simlt$1).   Such a scenario appears to
be supported by several studies
(Faber et~al. 2007\nocite{Fab07}; Bell et~al. 2006\nocite{Bell06})
that find an increase in the mass density function of the 
so-called ``red-sequence'' by factor of $\sim$2
since $z$$\sim$1.\footnote{Although this might not be strictly 
true for the {\it most} massive galaxies (Cimatti, Daddi \& Renzini 2006\nocite{CDR06}; 
Ferreras et~al. 2009\nocite{Ferr09})}.
Counter to this thought, the various tight scaling
relations adhered to by elliptical galaxies
(e.g. Fundamental Plane - Djorgovski \& Davis 1987\nocite{DD87}; the 
colour-magnitude and Mg-$\sigma$ relationships - Bender, Burstein \& Faber
1993\nocite{BBF93}; Kuntschner 2000\nocite{K00}) argues for a very early and
coordinated formation of their stars.

This apparent contradiction can be resolved by separating the epoch of
mass assembly from that of star formation, by assuming that the final 
mergers that lead to massive early-type galaxies are between 
gas-free progenitors -- so-called ``dry mergers''.  Dry mergers and
their implication for the formation of early-type galaxies have, 
for good reason, received an inordinate degree of recent attention
(e.g. Bell et~al. 2004\nocite{Bell04};
Tran et~al. 2005\nocite{Tran05}; Khochfar \& Burkert 2005\nocite{KB05}; 
Gonz\'alez-Garc\'{\i}a \& van Albada 2005\nocite{GGvA05}; 
Faber et~al. 2007\nocite{Fab07}; Naab, Khchofar \& Burkert 
2006\nocite{NKB06}; Bell et~al. 2006\nocite{Bell06}; 
Boylan-Kolchin et~al. 2006\nocite{BK06}).  

van~Dokkum (2005; vD05, hereafter) analysed extremely deep images of red
galaxies at redshifts $z$$\sim$0.1 in the NOAO Deep Wide-Field Survey
(Jannuzi \& Dey 1999\nocite{JD99}) and MUSYC \citep{Gaw06} Survey
and found that, amongst the bulge-dominated galaxies, 71\% show tidal
features over a scale of $\sim$50~kpc. Because those features are red and
diffuse, he concluded that, most likely, they are the
consequence of interactions without associated star formation --
i.e., dry mergers.  vD05 also found that $\sim$50\%
of the systems undergoing ongoing interactions have luminosity
ratios $<$1:4 -- i.e., they are major mergers. He concluded that if the
observed galaxies in ongoing mergers are representative of the progenitor's remnants, then
$\sim$35\% of the bulge-dominated galaxies have experienced a major dry
merger in the recent past, underlying the importance of this mechanism
in the evolution of early-type galaxies. vD05 further suggests
that the observed increase in the stellar mass density of the
red sequence since z$\sim$1 could be the exclusive consequence of this
process.  However, this claimed relative importance has been questioned 
(Scarlata et~al. 2007\nocite{Scar07}; Brown et~al. 2008\nocite{Brown08};
Donovan, Hibbard \& van Gorkom 2007\nocite{DHvD07}).

\citet{DHvD07} examined the possibility that ``wet'' (gas-rich)
ellipticals might be identified as ``dry'' by the vD05 criteria. 
They found several examples of
galaxies matching the dry merger criteria that contained
significant amounts of neutral hydrogen ($\simgt$10$^8$~M$_{\odot}$),
as well as a significant degree of star formation in some
cases. This possibility was reinforced by the work by \citet{Fel08},
where it was claimed that only a merger involving a dynamically 
cold disk could produce several of the observed tidal 
tail morphologies in the vD05 sample (specifically, the arm- and loop-like  
tails found in 60\% of the major and 80\% of the minor 
red-red galaxy mergers in the sample).
In \citet{Kaw06}, we showed, using cosmological simulations, that
red, diffuse tails similar to those seen by vD05 could be the
consequence of minor mergers.   Furthermore, numerical simulations have 
shown the difficulty in reproducing 
the anisotropy and ellipticity distribution of massive early-type galaxies 
with major dry mergers (e.g, Cox et~al. 2006\nocite{Cox06}; 
Burkert et~al.\ 2008\nocite{B08}), while conversely, 
several minor dry mergers appear to do a much better job in this regard.

If these mergers are gas-free and/or if the merger remnants are the
consequence of minor instead of major mergers, the dry mergers might
not be the only, or even the dominant, mechanism producing the observed
evolution in the mass density of the red sequence as proposed by vD05.
Furthermore, if the mergers are not entirely gas-free, and a small amount
of star formation accompanied the interaction, the (nearly)-dry
mergers might also explain the presence of the trace population of
young stars seen in a large fraction of early-type galaxies and, therefore, 
the
dispersion in the $"$mean$"$ ages observed in several studies (Gonz\'alez 
1993\nocite{G93}; Trager et al.\ (2000)\nocite{T00b}; Caldwell et~al. 2003\nocite{CRC03}; 
S\'anchez-Bl\'azquez et~al. 2006\nocite{SB06c}).

If mergers are indeed central to the process of early-type galaxy
formation, then the nature of the progenitors and the characteristics
of the encounter become essential in describing the evolution of these
systems. There is not yet a systematic study of the nature of the
progenitors and of the merger remnant galaxies. Tran et~al. (2005)\nocite{Tran05}
performed spectroscopy of nine merging galaxy pairs identified by van
Dokkum et~al. (1999)\nocite{vD99} in the cluster MS1054-03 at
$z$=0.83, confirming that at least six were bound systems. They did
not, however, perform a detailed analysis of the stellar content of
these galaxies, due to the insufficient signal-to-noise ratio of the
spectra (although they did show that the 4000\AA\ break was strong,
characteristic of a population dominated by old G- and K-stars).

In what follows, we present a spectroscopic analysis of a sample of
galaxies extracted from the vD05 catalogue, in order to analyse their
stellar populations, the nature of any emission lines, and their
kinematical properties.  The questions that we wish to address are:
\it (1) Do the merger remnants contain a trace population of young
stars? \rm A large number of elliptical galaxies contain such a trace
population (e.g. Trager et~al. 2000\nocite{T00}; Thomas
et~al. 2005\nocite{Tho05}; Caldwell et~al. 2003; S\'anchez-Bl\'azquez
et~al. 2006b). The origin of this young population remains unclear --
mergers, stars formed from gas re-accreted in a galactic cooling flow
(Mathews \& Brighenti 1999\nocite{MB99}; Trager et~al. 2000) or from
satellite accretion, or star formation from gas released to the
interstellar medium by old stars (Faber \& Gallagher
1976\nocite{FG76}; Ciotti et~al. 1991\nocite{Cio91}) are but a few of
the proposed possibilities.  \it (2) Are these merger remnants
rotationally supported? \rm It has been claimed that bright early-type
galaxies supported by velocity anisotropies are the result of
dry-mergers while rotationally-supported ellipticals are the result of
mergers with gas (e.g. Cox et~al. 2006, and references therein). By
determining the degree of rotational support in these systems, we can
confront this suggestion directly.

Section~\ref{sec.observations} describes the observations and data 
reduction, Section~\ref{sec:analysis} the analysis of our spectra, 
including the determination of kinematical parameters and the measurement 
of line-strength indices, and Section~\ref{sec:results} the results of 
this analysis.
Section~\ref{sec:summary} summarises our conclusions.

\section[]{Observations and Data Reduction}
\label{sec.observations}

The galaxies analysed here are a subset of the ones presented
in vD05. This sample was selected from the 
Multiwavelength Survey by Yale-Chile (MUSYC; Gawiser et~al. 2006), and
NOAO Deep Wide-Field Survey (NDWFS; Jannuzi \& Dey 1999\nocite{JD99}).
Galaxies with magnitudes R$<$17, and colours 1.6$\le$(B$-$R)$\le$2.2 and 
with (B$-$R)$>$1.6+0.12$\times$(R$-$15), were included in our analysis. 
This corresponds, roughly, to red, early-type galaxies 
with luminosities L$>$L$^{*}$ at redshifts 0.05$\simlt$$z$$\simlt$0.2.
Galaxies which were likely members of known clusters were also removed 
from the sample.
We only observed those galaxies morphologically classified as 
bulge-dominated ellipticals or lenticulars.

vD05 found that half of the sample selected in this manner contains
low surface brightness features indicative of interactions, such as
plumes and tails.  In the original sample of vD05, consisting in 126
galaxies, 44 (35\%) show clear signs of past interactions and in an
additional 23 cases (18\%) the interaction is still in progress. Only
59 galaxies (47\%) appear undisturbed, showing no unambiguous tidal
features at the surface brightness limit of the survey.  In his
catalogue, vD05 flags those galaxies as ``strong'', ``weak'',
``none'', or ``ongoing'', depending upon the strength of the tidal
features, and in the case of ``ongoing'', indicating an ongoing
interaction with another galaxy\footnote{Note that vD05 did not
  classify all the galaxy pairs as ongoing mergers but only those
  showing some morphological signs of interactions, such as tidal
  bridges, double nuclei in a common envelope, or tidal tails or fans
  with disturbed isophotes.}. In the ``strong'' category are the
highly deformed merger remnants, whereas the ``weak'' class indicates
more subtle features. vD05 suggests that the red ongoing mergers and
the tidal features are very likely the consequence of the same
physical process seen at different times. Galaxies with strong tidal
features are seen shortly after the merger event and galaxies with
weak features are observed at later times.

\begin{figure}
\resizebox{0.5\textwidth}{!}{\includegraphics[angle=-90]{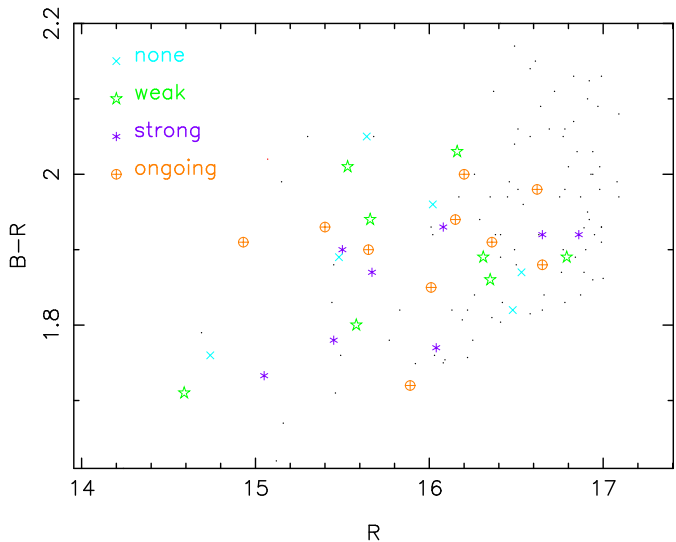}}
\caption{Colour-magnitude diagram of the 126 galaxies observed by vD05. We have marked, with different colors, 
the galaxies selected for this study. Of those, we have separated, with different symbols, galaxies
in different merger stages, as indicated in the inset.\label{sample.selection}}
\end{figure}

We selected, from the original sample, a sub-sample of galaxies
comprising those catalogued as ``ongoing'', ``strong'', ``weak'', 
and also ``none''
(unperturbed), the latter being used 
as a control sample. Table~\ref{table:sample}
summarises the main properties of the sample. 
Because we wanted to
compare galaxies with similar properties in different stages of the
merger, we tried to select galaxies with similar luminosity in the
R-band, concentrating on the the brightest of the vD05 sample.
 Figure~\ref{sample.selection} shows a colour-magnitude diagram with the 
original sample from vD05. Overplotted are the galaxies selected for this study.

Long-slit spectroscopy was obtained along the major axis\footnote{The
  position angle was obtained by fitting ellipses to the images as
  described in Section~\ref{sec:shape}},\footnote{In the ongoing
  mergers where the two galaxies met the criteria to belong to the
  vD05 sample (2-3070/2-3102; 17-681/17-596; 19-2206/19-2242), the
  slit was oriented to pass through the two galaxies.}  using the ISIS
double spectrograph mounted at the f/11 Cassegrain focus on the
William Herschel Telescope at the Roque de los Muchachos Observatory
(La Palma, Spain) on two different runs, May 19-20 2006 and May
28-June 1 2008. We made use of the R600B grating with a 2-arcsec slit
and the EEV12 detector, which provides a resolution
FWHM=4\AA~($\sigma=102$ kms$^{-1}$) in the unvignetted spectral range
3700-5260\AA. In the red arm, we used the R600R grating with the
MARCONI2 detector in the range 8100-9320\AA.  With our slit width,
this provides us with a spectral resolution of FWHM=3.6\AA.
 We only present the data from the blue arm
here, as the analysis of the Calcium Triplet will be the subject of a
future paper.  Typical exposure times for each galaxy were $\sim$2
hours, although this varied depending upon the weather conditions.

Short exposures of 40 template stars from the MILES
\citep[][]{SB06_miles} stellar library were obtained to
ensure that our data were in the same spectrophotometric system of 
that of the models employed  during the analysis.
Flat-field images were
obtained at each position of the telescope to remove fringing
in the red arm and to trace the relative response curve of the
dichroic that depends upon the position of the telescope.

\begin{table*}
\centering
\begin{tabular}{lrrrrrrrr}
\hline\hline
galaxy    & Flag      &  redshift  & \multicolumn{1}{c}{r$_{\rm eff}$}& \multicolumn{1}{c}{r$_{\rm eff}$}& \multicolumn{1}{c}{n} & \multicolumn{1}{c}{b/a}    
& \multicolumn{1}{c}{v$_{\rm max}$} & \multicolumn{1}{c}{$\sigma$} \\
          &           &             & \multicolumn{1}{c}{($``$)}& \multicolumn{1}{c}{kpc}&&& \multicolumn{1}{c}{kms$^{-1}$}& 
\multicolumn{1}{c}{kms$^{-1}$}\\
\hline
1-1403    & none      &  0.132        &  4.6$\pm$0.8  & 10.9$\pm$1.9  &  4.07$\pm$0.53&   0.68$\pm$0.02     &  20$\pm$14 & 292.6$\pm$5.5\\  
11-1014   & none      &  0.099        &  2.2$\pm$0.1  & 4.1$\pm$0.2   &  2.18$\pm$0.48  &   0.96$\pm$0.03     & 50$\pm$13 & 251.1$\pm$5.5\\
17-2031   & none      &  0.099        &  1.7$\pm$0.4  & 3.2$\pm$0.7   &  2.13$\pm$2.36  &   0.62$\pm$0.05     &  60$\pm$17 & 188.1$\pm$7.7\\
18-2684   & none      &  0.084        &  2.7$\pm$0.2  & 4.3$\pm$0.3   &  3.14$\pm$0.32  &   0.78$\pm$0.01     &  25$\pm$11 & 252.5$\pm$4.3\\
2-5013    & none      &  0.133        &  2.8$\pm$0.1  & 6.5$\pm$0.2   &  2.40$\pm$0.80  &   0.90$\pm$0.02      & 60$\pm$9 & 139.1$\pm$7.3\\
22-991    & none      &  0.085        &  2.4$\pm$0.3  & 3.7$\pm$0.5   &  5.22$\pm$1.80  &   0.70$\pm$0.01     & 25$\pm$26 & 175.6$\pm$4.5\\
\hline
10-232    & weak      &  0.126        &  1.5$\pm$1.2 &  3.4$\pm$2.7  &   2.18$\pm$0.42 &   0.90$\pm$0.05  &  0$\pm$10  & 119.8$\pm$10.0\\
12-1734   & weak      &  0.116        &  1.6$\pm$0.6 &  3.3$\pm$1.2  &   2.06$\pm$0.59 &   0.80$\pm$0.04  & 30$\pm$32  & 166.2$\pm$4.7\\
16-650    & weak      &  0.128        &  3.4$\pm$0.1 &  7.7$\pm$0.2  &   2.05$\pm$0.60 &   0.66$\pm$0.04  & 10$\pm$17  & 197.2$\pm$6.9\\
22-790    & weak      &  0.085        &  3.3$\pm$0.2 &  5.3$\pm$0.3  &   2.39$\pm$0.33 &   0.42$\pm$0.05  & 100$\pm$12 & 207.3$\pm$5.4\\
6-1553    & weak      &  0.028        &  2.6$\pm$0.4 &   1.5$\pm$0.7 &   2.37$\pm$0.80 &   0.74$\pm$0.04  & 140$\pm$15 &  148.0$\pm$1.7\\
6-1676    & weak      &  0.099        &  2.0$\pm$0.1 &  3.7$\pm$0.2  &   3.35$\pm$0.53 &   0.92$\pm$0.02  &  73$\pm$9 &  147.0$\pm$3.9\\
7-2322    & weak      &  0.130        &  5.2$\pm$1.1&  11.9$\pm$2.5  &   4.09$\pm$1.01 &   0.82$\pm$0.04  &  25$\pm$11 &  278.3$\pm$8.8\\
9-2105    & weak      &  0.084        &  2.1$\pm$0.1 &  3.3$\pm$0.1  &   2.56$\pm$0.88 &   0.95$\pm$0.02  &  40$\pm$19 &  199.0$\pm$18.6\\
\hline
10-112    & strong    &  0.128        & 2.3$\pm$0.6 & 5.2$\pm$1.3  &  1.95$\pm$2.50  &   0.69$\pm$0.07 & 160$\pm$16 & 164.4$\pm$6.6\\
1256-5723 & strong    &  0.099        & ----        &  ----          &  ----           & 76$\pm$18       &  ----     & 226.2$\pm$6.0\\
13-3813   & strong    &  0.085        & 2.9$\pm$0.3 &  4.6$\pm$0.5   &  5.55$\pm$0.35  &  0.92$\pm$0.04  & 25$\pm$20 &   209.5$\pm$5.2\\
16-1302   & strong    &  0.084        & 3.1$\pm$0.3 &  5.0$\pm$0.5   &  4.04$\pm$1.78  &  0.94$\pm$0.05  & 60$\pm$20  &  235.0$\pm$4.6\\
17-2134   & strong    &  0.084        & 3.3$\pm$0.7 &  5.2$\pm$1.1   &  2.43$\pm$0.05  &  0.61$\pm$0.03  & 45$\pm$9  &  179.1$\pm$6.7\\
8-2119    & strong    &  0.133        & 3.2$\pm$0.4 &  7.7$\pm$0.9   &  1.17$\pm$0.16  &  0.99$\pm$0.51  &140$\pm$20  &  178.2$\pm$8.2\\
3-601     & strong    &  0.153        & 2.5$\pm$0.8 &  6.6$\pm$2.1   &  2.73$\pm$0.90  &  0.60$\pm$0.02  & 60$\pm$9  &  195.3$\pm$5.5\\
5-994     & strong    &  0.077        & 3.6$\pm$0.1 &  5.3$\pm$0.1   &   2.36$\pm$0.70 &  0.83$\pm$0.02  &  0$\pm$30  &  158.6$\pm$5.5\\
9-3079    & strong    &  0.129        & 36.9$\pm$60.0& 85.1$\pm$138   &   6.46$\pm$4.0 &  0.50$\pm$0.03  & 45$\pm$80  &  279.3$\pm$13.0\\
\hline
1-2874    & ongoing   &  0.133        & 3.1$\pm$1.9 &  7.4$\pm$4.5  &    3.57$\pm$1.32 &  0.89$\pm$0.09 &  -- & 216.8$\pm$5.8\\
11-1278   & ongoing   &  0.078        & 2.1$\pm$0.1 &  3.0$\pm$0.1  &    1.85$\pm$1.19  &  0.68$\pm$0.04 & 90$\pm$21  & 74.4$\pm$33.4\\
11-1732   & ongoing   &  0.122        & 3.1$\pm$0.9 &  6.9$\pm$2.0  &      2.80$\pm$0.68    & 0.70$\pm$0.04 &  35$\pm$30 & 237.6$\pm$4.4\\
14-1401   & ongoing   &  0.099        & 2.1$\pm$0.1 &  3.9$\pm$0.2  &      2.82$\pm$0.29   & 0.98$\pm$0.09 &  -- & 198.8$\pm$8.9\\
17-596    & ongoing   &  0.084        & 32.5$\pm$40.7& 51.4$\pm$64.3 &      6.87$\pm$2.13   & 0.79$\pm$0.05 &  90$\pm$20 &  215.5$\pm$5.8\\
17-681    & ongoing   &  0.082        & 3.1$\pm$0.1  &  4.8$\pm$0.1 &     2.96$\pm$0.13    & 0.68$\pm$0.04 &   0$\pm$12 &  202.3$\pm$6.0\\
19-2242   & ongoing   &  0.121        & 2.3$\pm$0.5 &  5.1$\pm$1.1  &     2.69$\pm$0.67    &  0.77$\pm$0.09 & 25$\pm$25  &  233.5$\pm$3.2\\
19-2206   & ongoing   &  0.121        & 2.5$\pm$0.1 &  5.5$\pm$2.1  &     2.32$\pm$0.16    &  0.84$\pm$0.03 & 45$\pm$15  &  154.0$\pm$3.8\\
2-3070    & ongoing   &  0.102        & 2.4$\pm$0.1 &  4.5$\pm$0.2  &     2.15$\pm$0.19    &  0.81$\pm$0.01 & 25$\pm$11  &  191.6$\pm$8.5\\
2-3102    & ongoing   &  0.100        & 1.5$\pm$0.4 &  2.7$\pm$0.7  &     1.28$\pm$0.24    & 0.67$\pm$0.04 & 70$\pm$40  &  192.1$\pm$15.7\\
\hline
\end{tabular}
\caption{Main characteristics of the sample: 
Col.~1: galaxy name;
Col.~2:  flag describing the type of tidal features in the 
galaxy, from vD05: {\it strong}: strong tidal features
(these galaxies are generally highly-deformed merger remnants),
{\it weak}: indicates more subtle features, such as tidal tails or shells,
{\it ongoing}: ongoing interaction (both
the primary and secondary galaxy show clear distortions or 
tidal tails); {\it none}: undisturbed galaxy;
Col.~3: measured redshift; 
Col.~4: effective radius in arcsec;
Col.~5: effective radius in kpc; 
Col.~6: S\'ersic index;
Col.~7: Ratio between minor and mayor axis calculated with GALFIT;
Col.~8: maximum rotational velocity (calculated only in
those cases where a 
regular rotation curve is observed -- see appendix~\ref{rotationcurves});
Col.~9: velocity dispersion  measured within one effective radius; 
\label{table:sample}}
\end{table*}

\subsection{Data Reduction}

The standard data reduction procedures (flat-fielding, cosmic ray
removal, wavelength calibration, sky subtraction, and flux calibration) were
performed with \reduceme~ \citep{Car99}. This reduction package allows
a parallel treatment of data and error frames and, therefore, produces
an associated error spectrum for each individual data spectrum.
Initial reduction of the CCD frames involved bias and dark current
subtraction, the removal of pixel-to-pixel sensitivity variations
(using flat-field exposures of a tungsten calibration lamp), and
correction for two dimensional low-frequency scale sensitivity
variations (using twilight sky exposures).  The dichroics in the ISIS
spectrograph produce an intermediate frequency pattern
\citep[see][]{SB06a} which varies with the position of the telescope.
As we were aware of this pre-existing problem from our earlier work, 
we acquired flat field images for every
position of the telescope.  Using these images, the fringing pattern
was eliminated during the flat-fielding.  Prior to the wavelength
calibration, arc frames were used to correct the C-distortion in the
images.  This rotation correction guarantees alignment errors to be
below 0.1 pixel. Spectra were converted to a linear wavelength scale
using typically $\sim$100 arc lines, fitted by third-order polynomials.
The Root-Mean-Square (RMS) errors of the fits were $\sim$0.1~\AA. All
spectra were also corrected for S-distortion.  Atmospheric extinction
was calculated using the extinction curve provided by the Observatory.
To correct for the effect of interstellar extinction, we used the
curve of Fitzpatrick (1999)\nocite{Fit99}.  Foreground reddenings were
obtained using the dust maps by Schlegel, Finkbeiner \& Davis
(1998)\nocite{Sch98}. We used the IDL subroutines provided by the
authors to read the maps and extract and E(B$-$V) for each galaxy
depending upon their Galactic coordinates.  We corrected the continuum
shapes of our spectra using exposures of 8-10 (depending on the run)
standard stars. The final response curve was obtained as the average
of all the individual curves while the difference between curves was
used to derive an error in the indices due to flux calibration.

We extracted central spectra within an aperture of one effective 
radius (r$_{\rm eff}$).
Effective radii were calculated by fitting the two-dimensional 
spatial profile with a S\'ersic law --as explained in Sec.~\ref{sec:shape} --
 the values for which
are listed in Table~\ref{table:sample}.
To study the rotational support of these galaxies, we also extracted
spectra along the radius of each galaxy.  We binned in the spatial
direction to obtain a minimum signal-to-noise ratio per \AA~of 20 in
the region of H$\beta$.

\section{Measurements}
\label{sec:analysis}
\subsection{Kinematics, velocity dispersion, and emission line removal}
\label{kinematics}

Velocity dispersions and radial velocities for the galaxies were
measured using the IDL routine PPXF (Cappellari \&
Emsellem 2004).  This routine applies a maximum
penalised likelihood formalism to extract as much information as
possible from the spectra while suppressing the noise in the solution.

Surveys of large samples of bulge-dominated galaxies have revealed
that 50-60\% of these galaxies show weak optical emission lines
\citep{Cal84, Phi86, Goud94}. The measurement of several Lick/IDS
indices can be affected by the presence of these lines. The effect on
the Balmer lines is to lower the value of the index and, therefore,
increase the derived mean age.  To obtain emission line
amplitudes and to separate the relative contribution of the stellar
continuum from nebular emission in the spectra, we combine the PPXF
routine with GANDALF \citep{Sar06}. This latter algorithm
fits simultaneously the emission line kinematics using the optimal
template convolved with the broadening function derived from PPXF.
One of the critical steps for the accurate measure of both emission
lines and velocity dispersions is the calculation of a good template
that reproduces the observed spectra. To build these templates we use a
library of stellar population models by Vazdekis et~al. (2009, in
preparation) based on the MILES stellar library \citep{SB06_miles,
Cen07_miles}. We include models with a wide range of ages (from 0.5
to 17~Gyr) and metallicities ([Z/H]=$-1.68$ to $+$0.2).

To estimate errors in the measured parameters, we perform 
Monte Carlo simulations
repeating the measurement process for spectra perturbed with Gaussian
noise, the latter for which is derived from the \reduceme~ error spectra. 
Fifty simulations were performed and, each time, a new optimal template was
calculated.

The rotation curves are shown in Appendix~\ref{rotationcurves}. A
maximum rotational velocity (v$_{\rm max}$) was derived from these
curves as half the difference between the peaks on the rotation
curve.  An error estimation of this value was done 
using the derived error bars for the velocity values and the difference between the maximum rotational 
velocity measured at the two sides from the centre of the galaxy. 
Some of the galaxies do not show regular rotation curves. In
particular U-shaped curves, as predicted in merger simulations \citep{Com95} 
are observed in some of the ongoing mergers. In those
cases, a v$_{\rm max}$ was not derived.  The v$_{\rm max}$ values and
the velocity dispersion inside r$_{\rm eff}$ are shown in
Table~\ref{table:sample}.

\subsection{Structural parameters}
\label{sec:shape}
\begin{figure*}
\resizebox{\textwidth}{!}{\includegraphics[angle=-90]{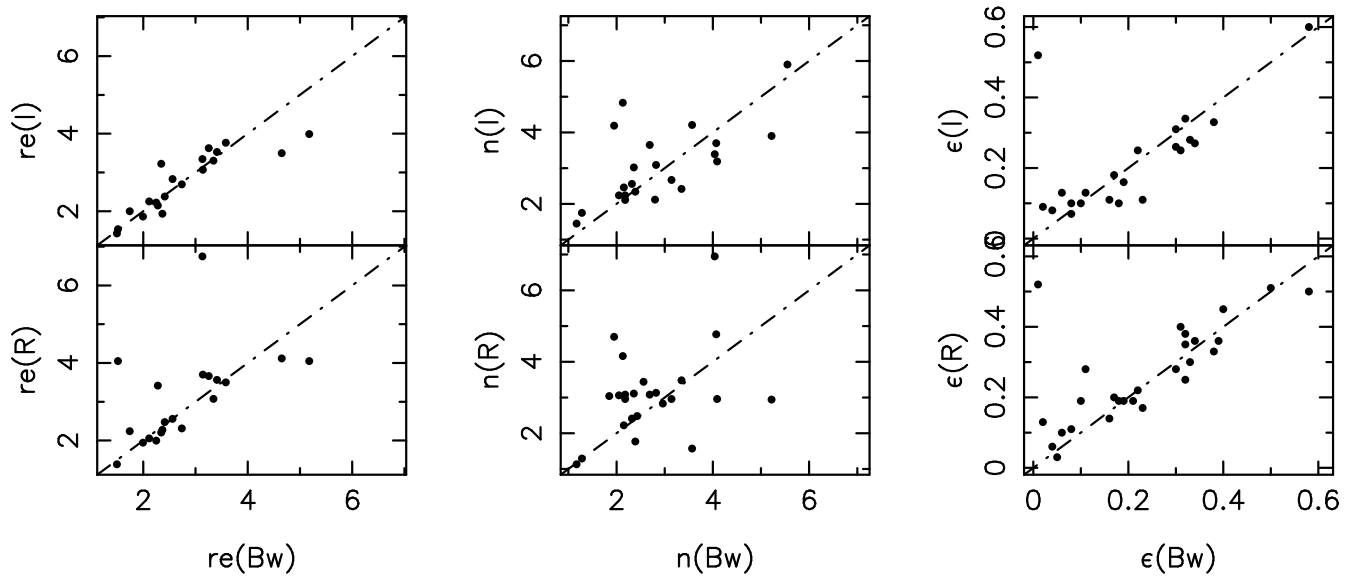}}
\caption{Comparison of the structural parameters derived in different band-passes.\label{comparison.bands}}
\end{figure*}

We used the code GALFIT (Peng et al.\ 2002)
to estimate structural parameters, i.e., ellipticities ($\epsilon$), 
effective radius (r$_{\rm eff}$) and S\'ersic indices, using the 
 publicly-available images from the Third Data Release
of the NOAO Deep Wide-Field Survey. The images were obtained with the
KPNO Mayall 4m telescope and MOSAIC-1 imager. We use the images in the
Bw-band (the Bw filter has a bluer central wavelength than the
standard Johnson B-band filter). 
GALFIT convolves S\'ersic (1968) r$^{1/n}$ galaxy models with the 
PSF of the images, and determines the best fit by comparing the 
convolved model with the galaxy surface brightness distribution
using a Levenberg-Marquardt algorithm to minimise the $\chi^2$
of the fit.
 The PSF has been obtained from isolated stars, as close as possible to the 
observed galaxies.
The S\'ersic index n measures the shape of the surface brightness
profiles, where n=1 represent an exponential and n=4 a de Vaucouleurs 
profile. During the fit, neighbouring galaxies were excluded using a mask, 
but in the case of closely neighbouring objects or interacting galaxies, both 
objects were fitted simultaneously.
The measured parameters with this process are listed in Table~\ref{table:sample}.
 We have tested the internal consistency of our data, comparing the size 
and shape of our galaxies in different bands, i.e., Bw, R and I. The seeing and 
depth of those images are slightly different, which allows us to test the 
robustness of our derived parameters. Of course, this test is only valid under
the assumption that the change in size and shape of the light profile of the galaxies
due to changes in the wavelength along the given filters is smaller than the intrinsic
error in estimating the structural parameters.
Fig.~\ref{comparison.bands} shows the comparison between the sizes ,  S\'ersic 
indices and ellipticities calculated in the different bands.
We estimate the error in the parameters as the mean difference between the parameters
 in the different bands.
 
\subsection{Line-strength indices}
\label{sec:linestrength}

We measure Lick/IDS line-strength indices as defined by 
Trager et~al. (1998)\nocite{T98} with the additional definitions
by Worthey \& Ottaviani (1997)\nocite{WO97} for the higher-order
Balmer lines H$\delta$ and H$\gamma$. Our wavelength range did not allow us
to measure lines such as Mgb, Fe5270, and Fe5335 for most of the galaxies.
In particular, we were not able to measure any feature with high sensitivity to 
Mg variations, which limits our ability to obtain an accurate
estimates of both [Mg/Fe] and the ages.
Errors in the indices were estimated from the uncertainties caused by 
photon noise, wavelength calibration, and flux calibration.

Lick/IDS indices depend on the broadening of the spectra due to the
blending of lines in the pass-band definition. In order to compare the
indices of different galaxies with each other these have to be
corrected to the same level of intrinsic Doppler and instrumental
broadenings. When comparing with stellar population models based on
the Lick/IDS system (Vazdekis 1999\nocite{V99}; Worthey
1994\nocite{w94}; Thomas et~al. 2003\nocite{TMB03}) one must follow
two steps: (1) degrade the spectra to the wavelength-dependent
instrumental resolution of this library, and (2) make a correction due
to the velocity dispersion of the galaxy.  The broadening correction
depends on the strength of the indices and is difficult to apply
accurately. In most cases, it does introduce systematic effects in the
final measurements which, in some cases, can dominate the final 
error (see, e.g., Kelson et~al. 2006)\nocite{Kel06}.

In this paper we do not use models based on the Lick system but the
new models by Vazdekis et~al. (2009). The latter reflect a 
significant improvement upon the earlier models of
the Vazdekis et~al. (2003)\nocite{v03}, due to the use of the new MILES stellar library
(S\'anchez-Bl\'azquez et~al.\ 2006\nocite{SB06_miles}) and its
associated extended 
coverage in stellar atmospheric parameter-space.
As the library is fully flux-calibrated the models not
only predict spectral features, but the entire spectrum from 3500 to
7500\AA.  The use of these models allows us to choose the resolution
at which we want to measure our indices. After careful consideration,
we decided to measure all our indices in spectra with a total
broadening of 300~km~s$^{-1}$.  All the spectra were broadened by 
$\sigma_{\rm broad}$, such that
300=$\sqrt{\sigma_{gal}^2+\sigma_{ins}^2+\sigma_{\rm broad}^2}$, 
where $\sigma_{gal}$ is 
the velocity dispersion of the galaxy and $\sigma_{\rm ins}$ the instrumental 
resolution in km~s$^{-1}$.  By doing so, we avoid
any further correction due to the velocity dispersion of the galaxies
and, therefore, avoid the introduction of systematic errors, as
mentioned above.

Our sample contains galaxies with a range of redshifts, and so 
a fixed aperture samples
different regions of each. 
Despite having added spectra within 
one r$_{\rm eff}$ for all the galaxies, we still have 
to make a correction due to aperture effects, as the width of the slit is 
fixed to 2~arcsec.
To do this we have made used of the aperture corrections derived by 
S\'anchez-Bl\'azquez et~al. (2009)\nocite{SB09} obtained using a large sample of line-strength gradients from 
S\'anchez-Bl\'azquez et~al. (2006c; 2007)\nocite{SB06c}\nocite{SB07}.
The aperture correction of the atomic indices can be written 
as (J\o rgensen 1995)\nocite{Jor95}:

\begin{equation}
\log(I_{reff})=\log(I_{med})+\alpha \log \frac{r_{\rm med}}{r_{eff}}
\end{equation}

\noindent
where $r_{\rm med}=1.025 \times \sqrt(\frac{2 \times r_{\rm eff}}{\pi})$,
while that of the molecular indices and the higher-order Balmer lines 
can be written as:

\begin{equation}
(I_{reff})=(I_{med})+\alpha \log \frac{r_{\rm med}}{r_{eff}}
\end{equation}

\noindent
This aperture correction is very uncertain, as elliptical galaxies show a 
large scatter in their gradients and there are no clear correlations between 
the line-strength gradients and other parameters of the galaxies. However, 
the aperture corrections for our 
galaxies are very small and none of our results change should the aperture
correction not be applied.
Table~\ref{tab:alpha} shows the $\alpha$ coefficient for all the indices, 
while Table~\ref{line-strength} in appendix~\ref{sec:linestrength} shows the fully-corrected indices 
at $\sim$300~kms$^{-1}$ resolution.

\begin{table}
\centering
\begin{tabular}{lrr}
\hline\hline
Index       &  $\alpha$       & Type\\
\hline
H$\delta_A$ & $-0.721\pm 0.508$ &  1\\
H$\delta_F$ & $-0.206\pm 0.249$ &  1\\
CN$_2$      & $ 0.059\pm 0.030$ &  2\\
Ca4227      & $ 0.047\pm 0.075$ &  1\\
G4300       & $ 0.027\pm 0.022$ &  1\\
H$\gamma_A$ & $-0.850\pm 0.422$ &  2\\
H$\gamma_F$ & $-0.436\pm 0.321$ &  2\\
Fe4383      & $ 0.061\pm 0.033$ &  1\\
Ca4455      & $ 0.097\pm 0.075$ &  1\\
Fe4531      & $ 0.040\pm 0.021$ &  1\\
C4668       & $ 0.119\pm 0.049$ &  1\\
H$\beta$    & $-0.050\pm 0.190$ &  1\\
\hline  
\end{tabular}
\caption{$\alpha$ coefficient and corresponding error for the 
aperture correction. The final column indicates whether the correction 
is additive (1) or multiplicative (2).\label{tab:alpha}}
\end{table}

\section{Analysis}
\label{sec:results}

\subsection{Nature of the emission lines}

Active nuclei may play a role during mergers, preventing the gas from
cooling and forming stars and, therefore, leaving the remnant 
red (Kawata et~al. 2006\nocite{Kaw06}; Springel, di~Matteo \& 
Hernquist 2005\nocite{SdMH05}).  It is therefore interesting
to compare the degree of nuclear activity in the merger remnants
with those of the undisturbed galaxies.

To examine the possible presence of nuclear activity in our sample, we
adopt here the same criteria as in \citet{Sar06} and consider that
emission is present when the line protrudes above the noise in the
stellar spectrum by more than a factor of four times for all the lines,
except for
H$\beta$, for which we lower this to three.  \citet{BPT81} proposed
a suite of three diagnostic diagrams to classify the dominant energy
source in emission line galaxies. These diagrams are based on the four
optical line ratios [OII]/H$\beta$, [NII]/H$\alpha$, [SII]/H$\alpha$,
and [OI]/H$\alpha$.  We do not possess sufficient lines in our wavelength
range to plot our galaxies in several of the classical diagnostic
diagrams.  However, we can infer approximately the nature of
ionisation in several of them.  Yan et~al. (2006)\nocite{Yan06} studied
the source of ionisation in a sample of red-sequence galaxies from the
Sloan Digital Sky Survey (SDSS). They showed that galaxies with
high-values of [OII]/H$\alpha$ have relatively uniform line ratios in
[OII]/H$\alpha$, [OIII]/H$\beta$, [NII]/H$\alpha$, [OI]/H$\alpha$, and
[OIII]/[OII], and were identifiable as LINER-type objects. Galaxies with lower
values of [OII]/H$\beta$ could be star forming galaxies, transition
objects, or Seyferts.  The wavelength range of our spectra does not
allow us to measure H$\alpha$ so, instead, we use H$\beta$.  Yan 
et~al. (2006) calculate a median value of H$\alpha$/H$\beta$=4.46 before
applying any reddening correction to 
their sample of red-sequence galaxies.
The demarcation proposed by these
authors to separate star forming galaxies from other categories is:
EW[OII]$<$18~EW(H$\beta$)$-6$.  To separate between LINERs and Seyferts
we use the theoretical line derived by \citet{KE08} based on the ratio
[OIII]/[OII], which is an indicator of the ionisation parameter of the
gas. Roughly, Seyfert galaxies show ratios [OIII]/[OII] $>$ 1, while
lower ratios are characteristic of LINERs.  Table~\ref{emission:table}
shows  the equivalent widths of the different emission lines calculated for our sample of
galaxies. 

We see an enhanced occurrence of emission lines in merger
remnants and ongoing mergers with respect to the sample of undisturbed
galaxies, although the number of galaxies is small to make definitive
conclusions.

It can be seen in Table~\ref{emission:table} that most galaxies
showing emission lines have line-ratios compatible with being LINERs or Seyferts.
 None of the spectra present line ratios
 consistent with ongoing star formation, although $\sim$35\% of the sample
 showing emission could not be accurately classified due to
 their insufficient detection.  Our classifications would not be
 affected for smaller assumed values of H$\alpha$/H$\beta$, however if the
 ratio was significantly larger, perhaps due to the presence of
 dust, the incidence of star forming galaxies could be larger.
We conclude that, under the limitations of our sample and assumptions,
we do not detect an enhancement of
{\it current} star formation line-ratios in galaxies now experiencing a
merger (or with strong signs of recent interactions).

\begin{figure}
\resizebox{0.45\textwidth}{!}{\includegraphics[angle=-90]{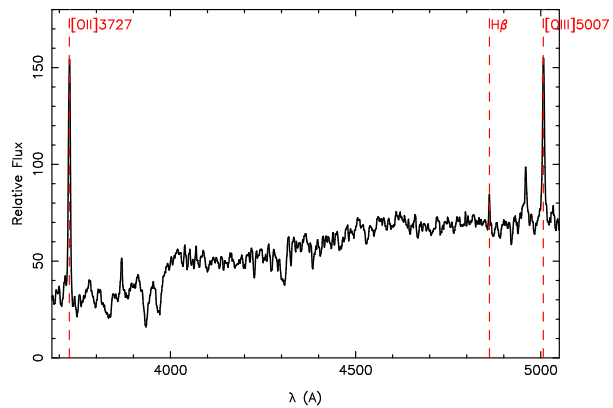}}
\caption{Spectra of 10-232, a galaxy with LINER-type emission.\label{example}}
\end{figure}

An example of a galaxy spectrum with
LINER-type emission is shown in Figure~\ref{example}.  The origin of
emission lines in LINERs, (low-ionisation narrow emission line regions
-- Heckman 1980) is still unknown.  An AGN power source, fast shocks,
photoionisation by hot stars, or photoionisation by an old metal-rich
population, are several of the proposed possibilities (see Kewley
et~al. 2006 for a discussion). Most likely, the current definition of
LINER encompasses more than one type of ionisation mechanism.
However, in a sample of red galaxies extracted from SDSS, Graves
et~al. (2006)\nocite{Grav07} found that those with LINER-like emission were
systematically 10-40\% younger than their emission-free counterparts,
suggesting a connection between the mechanism powering the emission
(the possibilities being AGN, post-AGB stars, shocks, or cooling
flows) and more recent star formation in the galaxies.  The enhanced
emission in merger remnants in our sample suggests that mergers might be the triggering 
mechanism for both star formation and emission, producing gravitational torques capable 
of driving radial flows towards the centre of the galaxies (Hernquist 1989; Barnes \& Hernquist 1991).

\begin{table*}
\begin{tabular}{lrrrrr}
\hline\hline
gal       &       &   EW[OII] &  EW H$\beta$& EW([OIII]) & class\\
          &       &    (\AA)  &   (\AA)     &  (\AA)     &      \\
\hline
1-1403    & none  &   No    &  No   &  No     & No emission \\
11-1014   & none  &   No    &  No   &  No     & No emission \\
17-2031   & none  &   No    &  No   &  No     & No emission\\
18-2684   & none  &   --    &  No   &  No     & No emission\\
2-5013    & none  &   7.9   &  No   &  No     & LINER \\
22-991    & none  &   No    &  No   &  No     & No emission\\
\hline
10-232    & weak  &   6.21    &  2.97   &  12.52     & Seyfert\\ 
12-1734   & weak  &   No      &  No   &    No     & No emission\\
16-650    & weak  &   0.58    &  --   &   ---    &   ?\\
22-790    & weak  &   --      &  No  &     No     &  No emission\\
6-1553    & weak  &   --      &  No  &     0.76     & ?\\
6-1676    & weak  &   0.71    &  No   &   1.25   & Seyfert\\
7-2322    & weak  &   0.45    &  --   &  --  &       ?\\       
9-2105    & weak  &   --    &   1.72  &   4.71   &   ? \\
\hline
10-112    & strong&   0.47   &  No  &  No     & LINER \\
1256-5723 & strong&  2.80    &  No  &  1.15     & LINER\\
13-3813   & strong&   --    &  No  &  No     &  No emission\\
16-1302   & strong&   --    &  0.93  &  1.74    &  ?\\
17-2134   & strong&   --    &  2.99  &   1.56    & ? \\
8-2119    & strong&   0.71    & 0.81  &   --    &  LINER/Seyfert\\ 
3-601     & strong&   0.70    &  --  &   --    &   ?  \\
5-994     & strong&   --    &  No   &  0.65    & ?\\
9-3079    & strong&   No    &  No  &   No    &  No emission \\ 
\hline
1-2874    & ongoing&   No   &  No   &  No    &   No emission\\
11-1278   & ongoing&   --   &  No   &  1.37    &    ? \\
11-1732   & ongoing&   3.5   & 0.77   & 0.85    &    LINER\\
14-1401   & ongoing&   0.80   &  No   &  1.2   &    Seyfert\\
17-596    & ongoing&   --   &  No   &   No   &    No emission\\
17-681    & ongoing&   --   &  No   &   No   &    No emission\\  
19-2242   & ongoing&   No   &  No   &   No   &    No emission\\
19-2206   & ongoing&  0.39   &  No   &   No   &    LINER\\
2-3070    & ongoing&   2.37   &  0.81  &   1.49   &    LINER\\
2-3102    & ongoing&   1.69   &  No   &  1.33    &   LINER  \\ 
\hline
\end{tabular}
\caption{Emission line detection in our sample of galaxies; First and 
second column give the name of the 
galaxy and the characteristics of the morphological perturbations. 
Columns 3, 4, and 5 indicate the detection (or not)
of emission in [OII]$\lambda\lambda$3727, H$\beta$, and 
[OIII]$\lambda\lambda$5007. We consider a detection when 
the line protrudes more than a factor of three above the 
noise.\label{emission:table}}
\end{table*}

\subsection{Rotational support}

Major dry and wet mergers, and minor mergers, can all produce red, 
bulge-dominated galaxies, but the kinematical structure of the remnant is
expected to be very different (Cox et~al. 2006\nocite{Cox06}; Bornaud, Jog \& Combes 2005\nocite{BJC05};
Naab \& Burkert 2003\nocite{NB03}).  Observations indicate that most luminous
galaxies show box-shaped isophotes and little rotation, while less
luminous spheroids show disk-shaped isophotes and exhibit rotation
along the photometric major axis (Davies et~al. 1983\nocite{Dav83};
Bender et~al. 1989\nocite{Ben89}; Faber et~al. 1997\nocite{Fab97};
Kormendy \& Bender 1996\nocite{KB96}).  Major dry mergers cannot
produce rotating remnants (Cox et~al. 2006, although see Gonz\'alez-Garc\'{\i}a \& Balcells 2005 for a different
point of view), while minor mergers or
mergers with a gaseous component can both produce 
rotationally~supported bulge dominated galaxies 
(Naab \& Burkert 2003; Cox et~al. 2006).

\begin{figure}
\resizebox{0.45\textwidth}{!}{\includegraphics[angle=-90]{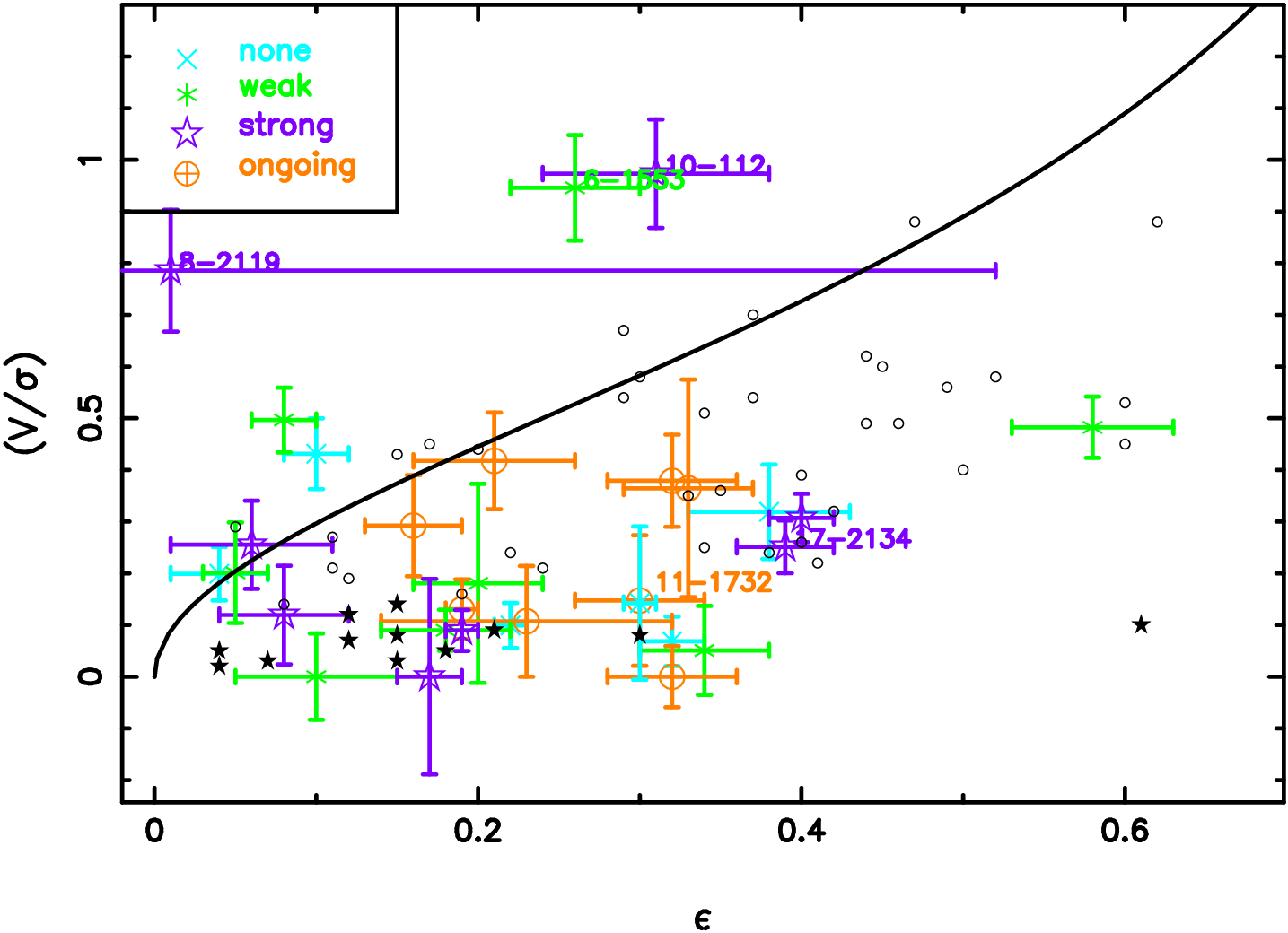}}\vspace{0.3cm}

\resizebox{0.45\textwidth}{!}{\includegraphics[angle=-90]{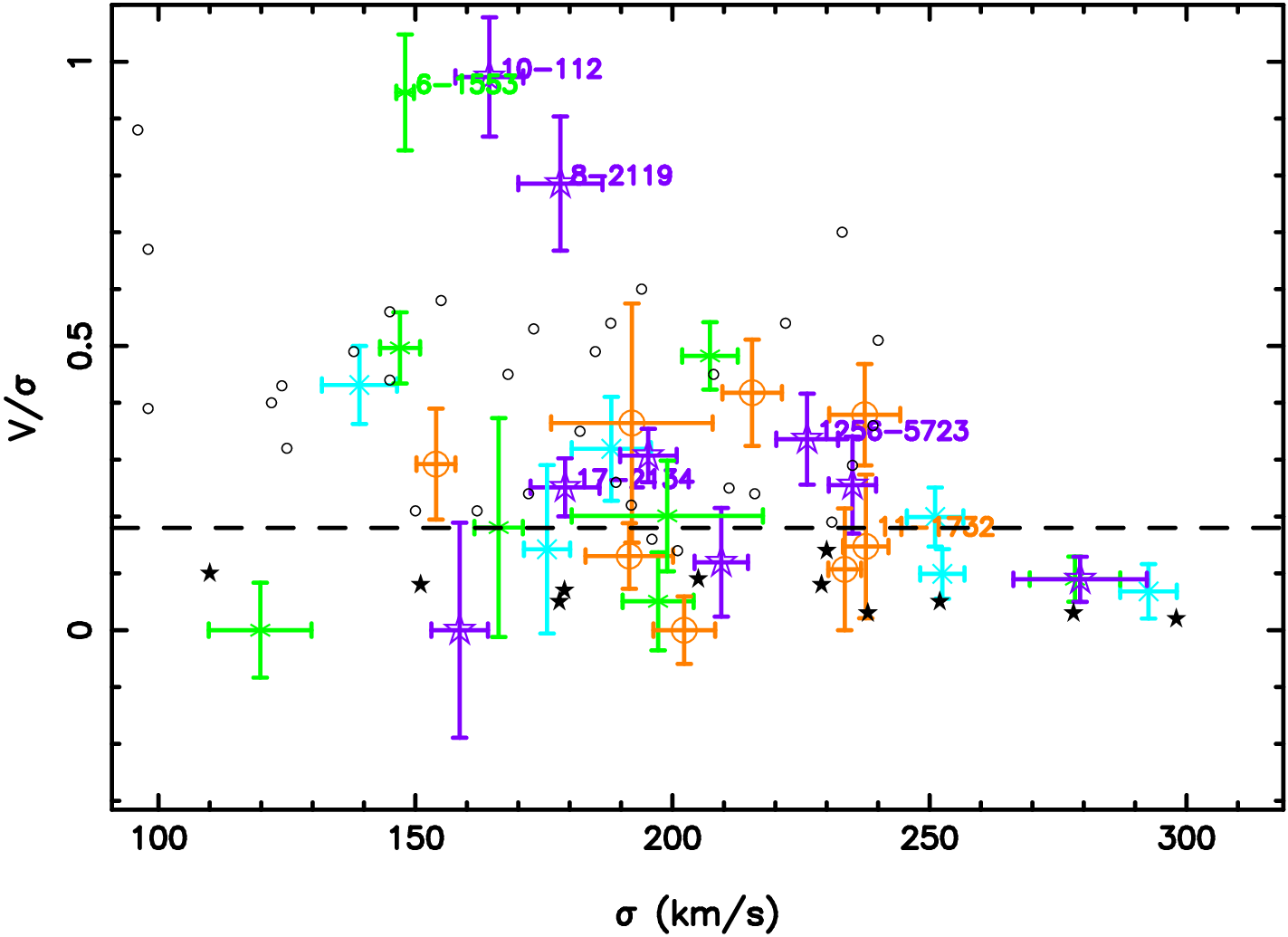}}
\caption{Top Panel: (V$_{\rm max}/\sigma$, $\epsilon$) diagram for our
  sample. Small open circles and filled stars correspond to the fast-
  and slow-rotators from the SAURON database, respectively. The line
  corresponds to the relation expected for an isotropic, oblate,
  edge-on rotators, in the revised formalism for integral field
  kinematics of Binney (2005).  Lower panel: (Vmax/$\sigma$, $\sigma$) 
  diagram for our sample. The dashed line show the approximate demarcation between 
 slow- and fast-rotators.\label{fig.binney}}
\end{figure}

Figure~\ref{fig.binney} shows the relation between v$_{\rm max}$ and
$\sigma$ (v/$\sigma$ in the plot) against ellipticity, the so-called
anisotropy diagram. This diagram quantifies the relative
contribution of ordered and random motions to the overall kinematics of
each galaxy.  The solid line indicates the expected relationship for an
oblate spheroid with an isotropic velocity distribution.  Recently,
the SAURON collaboration have shown that the anisotropy diagram is not
entirely adequate to describe the kinematic structure of spheroidal
galaxies (Emsellem et~al. 2007\nocite{Ems07}; Cappellari et~al. 2007\nocite{Cap07}).  They define
a new quantity, $\lambda_R$, that is linked to the baryonic angular
momentum.  Using this parameter, they separate the galaxies
into slow- ($\lambda_R < 0.1$) and fast-rotators ($\lambda_R> 0.1$).
2D spectroscopy is necessary to measure the $\lambda_R$ parameter, but
Cappellari et~al. (2007) show that fast- and slow-rotators occupy
different regions of the anisotropy diagram.  To compare with this
study, we have plotted in Figure~\ref{fig.binney}, the slow- and fast-rotators
from the SAURON database (Emsellem et~al. 2007).

As can be seen, our sample shows different degrees of rotational
support. 
 Two out of eight galaxies with strong signs of recent interactions --
for which we could measure ellipticities -- are in the same zone 
of the diagram as the SAURON fast rotators. Two others deviate from the 
region populated by early-type galaxies towards much higher rotational 
velocities while another two occupy the same regions as the slow-rotators
from the SAURON sample. The rest of the galaxies in this sub-sample (two)
lie in the transition region between slow- and fast-rotators.
Clearly, the  nature of the merger remnants is not unique, reflecting, 
most likely, the different nature of the progenitors.
Half of the ``weak''
and ``undisturbed'' galaxies, and half of the systems in ``ongoing'' 
mergers, are also compatible with being supported by rotation

Among ellipticals, those with higher
$\sigma$ tend to be supported by velocity anisotropies while those
with small $\sigma$ tend to be supported 
by rotation (Bender et~al. 1998)\nocite{Ben98}.  Therefore, the
percentage of rotationally-supported galaxies might depend on the
$\sigma$ of each sub-sample.  To explore if this is the case, we have
also plotted in Fig.~\ref{fig.binney} the relation between (v/$\sigma$)
and $\sigma$. It can be seen that, in agreement with the SAURON
sample, low-rotators span a wide range of $\sigma$  and  there are
no fast rotators galaxies with $\sigma> 250$ kms$^{-1}$.  The
distribution of galaxies in the v/$\sigma$ diagram {\it is} very similar to
that of the SAURON galaxies, independent of the morphological
disturbances of the sample The only exceptions are three merger
remnants (two in the ``strong'' group and one in the ``weak'')
showing a very high value of v/$\sigma$.   
The very high (v/$\sigma$) values can be explained if there is a decrease
in the central $\sigma$ due to the formation of a cold disc 
(so-called $\sigma$-drops). The standard explanation for these drops is
the presence of cold central stellar discs originating from gas inflow
(Emsellem et~al. 2001\nocite{Ems01}; Wozniak et~al. 2003\nocite{Woz03}).
\cite{WC06} predict that these features last until the fresh gas ceases
flowing into the centre to form stars. Therefore, these are not 
long-lived.  A scenario where  these features get formed during the merger process, when
there is enough gas, but do not last
very long,  
could explain  why they do not appear in the SAURON
sample.

To summarize, with the exception of three galaxies with very high
v/$\sigma$, the distribution of merger remnants in the anisotropy
diagram is not different to that of a sample of normal early-type
galaxies. We do not find that {\it all} of our remnants are supported by
velocity anisotropies, as would be expected if they were all remnants
of major dry mergers (Cox et~al. 2006), but some of them are.

\nocite{Binney05}

\begin{figure}
\resizebox{0.45\textwidth}{!}{\includegraphics[angle=-90]{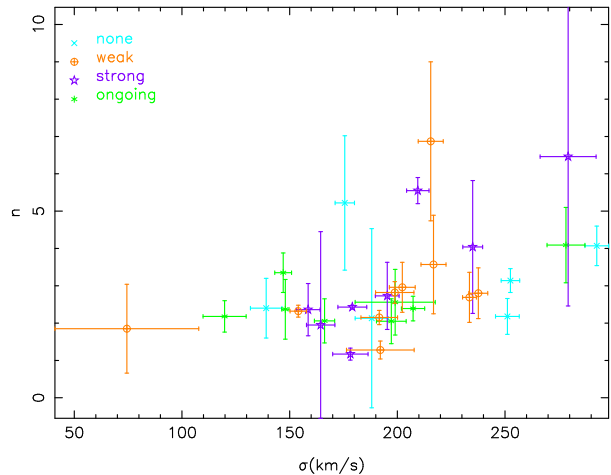}}
\caption{Relation between the S\'ersic index (n) and the central velocity 
dispersion for our sample of galaxies.\label{fig:ser_sigma}}
\end{figure}

Figure~\ref{fig:ser_sigma} shows the relation between the S\'ersic index and the 
central velocity dispersion for our sample of galaxies. As can be seen, we also 
do not find any difference between the distribution of S\'ersic indices for the 
galaxies in different merger stages and the control sample.
\subsection{Index-$\sigma$ relations}
\label{sec:indices}

Elliptical galaxies show tight correlations between line-strength indices 
and the central velocity dispersions;
Figure~\ref{index.sigma.fig} show this relation 
for our sample.
We also show a linear fit to the sample of undisturbed
galaxies.  As can be appreciated, 4 of the 9 galaxies
with strong morphological signs of recent merger activity
show systematically weaker metallic lines and stronger Balmer lines
at a given velocity dispersion than those undisturbed or with only
weak signs of recent interactions.  One of the galaxies in an ongoing
merger, 11-1732, also shows these strong deviations from the fit defined
by the undisturbed galaxies.

\begin{figure*}
\resizebox{0.4\textwidth}{!}{\includegraphics[angle=-90]{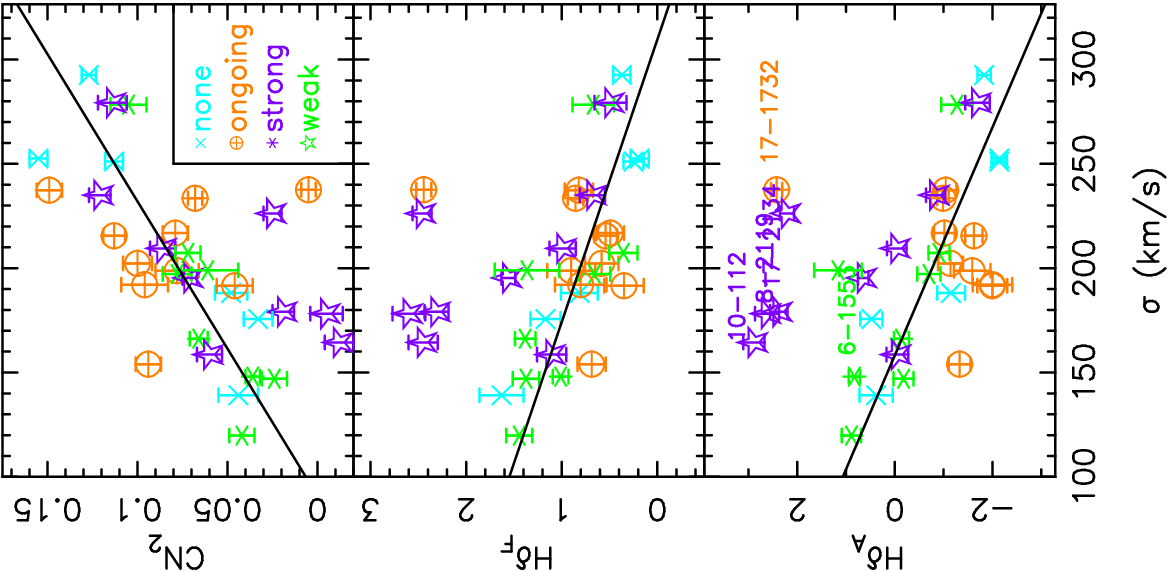}}
\hspace{0.3cm}
\resizebox{0.4\textwidth}{!}{\includegraphics[angle=-90]{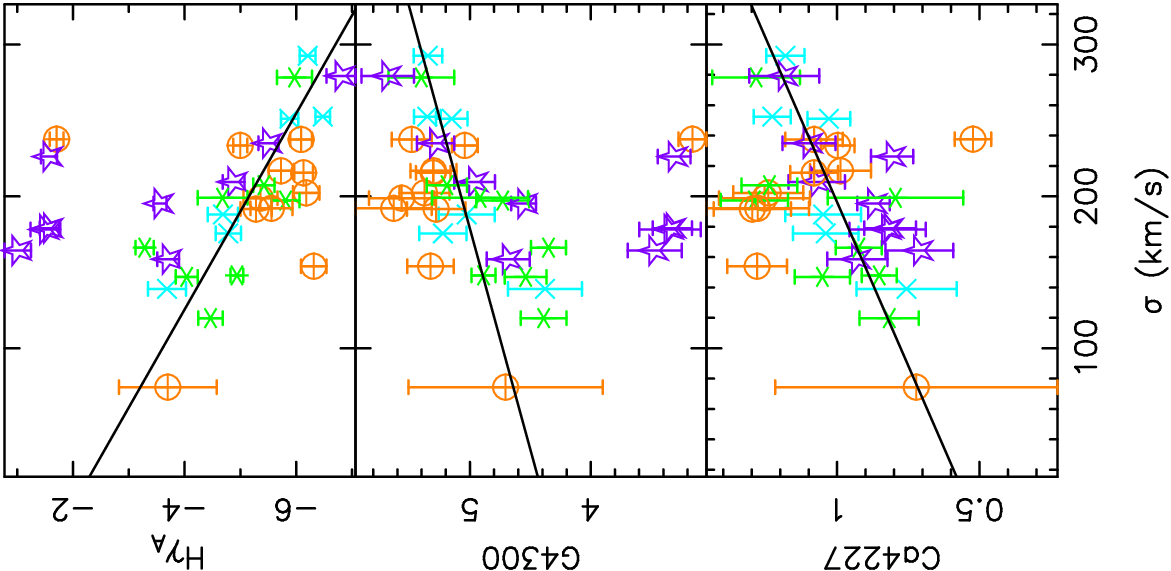}}
\caption{Relation between line-strength indices and velocity dispersion 
measured in an aperture of one effective radius. Galaxies with no or weak 
signs of recent interactions are represented with blue and green crosses, 
respectively; galaxies with strong signatures of recent interactions are 
plotted with purple stars, while ongoing mergers are represented in orange. 
A linear fit to the galaxies without any signs of recent interactions 
(labeled as ``none'') is also shown.\label{index.sigma.fig}}
\end{figure*}
\addtocounter{figure}{-1}
\begin{figure*}
\resizebox{0.4\textwidth}{!}{\includegraphics[angle=-90]{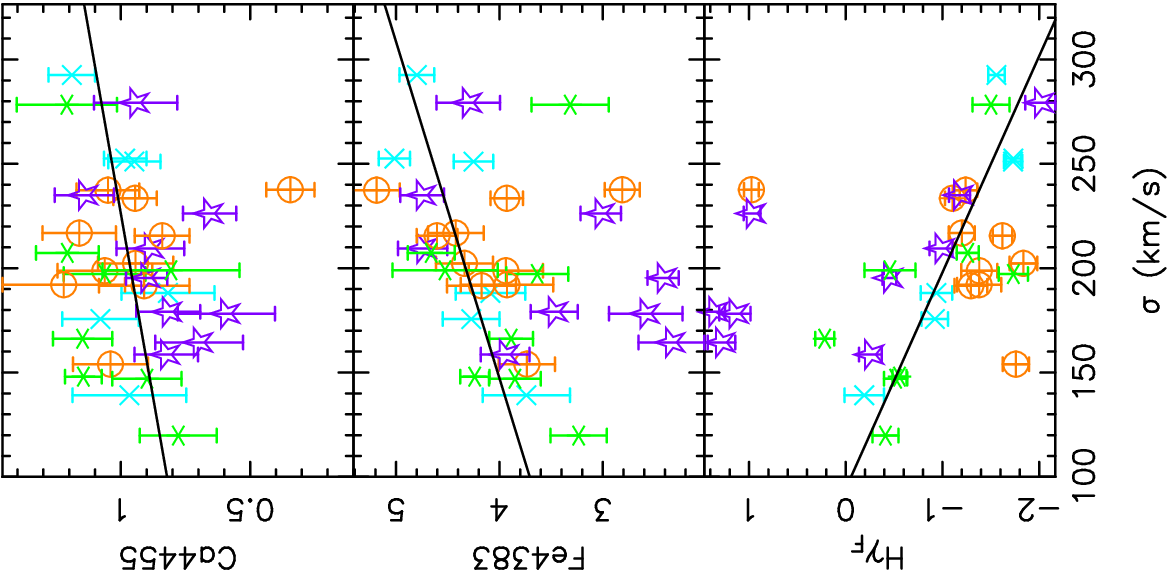}}
\hspace{0.3cm}
\resizebox{0.4\textwidth}{!}{\includegraphics[angle=-90]{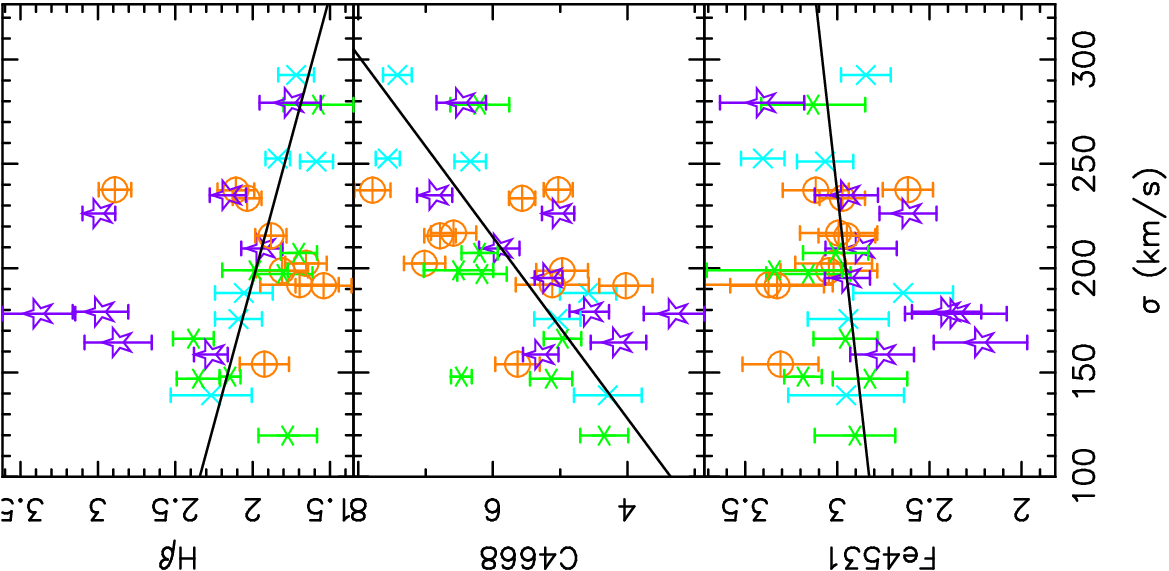}}
\caption{Continued}
\end{figure*}

\subsection{Comparison with SSPs}

\begin{figure*}
\resizebox{0.4\textwidth}{!}{\includegraphics[angle=-90]{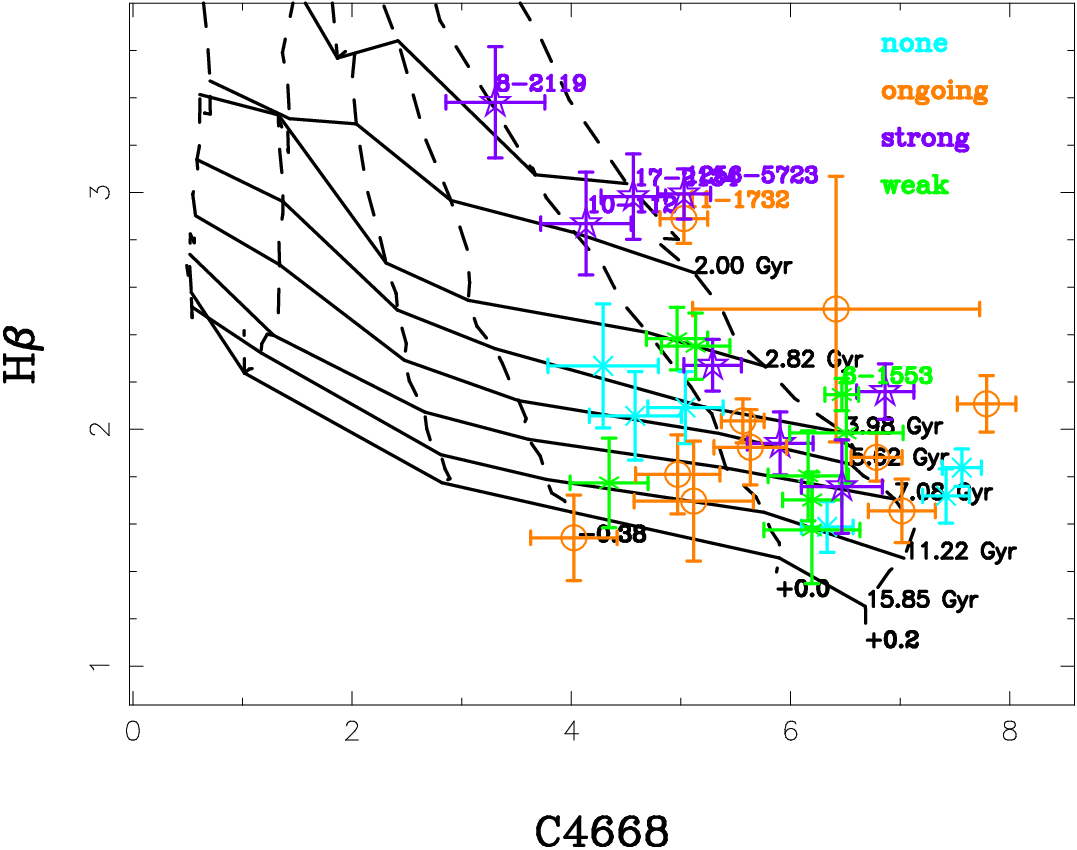}}\hspace{0.3cm}
\resizebox{0.4\textwidth}{!}{\includegraphics[angle=-90]{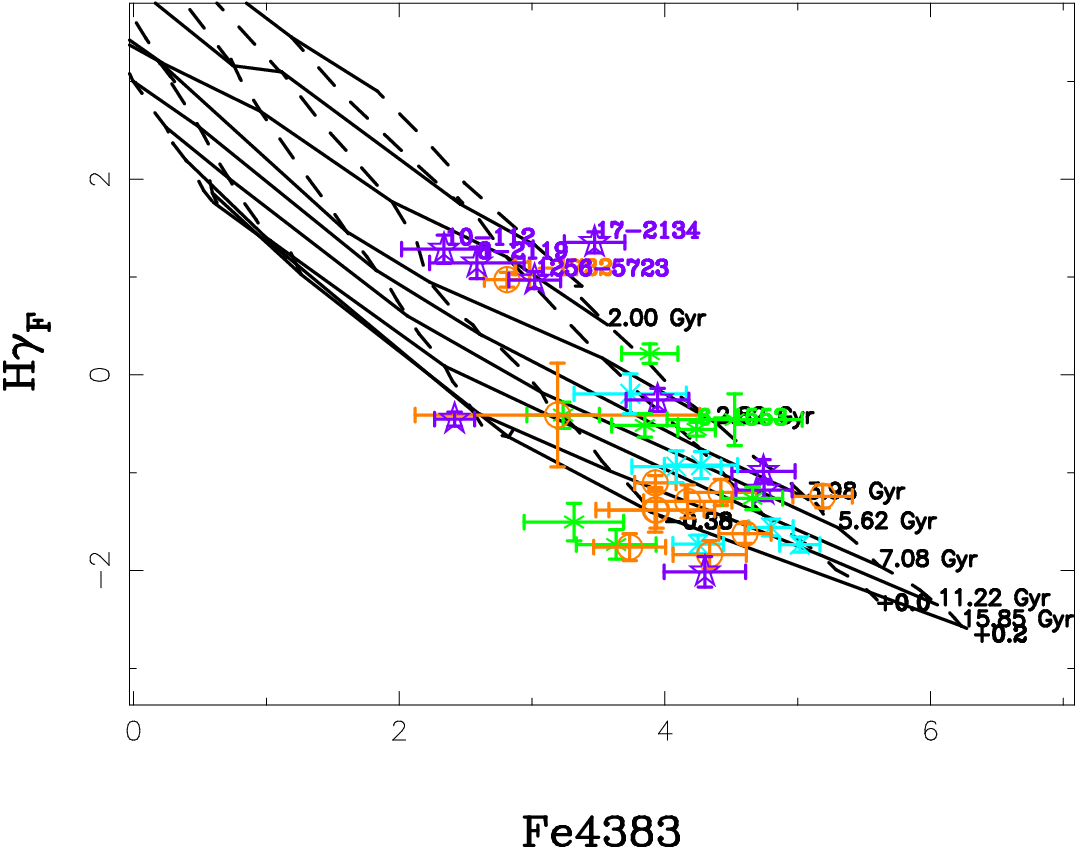}}\vspace{0.3cm}
\resizebox{0.4\textwidth}{!}{\includegraphics[angle=-90]{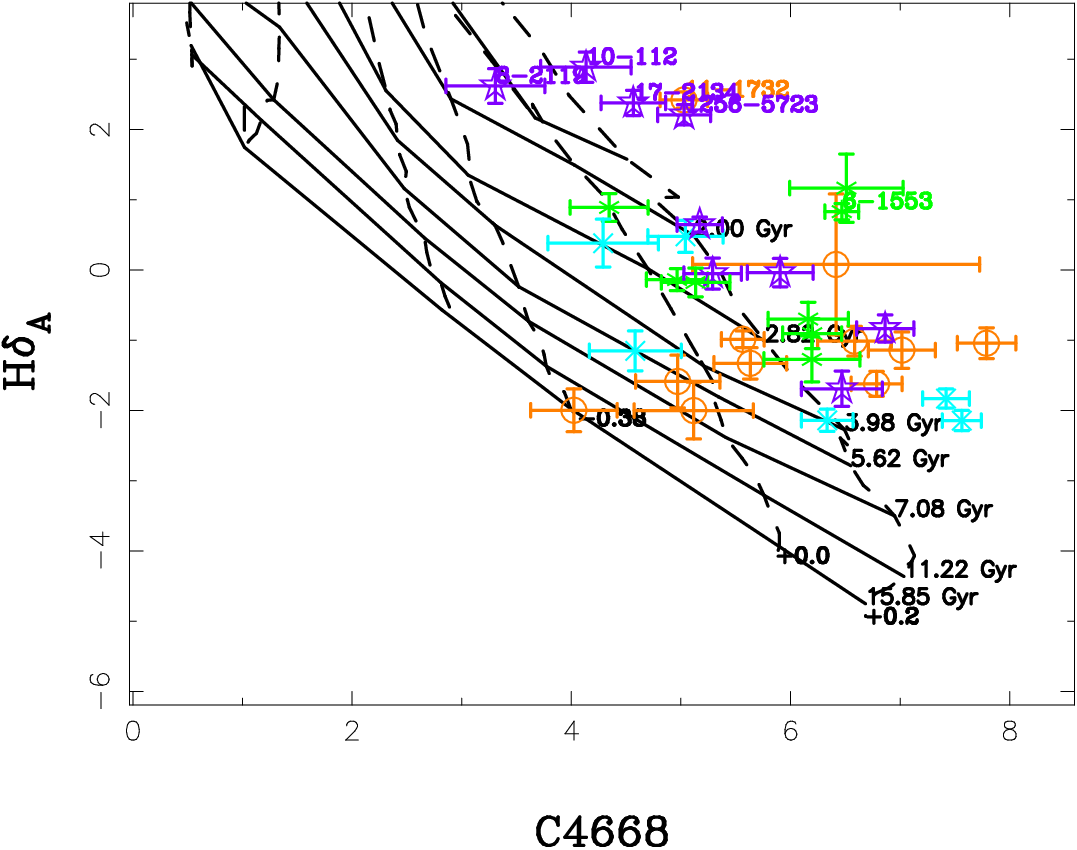}}\hspace{0.3cm}
\resizebox{0.4\textwidth}{!}{\includegraphics[angle=-90]{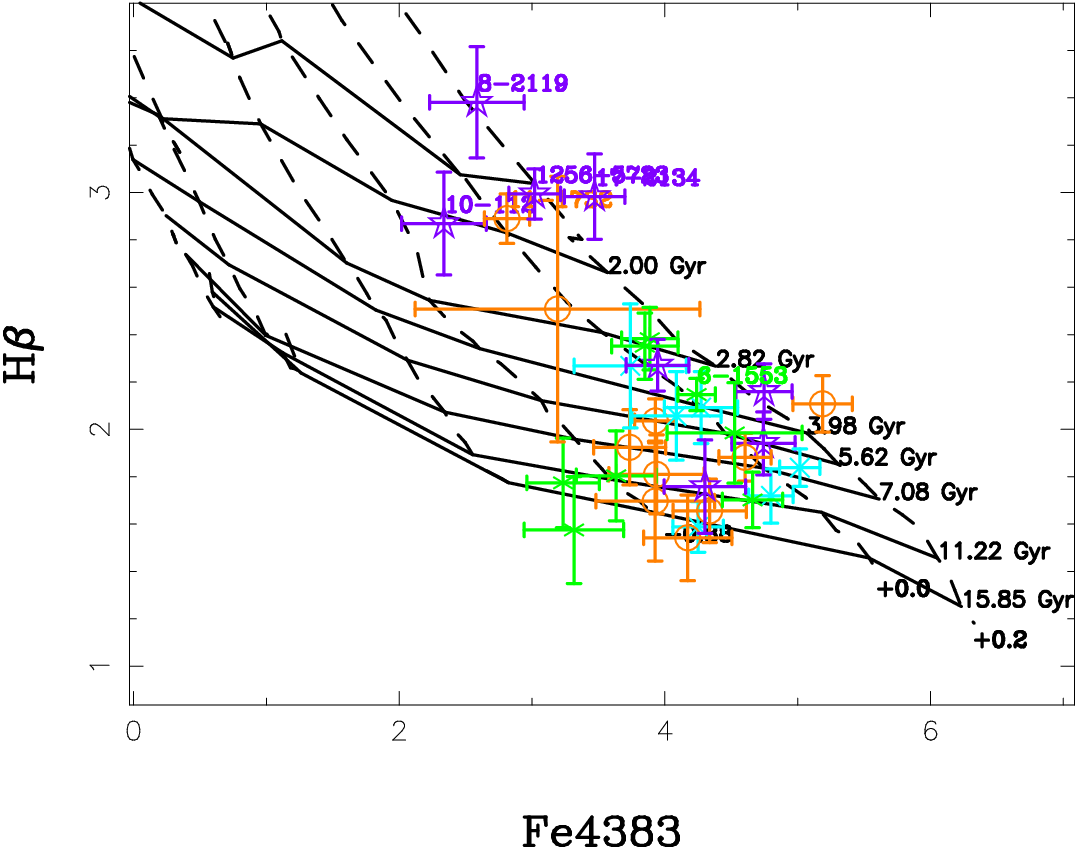}}
\caption{Index-index diagram combining a metal-sensitive index with a Balmer line, more sensitive
to age variations. 
Overplotted
are the models of Vazdekis et~al. (2009) for different values of 
age and metallicity.  Symbols are the same as in Figure~4.
As can be seen, galaxies with strong signatures of having experienced 
a recent merger, plotted in purple,
deviate systematically from the relation defined by the control 
sample. The deviation
is toward strong Balmer lines and weaker 
metallic lines, which is expected in the 
presence of a young population.
\label{index.index.fig}}
\end{figure*}

The deviations from the index-$\sigma$ relations observed  for half 
of the  merger
remnants with strong signs of recent interactions -- i.e., weaker
metal-lines and stronger Balmer lines than the undisturbed galaxies -- can
be the consequence of a younger population, a more metal-poor
population, or a combination of both. This is due to the
age-metallicity degeneracy that affects the line-strength indices, as
well as broad-band colors (Worthey 1994)\nocite{w94}. However, some (Balmer) lines
are more sensitive to age variations than to metallicity variations
and, therefore, by combining them with a non-Balmer line, one can
partially break the degeneracy between these two parameters.
Figure~\ref{index.index.fig} shows several index-index diagrams
combining different metal-indices with Balmer lines.  Overplotted are
the single-stellar population (SSP) models of Vazdekis et~al. (2009) for
populations of different ages (horizontal lines) and metallicities
(vertical lines), as indicated in the labels.

It can be seen from those figures that $\sim$half of the  galaxies with strong
signs of recent interactions are, indeed, younger\footnote{When we use
the term ``younger'' (or more metal-rich), we mean they they
have a lower single-stellar population (SSP)-equivalent age (higher
SSP-equivalent metallicity). As we show later in the paper, this
does not mean that the galaxy as a whole is younger, because even a
small contaminating population 
of young stars can bias this age towards low
values.} than those undisturbed or with weak signs of recent
interactions. However, they are as metal-rich as the other galaxies.

This behaviour indicates that at least some of the ``red'' mergers
were not completely dry.  11-1732, a galaxy in an ongoing merger, also
shows a young component, indicating that, most likely, in this merger
some star formation has been triggered.  All the galaxies with a mean
younger population show LINER-type emission lines.

From these index-index diagrams, ages and metallicity could be
derived.  The primary drawback to deriving an accurate age from the
integrated light of a single-age single-abundance stellar population
is the complication of abundance ratio effects, i.e., the fact that
the ratio between different chemical elements is not always the same
in the systems under study as in the solar neighbourhood (Worthey
1998). In particular, giant elliptical galaxies show enhanced levels
of $\alpha$-elements with respect to Fe.  Several authors have developed
methods to correct for these effects, and significant progress has been
made (Tantalo et~al. 1998\nocite{Tan98}; Trager et~al. 2000a\nocite{T00a};
Thomas et~al. 2003; Cohelo et~al. 2007\nocite{Coel07}; Dotter et~al. 2007\nocite{Dott07}; 
Lee et~al. 2008\nocite{Lee08}).  However, in order to apply these methods, lines with
sensitivities to different chemical element variations are needed.  In
our case, the derivation of accurate ages is compromised as our
wavelength range did not allow us to measure any of the Mg-sensitive
Lick-indices.  In any case, none of the conclusions of this paper
requires the derivation of accurate values of age and metallicity.
 
Furthermore, the ages and chemical abundances derived from these
models are SSP-equivalent parameters, ie., the stellar population of
the galaxies if all the stars were coeval and chemically
homogeneous. We know that this is likely not the case for our objects.
It is well-known that even very small amounts of recent star formation
can bias the final mean ages derived with single-stellar populations
(e.g., Trager et~al. 2000b).  The exact amount depends on the age of
the burst.  If the tidal tails are produced in a major merger event,
those would last for only a few hundred Myrs (vD05), meaning that the
young population would be quite young.  A small fraction of
these very young stars would change the measured indices dramatically.
On the contrary, if the tidal tails have been produced in minor
mergers, then they are believed to last longer ($\sim$1$-$2~Gyr), 
and the fraction of new stars formed may be larger.

To quantity the degree of ``pollution'' from
such new stars that could be present in our galaxies,
we have added to an underlying population of age 11~Gyr (corresponding,
roughly to a redshift formation of three in our cosmology),
different bursts of young stars with ages from 0.1 to 1.3~Gyr. We have
also varied the light-fraction of the burst from 1 to 30\% in the
V-band.  This translates into different mass fractions of the burst,
depending on the age of the young population.

\begin{figure*}
\resizebox{0.4\textwidth}{!}{\includegraphics[angle=-90]{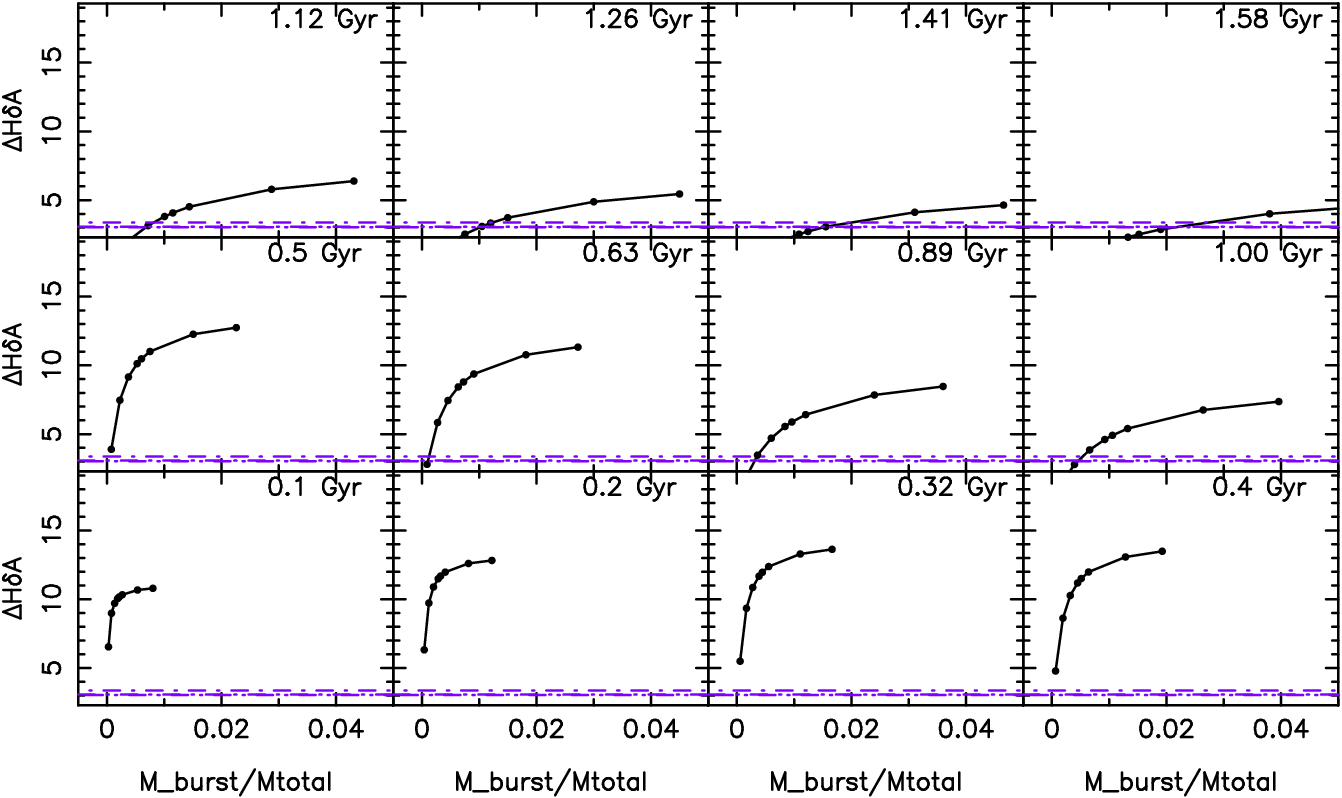}}
\resizebox{0.4\textwidth}{!}{\includegraphics[angle=-90]{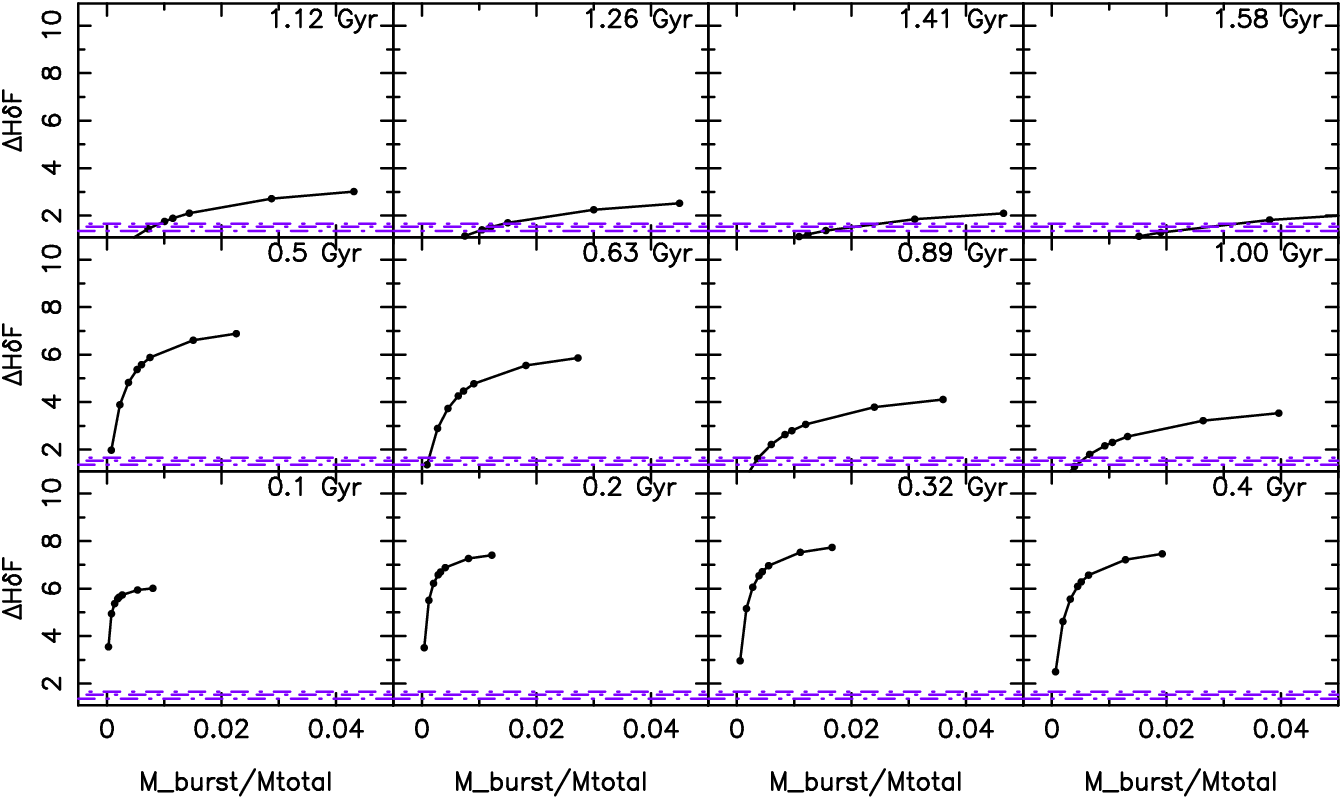}}
\resizebox{0.4\textwidth}{!}{\includegraphics[angle=-90]{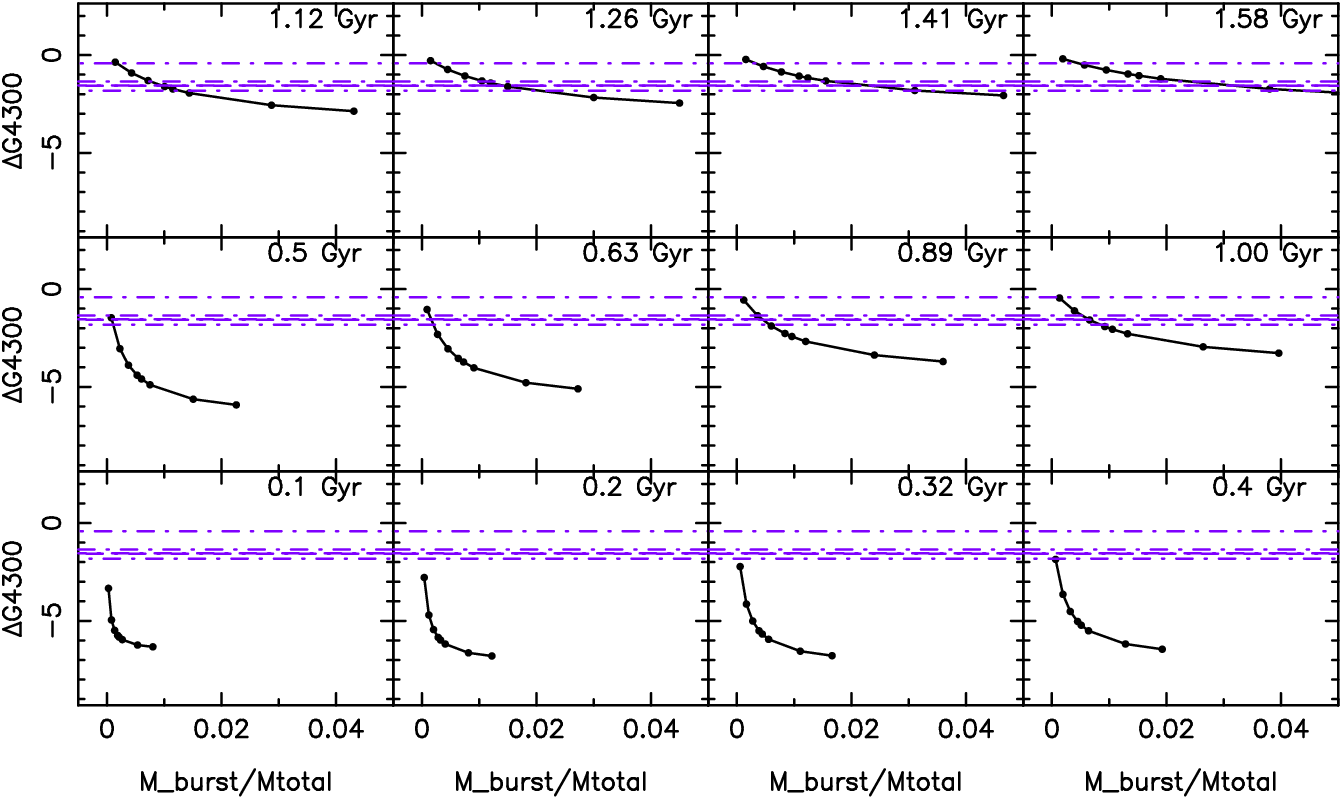}}
\resizebox{0.4\textwidth}{!}{\includegraphics[angle=-90]{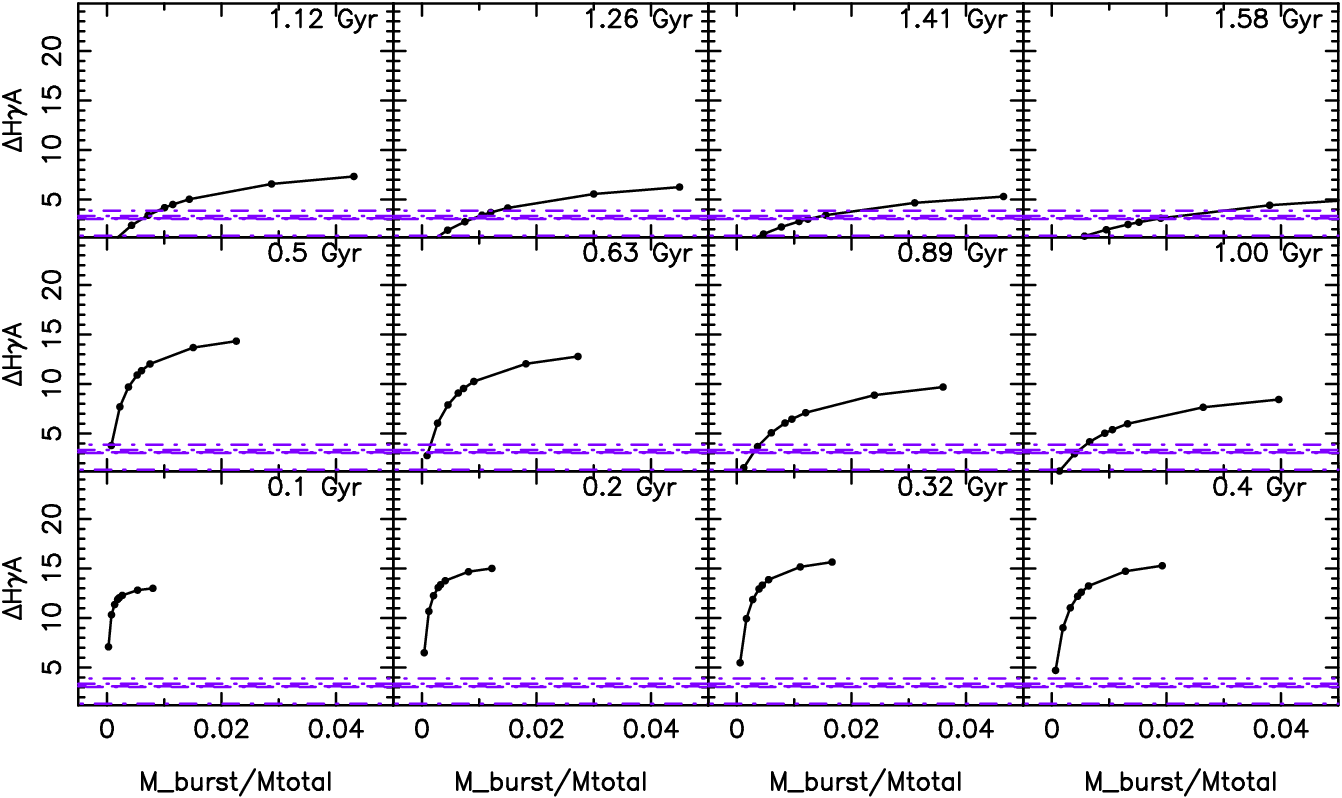}}
\resizebox{0.4\textwidth}{!}{\includegraphics[angle=-90]{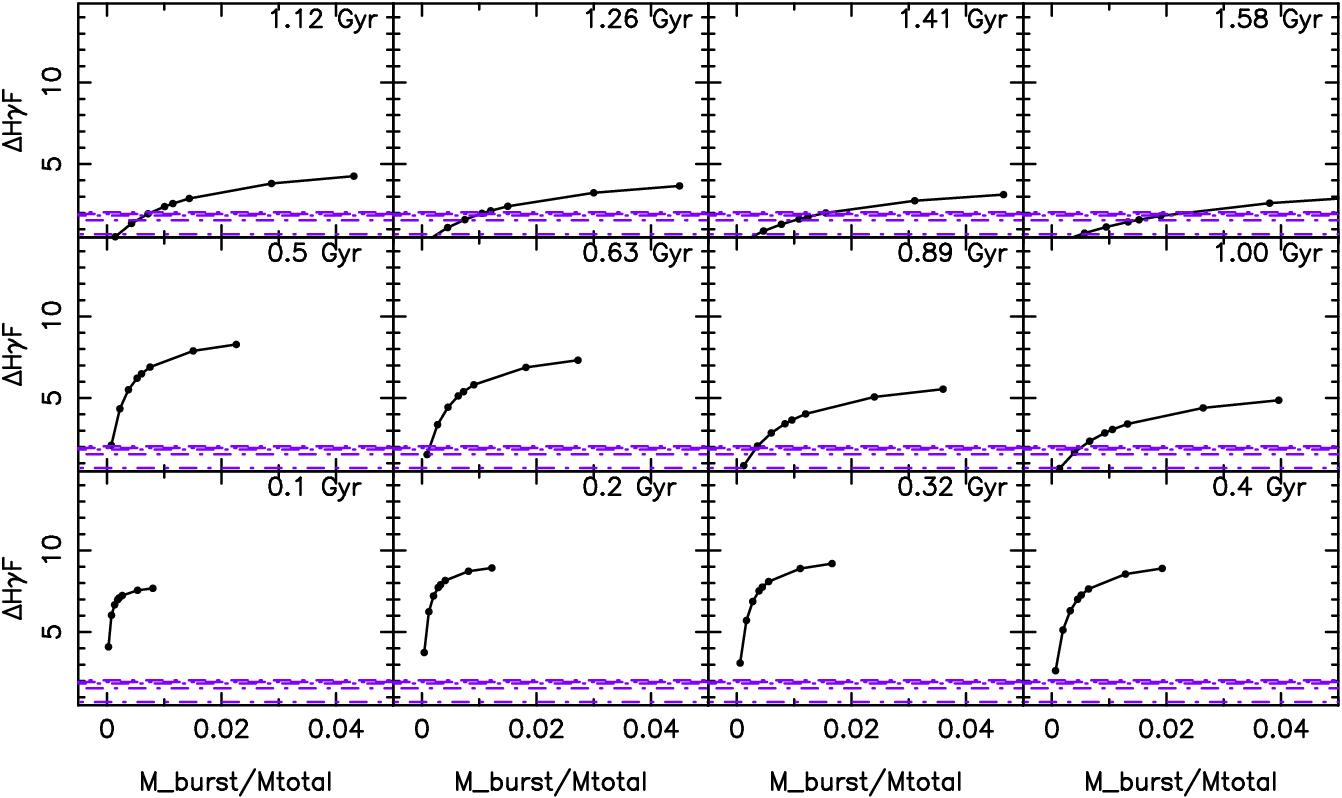}}
\resizebox{0.4\textwidth}{!}{\includegraphics[angle=-90]{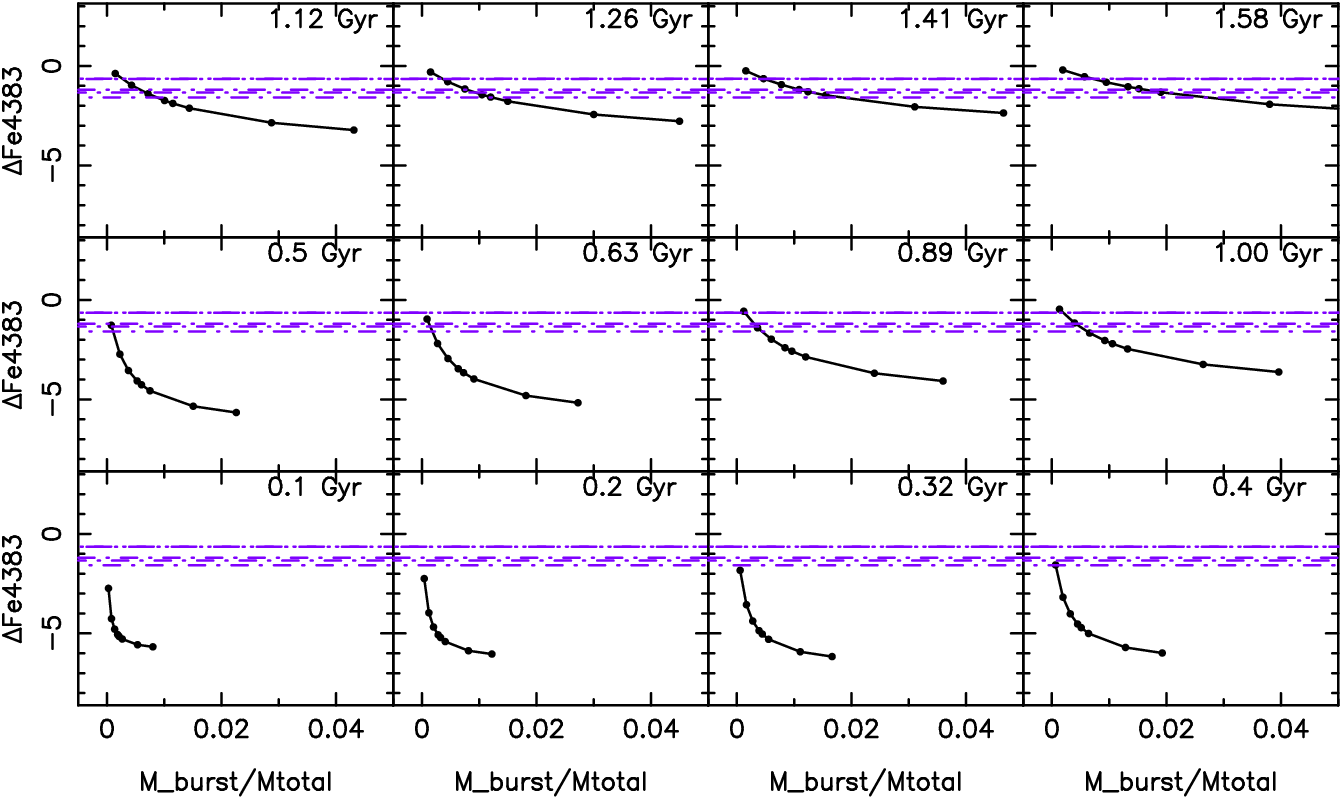}}
\resizebox{0.4\textwidth}{!}{\includegraphics[angle=-90]{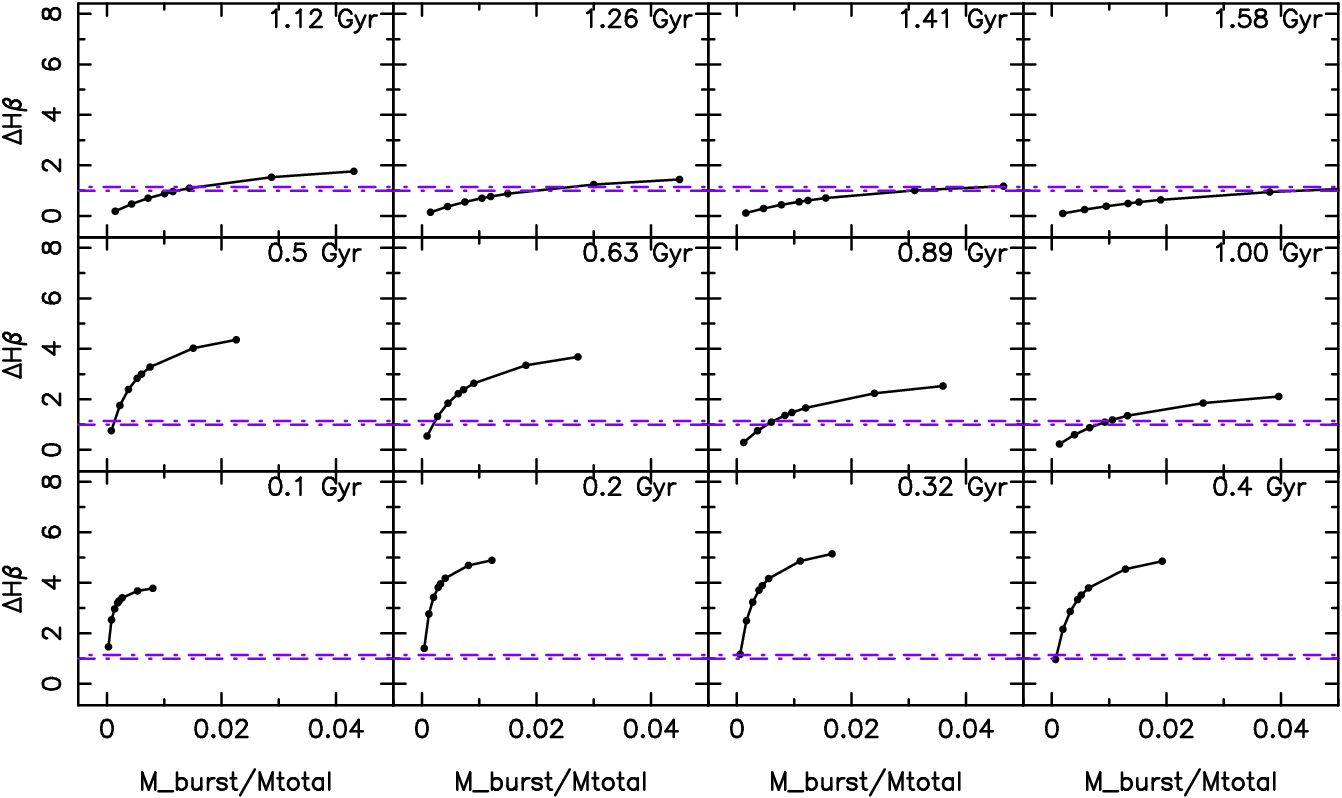}}
\caption{Variation of line-strength when adding to an underlying
old population of age 11~Gyr
and solar metallicity, a younger population of the age and mass
fraction indicated in the insets . Purple lines represent 
the residuals with respect to the index-$\sigma$ relation of the galaxies with 
strong signs of recent interactions. Only those values that are 3 times greater
than the errors are shown.\label{fig:burst}}
\end{figure*}

Figure~\ref{fig:burst} shows the variation of the line-strength indices
as a function of the mass fraction of the burst. We built these
diagrams using stellar population models by Vazdekis et~al. (2009)
with solar metallicity for both the underlying ``old'' and superimposed
``new'' burst populations.  
If we make the assumption that most of the stars in our sample are old 
and with metallicities around solar, and that the deviations from the 
index-$\sigma$ relation defined by the unperturbed sample are due to 
a young component formed in the merger, we can calculate the amount of new
stars as a function of the mass fraction of the young population and of
its age.

In the figure, we have indicated, with horizontal lines, the
deviations from the index-$\sigma$ relation ($I_{\rm gal}-I_{fit}$)
for the galaxies with strong signs of recent interactions, that deviate
more from the index-$\sigma$ relations 
(8-2119,1256-5723,10-112 and 17-2134).  These
galaxies will give us an upper limit to the amount of young stars in
our sample.

It can be seen that, if the young population is younger
than 0.1~Gyr, the mass fraction has to be lower than 1\% in order to
reproduce the observed indices.  If the tidal features have lasted for
1.5~Gyr, the observed variations in the indices from the
index-$\sigma$ relation are reproduced with mass fractions always lower
than 2\%.  The masses of the sample of galaxies with strong signs of
recent interactions oscillate between $\sim$2$\times$10$^{10}$ and
$\sim$1.1$\times$10$^{11}$~M$_{\odot}$ -- 1\% of the total mass
translates into a $\sim$10$^{8}$-10$^{9}$~M$_{\odot}$ burst of star
formation.  These numbers are compatible with the masses of gas found
by Donovan et~al. (2007)\nocite{DHvD07} of 1-6$\times$10$^{8}$~M$_{\odot}$ for their
sample of red galaxies selected in a similar way as vD05. This amount
of gas can come from a dwarf galaxy in a minor interaction (the total
HI mass of a dwarf galaxy is on the order of 10$^{7}$-10$^{8}$
M$_{\odot}$ (Grebel, Gallagher \& Harbeck 2003\nocite{GGH03}) or from a gas poor system (such as a
low-luminosity elliptical), with typical masses of a few times
10$^{8}$~M$_{\odot}$ (Phillips et~al. 1986)\nocite{Phi86}.

This means that, in terms of new stars formed, these mergers are
fairly dry. Only a small fraction of stars are formed.  However, 
because these
mergers appear to be very common (72\% of the bulge-dominated red
galaxies from vD05 sample show morphological disturbances) , they
offer a good explanation for the number of elliptical galaxies showing
mean-young stellar populations.

\section{Discussion}
\label{sec:summary}

In this paper we have characterised the properties of a sample of
galaxies chosen to be remnants of dry mergers or red galaxies in the
process of merging from the sub-sample of vD05.  We have observed
galaxies classified as having ``weak'' and ``strong''
morphological disturbances as well as galaxies in ongoing
interactions. We have compared their properties with an undisturbed
control sample extracted from the same field.

We found a  large variety in the properties of the galaxies
with signs of having experienced recent interactions. Independently 
of the strength of the morphological distortions, our sample 
contains both, slow- and fast-rotators and galaxies with 
S\'ersic indices varying between $\sim$2 and 4.
The only difference between our merger remnants and the control 
sample is that  galaxies with strong signs of recent interactions show, more 
often, the presence of young stellar populations.
However, the line-strength indices can be
explained as a composition of an old population ($\sim$11~Gyr) and a
trace component of young stars ($\simlt$2\%, by mass) if the stars are
younger than 1-2~Gyr (which is the maximum time the morphological
signatures are expected to survive).
Given the stellar mass of our
galaxies (between 2$\times$10$^{10}$ and $\sim$10$^{11}$M$_{\odot}$),
this translates into masses between 1-6$\times$10$^{8}$M$_{\odot}$. 
This agrees with
recent results by Donovan et al. (2007) who found HI at the level of
10$^8$M$_{\odot}$ in a sample of galaxies selected to pass the same
colour selection criteria as that in vD05. Whitaker \& van~Dokkum
(2008)\nocite{WvD08} argued that the Donovan et~al. (2007) sample 
was not representative of the red-merger population and that the 
gas-to-stellar mass ratio in the vD05 sample was, at least, two 
orders-of-magnitude lower than in the Donovan et~al. sample 
(M$_{gas}$/M$_{star} < 3\times 10^{-4}$). They also concluded that  
the gas content in the vD05 sample was independent of the tidal distortion
parameter, ``strong'', ``weak'', or ``none''. They were all, essentially, 
gas-free galaxies. Our results are in apparent contradiction to this
conclusion, however it is difficult to assess the veracity of this discrepancy,
as the Whitaker \& van~Dokkum derivation of the gas content was 
indirect, inferred via the appearance of dust in HST images and,
as such, are likely to be lower limits to the true gas content.

If  we assume that galaxies in ongoing mergers are representative
of the progenitors of our merger remnants, we can study this 
issue in our ongoing sample. In our sample of ongoing mergers, we have three
major merger systems\footnote{Defined as those with a luminosity
ratio between the galaxies below 1:3}, 19-2242/19-2206;
17-596/17-681, and 2-3070/2-3102; the rest (four galaxies) are in
ongoing minor mergers.  Except in the case of 2-3079/2-3102, the other
two major mergers show clear tidal tails and loops, which are
difficult to
reproduce without the presence of a cold component in one of the
galaxies. However, none of these galaxies show a young stellar
component.  \citep{MH94} showed that in mergers between
bulge-dominated galaxies, most of the star formation happens when the
galaxies coalesce. This could explain the lack of young stars in these
systems.  Among the other four galaxies in minor mergers, the only galaxy
with a clear young component is 11-1732.  We do not find any
systematic difference between the stellar populations of those
galaxies involved in major and minor mergers.  We note that most of the
galaxies in ongoing interactions are systematically below the
index-index relations in the Balmer lines and above in the metal
lines. We speculate that this may be due to a decrease in the central
velocity dispersion in the first stages of the merger, although this
would have to be modeled numerically.

Regarding the possible differences  between galaxies with ``strong'' and ``weak''
morphological disturbances, vD05 proposed that the intensity of the disturbances in the final
remnant was the consequence of an evolutionary sequence. In this way,
galaxies with ``weak'' morphological disturbances would be the same
objects as galaxies with ``strong'' ones in a later stage of the
merger.  
On the other hand, Feldman et~al. (2008)\nocite{Fel08} proposed, based on
the morphology of the disturbances, that galaxies with 
``weak'' perturbations could be the
``true'' remnants of major dry mergers (while those remnants with
``strong'' disturbances were more compatible with being wet).  Our
analysis of stellar populations supports both views; in the first case,
if the mass fractions of the young populations are of the order of
1-2\%, these would not be visible any longer after a short period of time
($<$1~Gyr).  On the other hand, the stellar populations are also
compatible with a truly dry merger, as they are almost
indistinguishable from the control sample.

 The scenario of the minor mergers can also be discussed for 
those galaxies with ``strong''  signs of recent interactions.
vD05 argue that the progenitors of the remnants are typically major
mergers rather than low accretion events, because low accretion events
would produce much fainter debris, undetectable in the MUSYC and NOAO
surveys. Also because the fraction of tidally disturbed galaxies that
he found amongst disk-dominated galaxies was much lower than among
bulge-dominated galaxies (only 8\% of the disk dominated galaxies show
tidal features as opposed to 62\% of the bulge-dominated galaxies),
he claims this is consistent with the idea that the events responsible
for the tidal features are strong enough to destroy any dominant disk
component.  Furthermore, the features vD05 detected are red, suggesting
that the progenitors had old stellar populations. 
However, Kawata et
al. (2006)\nocite{Kaw06} and Feldmand et~al. (2008) argue that the types of
structures observed in the MUSYC and NOAO images can be explained by
minor interactions.
Our analysis of stellar populations suggests, however, that 
major mergers are more likely mechanism. This is because 
the metallicitiy of galaxies with young stellar populations is 
quite large.
Dwarf galaxies have lower metal content than more
massive ones (Skillman, Kennicutt \& Hodge 1989\nocite{Skill89};
 Tremonti et~al. 2004\nocite{Trem04}).  If the new
stars formed from gas coming from a dwarf galaxy we would expect to
measure, at least for a short period of time ($\sim$1~Gyr), a
lower-mean metallicity in the integrated spectrum (Serra \& Trager
2007\nocite{ST07}), which is not observed. However, the sample 
of galaxies with young stellar components is small and this analysis 
need to be done in a larger sample to extract firmer conclusions.

It has been suggested that there is a transition mass above which
red-galaxies formed, exclusively, via dry interactions (e.g., Faber
et~al. 2007).  The break point is at 1-2$\times$10$^{11}$M$_{\odot}$,
or M$_B$=$-20$ to $-21$, where boxy and disky ellipticals coexist
(Faber et~al. 2007; Lauer et~al. 2007\nocite{Lau07}). This mass corresponds to
$\sigma\sim$250~kms$^{-1}$ (Faber \& Jackson 1976\nocite{FJ76}). Despite the small number
of galaxies in our sample with $\sigma$ in excess of 250~km~s$^{-1}$,
we found that they are all slow-rotators and show old stellar
populations.  However, the scenario of major dry mergers to produce
boxy, non-rotating, and very isotropic early-type galaxies, have been
called into question in several works (Cox et~al. 2006\nocite{Cox06}; Burkert et~al. 2008\nocite{B08}).
 In particular, is difficult to
reproduce the small anisotropies and ellipticities observed in massive
elliptical galaxies in dissipationless simulations of mergers between
two galaxies of similar mass.  On the contrary, an accumulation of
minor dry mergers can produce the observed remnants (Burkert
et~al. 2008).  One possibility is that most of the interactions
suffered by these galaxies are minor, but that the relative
contribution of the young stars to the galaxy is lower in the most
massive systems.

\section{Conclusions}
In this paper we have studied the properties of a sample of
red merger remnants  or red galaxies in the
process of merging extracted from  vD05.  We have observed
galaxies classified as having ``weak'' and ``strong''
morphological disturbances as well as galaxies in ongoing
interactions. We have compared their properties with an undisturbed
control sample extracted from the same field.
Our main results can be summarized as follows:

\begin{itemize}

\item There is rich mixture in the properties of the galaxies
with signs of having experience recent interactions. 
Independently 
of the strength of the morphological distortions, our sample 
contains both, slow- and fast-rotators and galaxies with 
S\'ersic indices compatible with exponential and de Vaucouleurs profile.

\item Galaxies with strong signs of recent interactions show, more 
often, the presence of young stellar populations, indicating that, 
despite the red colour of the remnant, 
the mergers were not {\it completely} dry.

\item These  young components, however,  can be explained with a frosting $\sim$2\% in mass
of young stars on the top of an old population. Therefore, the mergers
are fairly dry in terms of stellar populations. However, these galaxies 
with young stellar components are all compatible with being supported 
by rotation. Therefore, the cold component is enough to produce 
a disc.

\item We  found a larger incidence of emission lines in our sample of
strong merger remnants compared with the control sample.  The emission
line ratios are compatible with being LINER- and Seyfert-type. The origin of
emission in LINER galaxies is not well-understood.  Graves et~al. (2006)
found that red-galaxies with LINER-like emission were systematically
10-40\% younger than their emission-free counterparts, suggesting a
connection between the mechanism powering the emission and more recent
star formation in the galaxies.  Based on our small sample, we suggest
that the mechanism triggering both is the fueling of gas towards the 
centre produced by the gravitational torques in galaxy mergers.

\item The previous results can be summarise saying that  
 the population of red merger remnants observed by vD05 is compatible 
with containing 
remnants of  mergers with and without gas. For the formers, we favour 
major mergers as the metallicity of the young component seems to be very 
large.

\end{itemize}
\section*{Acknowledgments}

PSB acknowledges the support
of a Marie Curie Intra-European Fellowship within the 6th
European Community Framework Programme.
BKG acknowledges the support of the UK's Science \& Technology
Facilities Council (STFC Grant ST/F002432/1) and the
Commonwealth Cosmology Initiative.
This paper is based on observations obtained at the William Herschel Telescope, operated by the Isaac Newton Group
in the Spanish Observatorio del Roque de los Muchachos of the Instituto de Astrof\'{\i}sica de Canarias. 
NC acknowledges financial support from the Spanish Programa Nacional de Astronom\'{\i}a y Astrof\'{\i}sica under grant AYA2006--15698--C02--02
\bibliographystyle{mn2e}
\bibliography{references}

\begin{thebibliography}{}

\bibitem[\protect\citeauthoryear{{Baldwin}, {Phillips} \&
  {Terlevich}}{{Baldwin} et~al.}{1981}]{BPT81}
{Baldwin} J.~A.,  {Phillips} M.~M.,    {Terlevich} R.,  1981, \pasp, 93, 5

\bibitem[\protect\citeauthoryear{{Bell}, {Phleps}, {Somerville}, {Wolf},
  {Borch} \& {Meisenheimer}}{{Bell} et~al.}{2006}]{Bell06}
{Bell} E.~F.,  {Phleps} S.,  {Somerville} R.~S.,  {Wolf} C.,  {Borch} A.,
  {Meisenheimer} K.,  2006, \apj, 652, 270

\bibitem[\protect\citeauthoryear{{Bell}, {Wolf}, {Meisenheimer}, {Rix},
  {Borch}, {Dye}, {Kleinheinrich}, {Wisotzki} \& {McIntosh}}{{Bell}
  et~al.}{2004}]{Bell04}
{Bell} E.~F.,  {Wolf} C.,  {Meisenheimer} K.,  {Rix} H.-W.,  {Borch} A.,  {Dye}
  S.,  {Kleinheinrich} M.,  {Wisotzki} L.,    {McIntosh} D.~H.,  2004, \apj,
  608, 752

\bibitem[\protect\citeauthoryear{{Bender}, {Burstein} \& {Faber}}{{Bender}
  et~al.}{1993}]{BBF93}
{Bender} R.,  {Burstein} D.,    {Faber} S.~M.,  1993, \apj, 411, 153

\bibitem[\protect\citeauthoryear{{Bender}, {Saglia}, {Ziegler}, {Belloni},
  {Greggio}, {Hopp} \& {Bruzual}}{{Bender} et~al.}{1998}]{Ben98}
{Bender} R.,  {Saglia} R.~P.,  {Ziegler} B.,  {Belloni} P.,  {Greggio} L.,
  {Hopp} U.,    {Bruzual} G.,  1998, \apj, 493, 529

\bibitem[\protect\citeauthoryear{{Bender}, {Surma}, {Doebereiner},
  {Moellenhoff} \& {Madejsky}}{{Bender} et~al.}{1989}]{Ben89}
{Bender} R.,  {Surma} P.,  {Doebereiner} S.,  {Moellenhoff} C.,    {Madejsky}
  R.,  1989, \aap, 217, 35

\bibitem[\protect\citeauthoryear{{Binney}}{{Binney}}{2005}]{Binney05}
{Binney} J.,  2005, \mnras, 363, 937

\bibitem[\protect\citeauthoryear{{Bournaud}, {Jog} \& {Combes}}{{Bournaud}
  et~al.}{2005}]{BJC05}
{Bournaud} F.,  {Jog} C.~J.,    {Combes} F.,  2005, \aap, 437, 69

\bibitem[\protect\citeauthoryear{{Boylan-Kolchin}, {Ma} \&
  {Quataert}}{{Boylan-Kolchin} et~al.}{2006}]{BK06}
{Boylan-Kolchin} M.,  {Ma} C.-P.,    {Quataert} E.,  2006, \mnras, 369, 1081

\bibitem[\protect\citeauthoryear{{Brown}, {Zheng}, {White}, {Dey}, {Jannuzi},
  {Benson}, {Brand}, {Brodwin} \& {Croton}}{{Brown} et~al.}{2008}]{Brown08}
{Brown} M.~J.~I.,  {Zheng} Z.,  {White} M.,  {Dey} A.,  {Jannuzi} B.~T.,
  {Benson} A.~J.,  {Brand} K.,  {Brodwin} M.,    {Croton} D.~J.,  2008, \apj,
  682, 937

\bibitem[\protect\citeauthoryear{{Burkert}, {Naab}, {Johansson} \&
  {Jesseit}}{{Burkert} et~al.}{2008}]{B08}
{Burkert} A.,  {Naab} T.,  {Johansson} P.~H.,    {Jesseit} R.,  2008, \apj,
  685, 897

\bibitem[\protect\citeauthoryear{{Caldwell}}{{Caldwell}}{1984}]{Cal84}
{Caldwell} N.,  1984, \apj, 278, 96

\bibitem[\protect\citeauthoryear{{Caldwell}, {Rose} \& {Concannon}}{{Caldwell}
  et~al.}{2003}]{CRC03}
{Caldwell} N.,  {Rose} J.~A.,    {Concannon} K.~D.,  2003, \aj, 125, 2891

\bibitem[\protect\citeauthoryear{{Cappellari}, {Emsellem}, {Bacon}, {Bureau},
  {Davies}, {de Zeeuw}, {Falc{\'o}n-Barroso}, {Krajnovi{\'c}}, {Kuntschner},
  {McDermid}, {Peletier}, {Sarzi}, {van den Bosch} \& {van de
  Ven}}{{Cappellari} et~al.}{2007}]{Cap07}
{Cappellari} M.,  {Emsellem} E.,  {Bacon} R.,  {Bureau} M.,  {Davies} R.~L.,
  {de Zeeuw} P.~T.,  {Falc{\'o}n-Barroso} J.,  {Krajnovi{\'c}} D.,
  {Kuntschner} H.,  {McDermid} R.~M.,  {Peletier} R.~F.,  {Sarzi} M.,  {van den
  Bosch} R.~C.~E.,    {van de Ven} G.,  2007, \mnras, 379, 418

\bibitem[\protect\citeauthoryear{{Cardiel}}{{Cardiel}}{1999}]{Car99}
{Cardiel} N.,  1999, PhD thesis, , Universidad Complutense de Madrid, Spain,
  (1999)

\bibitem[\protect\citeauthoryear{{Cenarro}, {Peletier},
  {S{\'a}nchez-Bl{\'a}zquez}, {Selam}, {Toloba}, {Cardiel},
  {Falc{\'o}n-Barroso}, {Gorgas}, {Jim{\'e}nez-Vicente} \&
  {Vazdekis}}{{Cenarro} et~al.}{2007}]{Cen07_miles}
{Cenarro} A.~J.,  {Peletier} R.~F.,  {S{\'a}nchez-Bl{\'a}zquez} P.,  {Selam}
  S.~O.,  {Toloba} E.,  {Cardiel} N.,  {Falc{\'o}n-Barroso} J.,  {Gorgas} J.,
  {Jim{\'e}nez-Vicente} J.,    {Vazdekis} A.,  2007, \mnras, 374, 664

\bibitem[\protect\citeauthoryear{{Cimatti}, {Daddi} \& {Renzini}}{{Cimatti}
  et~al.}{2006}]{CDR06}
{Cimatti} A.,  {Daddi} E.,    {Renzini} A.,  2006, \aap, 453, L29

\bibitem[\protect\citeauthoryear{{Ciotti}, {D'Ercole}, {Pellegrini} \&
  {Renzini}}{{Ciotti} et~al.}{1991}]{Cio91}
{Ciotti} L.,  {D'Ercole} A.,  {Pellegrini} S.,    {Renzini} A.,  1991, \apj,
  376, 380

\bibitem[\protect\citeauthoryear{{Coelho}, {Bruzual}, {Charlot}, {Weiss},
  {Barbuy} \& {Ferguson}}{{Coelho} et~al.}{2007}]{Coel07}
{Coelho} P.,  {Bruzual} G.,  {Charlot} S.,  {Weiss} A.,  {Barbuy} B.,
  {Ferguson} J.~W.,  2007, \mnras, 382, 498

\bibitem[\protect\citeauthoryear{{Cole}, {Lacey}, {Baugh} \& {Frenk}}{{Cole}
  et~al.}{2000}]{Cole00}
{Cole} S.,  {Lacey} C.~G.,  {Baugh} C.~M.,    {Frenk} C.~S.,  2000, \mnras,
  319, 168

\bibitem[\protect\citeauthoryear{{Combes}, {Rampazzo}, {Bonfanti}, {Pringniel}
  \& {Sulentic}}{{Combes} et~al.}{1995}]{Com95}
{Combes} F.,  {Rampazzo} R.,  {Bonfanti} P.~P.,  {Pringniel} P.,    {Sulentic}
  J.~W.,  1995, \aap, 297, 37

\bibitem[\protect\citeauthoryear{{Cox}, {Dutta}, {Di Matteo}, {Hernquist},
  {Hopkins}, {Robertson} \& {Springel}}{{Cox} et~al.}{2006}]{Cox06}
{Cox} T.~J.,  {Dutta} S.~N.,  {Di Matteo} T.,  {Hernquist} L.,  {Hopkins}
  P.~F.,  {Robertson} B.,    {Springel} V.,  2006, \apj, 650, 791

\bibitem[\protect\citeauthoryear{{Davies}, {Efstathiou}, {Fall}, {Illingworth}
  \& {Schechter}}{{Davies} et~al.}{1983}]{Dav83}
{Davies} R.~L.,  {Efstathiou} G.,  {Fall} S.~M.,  {Illingworth} G.,
  {Schechter} P.~L.,  1983, \apj, 266, 41

\bibitem[\protect\citeauthoryear{{Djorgovski} \& {Davis}}{{Djorgovski} \&
  {Davis}}{1987}]{DD87}
{Djorgovski} S.,  {Davis} M.,  1987, \apj, 313, 59

\bibitem[\protect\citeauthoryear{{Donovan}, {Hibbard} \& {van
  Gorkom}}{{Donovan} et~al.}{2007}]{DHvD07}
{Donovan} J.~L.,  {Hibbard} J.~E.,    {van Gorkom} J.~H.,  2007, \aj, 134, 1118

\bibitem[\protect\citeauthoryear{{Dotter}, {Chaboyer}, {Ferguson}, {Lee},
  {Worthey}, {Jevremovi{\'c}} \& {Baron}}{{Dotter} et~al.}{2007}]{Dott07}
{Dotter} A.,  {Chaboyer} B.,  {Ferguson} J.~W.,  {Lee} H.-c.,  {Worthey} G.,
  {Jevremovi{\'c}} D.,    {Baron} E.,  2007, \apj, 666, 403

\bibitem[\protect\citeauthoryear{{Emsellem}, {Cappellari}, {Krajnovi{\'c}},
  {van de Ven}, {Bacon}, {Bureau}, {Davies}, {de Zeeuw}, {Falc{\'o}n-Barroso},
  {Kuntschner}, {McDermid}, {Peletier} \& {Sarzi}}{{Emsellem}
  et~al.}{2007}]{Ems07}
{Emsellem} E.,  {Cappellari} M.,  {Krajnovi{\'c}} D.,  {van de Ven} G.,
  {Bacon} R.,  {Bureau} M.,  {Davies} R.~L.,  {de Zeeuw} P.~T.,
  {Falc{\'o}n-Barroso} J.,  {Kuntschner} H.,  {McDermid} R.,  {Peletier} R.~F.,
     {Sarzi} M.,  2007, \mnras, 379, 401

\bibitem[\protect\citeauthoryear{{Emsellem}, {Greusard}, {Combes}, {Friedli},
  {Leon}, {P{\'e}contal} \& {Wozniak}}{{Emsellem} et~al.}{2001}]{Ems01}
{Emsellem} E.,  {Greusard} D.,  {Combes} F.,  {Friedli} D.,  {Leon} S.,
  {P{\'e}contal} E.,    {Wozniak} H.,  2001, \aap, 368, 52

\bibitem[\protect\citeauthoryear{{Faber} \& {Gallagher}}{{Faber} \&
  {Gallagher}}{1976}]{FG76}
{Faber} S.~M.,  {Gallagher} J.~S.,  1976, \apj, 204, 365

\bibitem[\protect\citeauthoryear{{Faber} \& {Jackson}}{{Faber} \&
  {Jackson}}{1976}]{FJ76}
{Faber} S.~M.,  {Jackson} R.~E.,  1976, \apj, 204, 668

\bibitem[\protect\citeauthoryear{{Faber}, {Tremaine} \& {Ajhar}}{{Faber}
  et~al.}{1997}]{Fab97}
{Faber} S.~M.,  {Tremaine} S.,    {Ajhar} e.~a.,  1997, \aj, 114, 1771

\bibitem[\protect\citeauthoryear{{Faber}, {Willmer}, {Wolf}, {Koo}, {Weiner},
  {Newman} \& {et al.}}{{Faber} et~al.}{2007}]{Fab07}
{Faber} S.~M.,  {Willmer} C.~N.~A.,  {Wolf} C.,  {Koo} D.~C.,  {Weiner} B.~J.,
  {Newman} J.~A.,    {et al.} 2007, \apj, 665, 265

\bibitem[\protect\citeauthoryear{{Feldmann}, {Mayer} \& {Carollo}}{{Feldmann}
  et~al.}{2008}]{Fel08}
{Feldmann} R.,  {Mayer} L.,    {Carollo} C.~M.,  2008, \apj, 684, 1062

\bibitem[\protect\citeauthoryear{{Ferreras}, {Lisker}, {Pasquali}, {Khochfar}
  \& {Kaviraj}}{{Ferreras} et~al.}{2009}]{Ferr09}
{Ferreras} I.,  {Lisker} T.,  {Pasquali} A.,  {Khochfar} S.,    {Kaviraj} S.,
  2009, ArXiv e-prints

\bibitem[\protect\citeauthoryear{{Fitzpatrick}}{{Fitzpatrick}}{1999}]{Fit99}
{Fitzpatrick} E.~L.,  1999, \pasp, 111, 63

\bibitem[\protect\citeauthoryear{{Gawiser}, {van Dokkum} \&
  {Herrera}}{{Gawiser} et~al.}{2006}]{Gaw06}
{Gawiser} E.,  {van Dokkum} P.~G.,    {Herrera} D.,  2006, \apjs, 162, 1

\bibitem[\protect\citeauthoryear{{Gonz{\'a}lez}}{{Gonz{\'a}lez}}{1993}]{G93}
{Gonz{\'a}lez} J.~J.,  1993, Ph.D.~Thesis

\bibitem[\protect\citeauthoryear{{Gonz{\'a}lez-Garc{\'{\i}}a} \& {van
  Albada}}{{Gonz{\'a}lez-Garc{\'{\i}}a} \& {van Albada}}{2005}]{GGvA05}
{Gonz{\'a}lez-Garc{\'{\i}}a} A.~C.,  {van Albada} T.~S.,  2005, \mnras, 361,
  1043

\bibitem[\protect\citeauthoryear{{Goudfrooij}, {Hansen}, {Jorgensen} \&
  {Norgaard-Nielsen}}{{Goudfrooij} et~al.}{1994}]{Goud94}
{Goudfrooij} P.,  {Hansen} L.,  {Jorgensen} H.~E.,    {Norgaard-Nielsen} H.~U.,
   1994, \aaps, 105, 341

\bibitem[\protect\citeauthoryear{{Graves}, {Faber}, {Schiavon} \&
  {Yan}}{{Graves} et~al.}{2007}]{Grav07}
{Graves} G.~J.,  {Faber} S.~M.,  {Schiavon} R.~P.,    {Yan} R.,  2007, \apj,
  671, 243

\bibitem[\protect\citeauthoryear{{Grebel}, {Gallagher} III \&
  {Harbeck}}{{Grebel} et~al.}{2003}]{GGH03}
{Grebel} E.~K.,  {Gallagher} III J.~S.,    {Harbeck} D.,  2003, \aj, 125, 1926

\bibitem[\protect\citeauthoryear{{Jannuzi} \& {Dey}}{{Jannuzi} \&
  {Dey}}{1999}]{JD99}
{Jannuzi} B.~T.,  {Dey} A.,  1999, in {Weymann} R.,  {Storrie-Lombardi} L.,
  {Sawicki} M.,   {Brunner} R.,  eds, Photometric Redshifts and the Detection
  of High Redshift Galaxies Vol.~191 of Astronomical Society of the Pacific
  Conference Series, {The NOAO Deep Wide-Field Survey}.
pp 111--+

\bibitem[\protect\citeauthoryear{{Jorgensen}, {Franx} \&
  {Kjaergaard}}{{Jorgensen} et~al.}{1995}]{Jor95}
{Jorgensen} I.,  {Franx} M.,    {Kjaergaard} P.,  1995, \mnras, 276, 1341

\bibitem[\protect\citeauthoryear{{Kawata}, {Mulchaey}, {Gibson} \&
  {S{\'a}nchez-Bl{\'a}zquez}}{{Kawata} et~al.}{2006}]{Kaw06}
{Kawata} D.,  {Mulchaey} J.~S.,  {Gibson} B.~K.,    {S{\'a}nchez-Bl{\'a}zquez}
  P.,  2006, \apj, 648, 969

\bibitem[\protect\citeauthoryear{{Kelson}, {Illingworth}, {Franx} \& {van
  Dokkum}}{{Kelson} et~al.}{2006}]{Kel06}
{Kelson} D.~D.,  {Illingworth} G.~D.,  {Franx} M.,    {van Dokkum} P.~G.,
  2006, \apj, 653, 159

\bibitem[\protect\citeauthoryear{{Kewley} \& {Ellison}}{{Kewley} \&
  {Ellison}}{2008}]{KE08}
{Kewley} L.~J.,  {Ellison} S.~L.,  2008, \apj, 681, 1183

\bibitem[\protect\citeauthoryear{{Khochfar} \& {Burkert}}{{Khochfar} \&
  {Burkert}}{2005}]{KB05}
{Khochfar} S.,  {Burkert} A.,  2005, \mnras, 359, 1379

\bibitem[\protect\citeauthoryear{{Kormendy} \& {Bender}}{{Kormendy} \&
  {Bender}}{1996}]{KB96}
{Kormendy} J.,  {Bender} R.,  1996, \apjl, 464, L119+

\bibitem[\protect\citeauthoryear{{Kuntschner}}{{Kuntschner}}{2000}]{K00}
{Kuntschner} H.,  2000, \mnras, 315, 184

\bibitem[\protect\citeauthoryear{{Lauer}, {Gebhardt}, {Faber}, {Richstone},
  {Tremaine}, {Kormendy}, {Aller}, {Bender}, {Dressler}, {Filippenko}, {Green}
  \& {Ho}}{{Lauer} et~al.}{2007}]{Lau07}
{Lauer} T.~R.,  {Gebhardt} K.,  {Faber} S.~M.,  {Richstone} D.,  {Tremaine} S.,
   {Kormendy} J.,  {Aller} M.~C.,  {Bender} R.,  {Dressler} A.,  {Filippenko}
  A.~V.,  {Green} R.,    {Ho} L.~C.,  2007, \apj, 664, 226

\bibitem[\protect\citeauthoryear{{Lee}, {Worthey}, {Dotter}, {Chaboyer},
  {Jevremovic}, {Baron}, {Briley}, {Ferguson}, {Coelho} \& {Trager}}{{Lee}
  et~al.}{2008}]{Lee08}
{Lee} H.~.,  {Worthey} G.,  {Dotter} A.,  {Chaboyer} B.,  {Jevremovic} D.,
  {Baron} E.,  {Briley} M.~M.,  {Ferguson} J.~W.,  {Coelho} P.,    {Trager}
  S.~C.,  2008, ArXiv e-prints

\bibitem[\protect\citeauthoryear{{Mathews} \& {Brighenti}}{{Mathews} \&
  {Brighenti}}{1999}]{MB99}
{Mathews} W.~G.,  {Brighenti} F.,  1999, \apjl, 527, L31

\bibitem[\protect\citeauthoryear{{Mihos} \& {Hernquist}}{{Mihos} \&
  {Hernquist}}{1994}]{MH94}
{Mihos} J.~C.,  {Hernquist} L.,  1994, \apjl, 431, L9

\bibitem[\protect\citeauthoryear{{Naab} \& {Burkert}}{{Naab} \&
  {Burkert}}{2003}]{NB03}
{Naab} T.,  {Burkert} A.,  2003, \apj, 597, 893

\bibitem[\protect\citeauthoryear{{Naab}, {Khochfar} \& {Burkert}}{{Naab}
  et~al.}{2006}]{NKB06}
{Naab} T.,  {Khochfar} S.,    {Burkert} A.,  2006, \apjl, 636, L81

\bibitem[\protect\citeauthoryear{{Phillips}, {Jenkins}, {Dopita}, {Sadler} \&
  {Binette}}{{Phillips} et~al.}{1986}]{Phi86}
{Phillips} M.~M.,  {Jenkins} C.~R.,  {Dopita} M.~A.,  {Sadler} E.~M.,
  {Binette} L.,  1986, \aj, 91, 1062

\bibitem[\protect\citeauthoryear{{S{\'a}nchez-Bl{\'a}zquez}, {Forbes},
  {Strader}, {Brodie} \& {Proctor}}{{S{\'a}nchez-Bl{\'a}zquez}
  et~al.}{2007}]{SB07}
{S{\'a}nchez-Bl{\'a}zquez} P.,  {Forbes} D.~A.,  {Strader} J.,  {Brodie} J.,
  {Proctor} R.,  2007, \mnras, 377, 759

\bibitem[\protect\citeauthoryear{{S{\'a}nchez-Bl{\'a}zquez}, {Gorgas} \&
  {Cardiel}}{{S{\'a}nchez-Bl{\'a}zquez} et~al.}{2006}]{SB06c}
{S{\'a}nchez-Bl{\'a}zquez} P.,  {Gorgas} J.,    {Cardiel} N.,  2006, \aap, 457,
  823

\bibitem[\protect\citeauthoryear{{S{\'a}nchez-Bl{\'a}zquez}, {Gorgas},
  {Cardiel} \& {Gonz{\'a}lez}}{{S{\'a}nchez-Bl{\'a}zquez} et~al.}{2006}]{SB06a}
{S{\'a}nchez-Bl{\'a}zquez} P.,  {Gorgas} J.,  {Cardiel} N.,    {Gonz{\'a}lez}
  J.~J.,  2006, \aap, 457, 787

\bibitem[\protect\citeauthoryear{{Sanchez-Blazquez}, {Jablonka}, {Noll},
  {Poggianti}, {Moustakas}, {Milvang-Jensen}, {Halliday} \& {et
  al.}}{{Sanchez-Blazquez} et~al.}{2009}]{SB09}
{Sanchez-Blazquez} P.,  {Jablonka} P.,  {Noll} S.,  {Poggianti} B.~M.,
  {Moustakas} J.,  {Milvang-Jensen} B.,  {Halliday} C.,    {et al.} 2009, ArXiv
  e-prints

\bibitem[\protect\citeauthoryear{{S{\'a}nchez-Bl{\'a}zquez}, {Peletier},
  {Jim{\'e}nez-Vicente}, {Cardiel}, {Cenarro}, {Falc{\'o}n-Barroso}, {Gorgas},
  {Selam} \& {Vazdekis}}{{S{\'a}nchez-Bl{\'a}zquez} et~al.}{2006}]{SB06_miles}
{S{\'a}nchez-Bl{\'a}zquez} P.,  {Peletier} R.~F.,  {Jim{\'e}nez-Vicente} J.,
  {Cardiel} N.,  {Cenarro} A.~J.,  {Falc{\'o}n-Barroso} J.,  {Gorgas} J.,
  {Selam} S.,    {Vazdekis} A.,  2006, \mnras, 371, 703

\bibitem[\protect\citeauthoryear{{Sarzi}, {Falc{\'o}n-Barroso}, {Davies},
  {Bacon}, {Bureau}, {Cappellari}, {de Zeeuw}, {Emsellem}, {Fathi},
  {Krajnovi{\'c}}, {Kuntschner}, {McDermid} \& {Peletier}}{{Sarzi}
  et~al.}{2006}]{Sar06}
{Sarzi} M.,  {Falc{\'o}n-Barroso} J.,  {Davies} R.~L.,  {Bacon} R.,  {Bureau}
  M.,  {Cappellari} M.,  {de Zeeuw} P.~T.,  {Emsellem} E.,  {Fathi} K.,
  {Krajnovi{\'c}} D.,  {Kuntschner} H.,  {McDermid} R.~M.,    {Peletier} R.~F.,
   2006, \mnras, 366, 1151

\bibitem[\protect\citeauthoryear{{Scarlata}, {Carollo} \& {Lilly}}{{Scarlata}
  et~al.}{2007}]{Scar07}
{Scarlata} C.,  {Carollo} C.~M.,    {Lilly} S.~J. e.~a.,  2007, \apjs, 172, 494

\bibitem[\protect\citeauthoryear{{Schlegel}, {Finkbeiner} \&
  {Davis}}{{Schlegel} et~al.}{1998}]{Sch98}
{Schlegel} D.~J.,  {Finkbeiner} D.~P.,    {Davis} M.,  1998, \apj, 500, 525

\bibitem[\protect\citeauthoryear{{Serra} \& {Trager}}{{Serra} \&
  {Trager}}{2007}]{ST07}
{Serra} P.,  {Trager} S.~C.,  2007, \mnras, 374, 769

\bibitem[\protect\citeauthoryear{{Skillman}, {Kennicutt} \& {Hodge}}{{Skillman}
  et~al.}{1989}]{Skill89}
{Skillman} E.~D.,  {Kennicutt} R.~C.,    {Hodge} P.~W.,  1989, \apj, 347, 875

\bibitem[\protect\citeauthoryear{{Springel}, {Di Matteo} \&
  {Hernquist}}{{Springel} et~al.}{2005}]{SdMH05}
{Springel} V.,  {Di Matteo} T.,    {Hernquist} L.,  2005, \apjl, 620, L79

\bibitem[\protect\citeauthoryear{{Tantalo}, {Chiosi} \& {Bressan}}{{Tantalo}
  et~al.}{1998}]{Tan98}
{Tantalo} R.,  {Chiosi} C.,    {Bressan} A.,  1998, \aap, 333, 419

\bibitem[\protect\citeauthoryear{{Thomas}, {Maraston} \& {Bender}}{{Thomas}
  et~al.}{2003}]{TMB03}
{Thomas} D.,  {Maraston} C.,    {Bender} R.,  2003, \mnras, 343, 279

\bibitem[\protect\citeauthoryear{{Thomas}, {Maraston}, {Bender} \& {Mendes de
  Oliveira}}{{Thomas} et~al.}{2005}]{Tho05}
{Thomas} D.,  {Maraston} C.,  {Bender} R.,    {Mendes de Oliveira} C.,  2005,
  \apj, 621, 673

\bibitem[\protect\citeauthoryear{{Trager}, {Faber}, {Worthey} \&
  {Gonz{\'a}lez}}{{Trager} et~al.}{2000a}]{T00b}
{Trager} S.~C.,  {Faber} S.~M.,  {Worthey} G.,    {Gonz{\'a}lez} J.~J.,  2000a,
  \aj, 120, 165

\bibitem[\protect\citeauthoryear{{Trager}, {Faber}, {Worthey} \&
  {Gonz{\'a}lez}}{{Trager} et~al.}{2000b}]{T00}
{Trager} S.~C.,  {Faber} S.~M.,  {Worthey} G.,    {Gonz{\'a}lez} J.~J.,  2000b,
  \aj, 120, 165

\bibitem[\protect\citeauthoryear{{Trager}, {Faber}, {Worthey} \&
  {Gonz{\'a}lez}}{{Trager} et~al.}{2000c}]{T00a}
{Trager} S.~C.,  {Faber} S.~M.,  {Worthey} G.,    {Gonz{\'a}lez} J.~J.,  2000c,
  \aj, 119, 1645

\bibitem[\protect\citeauthoryear{{Trager}, {Worthey}, {Faber}, {Burstein} \&
  {Gonzalez}}{{Trager} et~al.}{1998}]{T98}
{Trager} S.~C.,  {Worthey} G.,  {Faber} S.~M.,  {Burstein} D.,    {Gonzalez}
  J.~J.,  1998, \apjs, 116, 1

\bibitem[\protect\citeauthoryear{{Tran}, {van Dokkum}, {Franx}, {Illingworth},
  {Kelson} \& {Schreiber}}{{Tran} et~al.}{2005}]{Tran05}
{Tran} K.-V.~H.,  {van Dokkum} P.,  {Franx} M.,  {Illingworth} G.~D.,  {Kelson}
  D.~D.,    {Schreiber} N.~M.~F.,  2005, \apjl, 627, L25

\bibitem[\protect\citeauthoryear{{Tremonti}, {Heckman}, {Kauffmann},
  {Brinchmann}, {Charlot}, {White}, {Seibert}, {Peng}, {Schlegel}, {Uomoto},
  {Fukugita} \& {Brinkmann}}{{Tremonti} et~al.}{2004}]{Trem04}
{Tremonti} C.~A.,  {Heckman} T.~M.,  {Kauffmann} G.,  {Brinchmann} J.,
  {Charlot} S.,  {White} S.~D.~M.,  {Seibert} M.,  {Peng} E.~W.,  {Schlegel}
  D.~J.,  {Uomoto} A.,  {Fukugita} M.,    {Brinkmann} J.,  2004, \apj, 613, 898

\bibitem[\protect\citeauthoryear{{van Dokkum}, {Franx}, {Fabricant}, {Kelson}
  \& {Illingworth}}{{van Dokkum} et~al.}{1999}]{vD99}
{van Dokkum} P.~G.,  {Franx} M.,  {Fabricant} D.,  {Kelson} D.~D.,
  {Illingworth} G.~D.,  1999, \apjl, 520, L95

\bibitem[\protect\citeauthoryear{{Vazdekis}}{{Vazdekis}}{1999}]{V99}
{Vazdekis} A.,  1999, \apj, 513, 224

\bibitem[\protect\citeauthoryear{{Vazdekis}, {Cenarro}, {Gorgas}, {Cardiel} \&
  {Peletier}}{{Vazdekis} et~al.}{2003}]{v03}
{Vazdekis} A.,  {Cenarro} A.~J.,  {Gorgas} J.,  {Cardiel} N.,    {Peletier}
  R.~F.,  2003, \mnras, 340, 1317

\bibitem[\protect\citeauthoryear{{Whitaker} \& {van Dokkum}}{{Whitaker} \& {van
  Dokkum}}{2008}]{WvD08}
{Whitaker} K.~E.,  {van Dokkum} P.~G.,  2008, \apjl, 676, L105

\bibitem[\protect\citeauthoryear{{White} \& {Frenk}}{{White} \&
  {Frenk}}{1991}]{WF91}
{White} S.~D.~M.,  {Frenk} C.~S.,  1991, \apj, 379, 52

\bibitem[\protect\citeauthoryear{{Worthey}}{{Worthey}}{1994}]{w94}
{Worthey} G.,  1994, \apjs, 95, 107

\bibitem[\protect\citeauthoryear{{Worthey} \& {Ottaviani}}{{Worthey} \&
  {Ottaviani}}{1997}]{WO97}
{Worthey} G.,  {Ottaviani} D.~L.,  1997, \apjs, 111, 377

\bibitem[\protect\citeauthoryear{{Wozniak} \& {Champavert}}{{Wozniak} \&
  {Champavert}}{2006}]{WC06}
{Wozniak} H.,  {Champavert} N.,  2006, \mnras, 369, 853

\bibitem[\protect\citeauthoryear{{Wozniak}, {Combes}, {Emsellem} \&
  {Friedli}}{{Wozniak} et~al.}{2003}]{Woz03}
{Wozniak} H.,  {Combes} F.,  {Emsellem} E.,    {Friedli} D.,  2003, \aap, 409,
  469

\bibitem[\protect\citeauthoryear{{Yan}, {Newman}, {Faber}, {Konidaris}, {Koo}
  \& {Davis}}{{Yan} et~al.}{2006}]{Yan06}
{Yan} R.,  {Newman} J.~A.,  {Faber} S.~M.,  {Konidaris} N.,  {Koo} D.,
  {Davis} M.,  2006, \apj, 648, 281

\end{thebibliography}

\appendix
\section{Line-strength indices}

\label{apendix:linestrength}
Table \ref{line-strength} shows the line-strength indices measured on the galaxy spectra.
\begin{table*}
\begin{tabular}{lrrrrrrrrrrrrr}
\hline\hline
\small
         &CN$_1$     & CN$_2$  &Ca4227&G4300&Fe4383&Ca4455&Fe4531&C4668 & H$\beta$& H$\delta_A$&H$\delta_F$&H$\gamma_A$ & H$\gamma_F$\\
\hline 
1-1403   &  0.089 & 0.127  &1.182&5.347&4.799&1.190&2.845&7.419&1.719&$-1.829$&0.371&$-6.206$&$-1.559$\\
         &  0.004 & 0.004  &0.067&0.118&0.168&0.090&0.135&0.213&0.115&$ 0.136$&0.091&$ 0.142$&$ 0.085$\\
11-1014  &  0.084 & 0.113  &1.029&5.150&4.251&0.948&3.066&6.335&1.588&$-2.139$&0.249&$-5.880$&$-1.731$\\
         &  0.004 & 0.005  &0.075&0.130&0.190&0.101&0.153&0.235&0.107&$ 0.158$&0.105&$ 0.154$&$ 0.095$\\
17-2031  &  0.024 & 0.048  &1.049&5.026&4.087&0.818&2.643&4.584&2.057&$-1.151$&0.814&$-4.687$&$-0.938$ \\
         &  0.008 & 0.009  &0.133&0.232&0.336&0.180&0.271&0.419&0.187&$ 0.284$&0.192&$ 0.269$&$ 0.163$\\
18-2684  &  0.118 & 0.155  &1.228&5.352&5.016&0.983&3.404&7.563&1.838&$-2.143$&0.186&$-6.477$&$-1.734$\\
         &  0.004 & 0.005  &0.065&0.110&0.151&0.082&0.118&0.178&0.079&$ 0.144$&0.096&$ 0.130$&$ 0.078$\\
2-5013   &  0.011 & 0.044  &0.756&4.379&4.113&0.967&2.952&4.723&1.760&$ 0.383$&1.629&$-3.640$&$-0.192$\\
         &  0.009 & 0.011  &0.176&0.306&0.422&0.220&0.314&0.504&0.262&$ 0.341$&0.230&$ 0.340$&$ 0.204$\\
22-991   &$-0.012$& 0.033  &1.040&5.224&4.273&1.079&2.940&5.042&2.092&$ 0.479$&1.170&$-4.783$&$-0.922$ \\
         &  0.007 & 0.008  &0.115&0.195&0.275&0.148&0.220&0.344&0.152&$ 0.228$&0.157&$ 0.227$&$ 0.137$\\
\hline
10-232   &  0.011 & 0.042  &0.817&4.390&3.690&0.778&2.904&4.610&1.774&$ 0.890$&1.444&$-4.496$&$-0.411$\\
         &  0.006 & 0.007  &0.104&0.189&0.273&0.149&0.218&0.356&0.189&$ 0.198$&0.136&$ 0.220$&$ 0.134$\\
12-1734  &  0.026 & 0.066  &0.924&4.350&3.887&1.148&2.956&4.965&2.383&$-0.136$&1.382&$-3.284$&$ 0.217$\\
         &  0.004 & 0.005  &0.081&0.145&0.212&0.115&0.174&0.279&0.132&$ 0.156$&0.107&$ 0.164$&$ 0.099$\\
16-650   &  0.045 & 0.078  &1.291&4.729&4.204&1.063&3.157&6.565&1.804&$-0.690$&0.657&$-5.776$&$-1.733$  \\
         &  0.007 & 0.008  &0.118&0.219&0.300&0.158&0.229&0.366&0.190&$ 0.239$&0.163&$ 0.250$&$ 0.151$\\
22-790   &  0.032 & 0.072  &1.238&5.190&4.660&1.207&3.009&6.198&1.702&$-0.905$&0.355&$-5.421$&$-1.262$  \\
         &  0.006 & 0.007  &0.098&0.166&0.227&0.121&0.177&0.269&0.117&$ 0.214$&0.147&$ 0.191$&$ 0.115$\\
6-1553   &$-0.004$& 0.036  &0.852&4.887&4.153&1.145&3.186&6.409&2.147&$ 0.831$&1.011&$-3.903$&$-0.556$\\
         &   0.030& 0.004  &0.061&0.099&0.141&0.071&0.102&0.155&0.068&$ 0.116$&0.081&$ 0.115$&$ 0.071$\\
6-1676   &  0.002 & 0.024  &1.052&4.539&3.957&0.899&2.823&5.366&2.351&$-0.178$&1.373&$-4.063$&$-0.516$\\
         &  0.006 & 0.007  &0.097&0.173&0.250&0.134&0.202&0.313&0.140&$ 0.205$&0.139&$ 0.196$&$ 0.119$\\
7-2322   &  0.068 & 0.105  &1.285&5.402&4.171&1.209&3.131&6.699&1.575&$-1.272$&0.670&$-5.516$&$-1.503$\\
         &  0.009 & 0.010  &0.154&0.271&0.374&0.194&0.282&0.438&0.226&$ 0.321$&0.217&$ 0.315$&$ 0.192$\\
9-2105   &  0.022 & 0.061  &0.795&4.916&4.818&0.806&3.341&6.626&1.985&$ 1.166$&1.363&$-4.403$&$-0.457$\\
         &  0.014 & 0.017  &0.238&0.397&0.508&0.265&0.367&0.517&0.212&$ 0.486$&0.337&$ 0.446$&$ 0.264$\\
\hline
10-112   &$-0.041$&$-0.012$&0.711&3.471&2.967&0.697&2.223&4.170&2.869&$ 2.888$&2.449&$-0.672$&$ 1.285$\\
         &  0.006 & 0.008  &0.119&0.224&0.319&0.170&0.253&0.413&0.217&$ 0.216$&0.152&$ 0.236$&$ 0.143$\\
1256-5723&$-0.005$& 0.025  &0.807&3.310&2.895&0.657&2.617&4.930&2.994&$ 2.209$&2.461&$-1.340$&$ 0.969$ \\
         &  0.004 & 0.005  &0.074&0.135&0.195&0.103&0.155&0.241&0.106&$ 0.144$&0.099&$ 0.142$&$ 0.087$\\
13-3813  &  0.038 & 0.086  &1.072&4.962&4.768&0.886&2.871&6.047&1.941&$-0.037$&0.991&$-4.845$&$-0.985$ \\ 
         &  0.006 & 0.007  &0.099&0.169&0.238&0.131&0.194&0.300&0.133&$ 0.201$&0.138&$ 0.196$&$ 0.119$\\
16-1302  &  0.076 & 0.121  &1.099&5.284&4.704&1.141&2.951&6.901&2.160&$-0.832$&0.677&$-5.162$&$-1.177$  \\
         &  0.005 & 0.006  &0.093&0.153&0.212&0.115&0.171&0.261&0.117&$ 0.193$&0.131&$ 0.180$&$ 0.109$\\
17-2134  &$-0.020$& 0.019  &0.814&3.326&3.589&0.817&2.407&4.673&2.983& $2.380$&2.308&$-0.858$&$ 1.355$\\
         &  0.005 & 0.006  &0.091&0.162&0.228&0.124&0.188&0.294&0.180& $ 0.179$&0.125&$ 0.173$&$ 0.106$\\
8-2119   &$-0.045$&$-0.005$&0.823&3.345&2.892&0.592&2.357&4.183&3.381&$ 2.620$&2.599&$-1.270$&$ 1.145$\\
         &  0.007 & 0.009  &0.134&0.254&0.356&0.188&0.276&0.451&0.235&$ 0.247$&0.171&$ 0.266$&$ 0.161$\\
3-601    &  0.026 & 0.071  &0.872&4.556&3.681&0.900&2.944&5.805& ---& $ 0.648$&1.557&$-3.724$&$-0.451$ \\
         &  0.003 & 0.004  &0.057&0.103&0.151&0.080&0.122&0.207& ---& $ 0.107$&0.073&$ 0.117$&$ 0.072$\\
5-994    &  0.017 & 0.060  &0.929&4.683&3.945&0.824&2.757&5.290&2.271&$-0.050$&1.105&$-3.705$&$-0.255$ \\
         &  0.006 & 0.007  &0.106&0.178&0.235&0.123&0.173&0.263&0.109&$ 0.222$&0.152&$ 0.195$&$ 0.117$\\
9-3079   &  0.019 & 0.114  &1.186&5.680&4.302&0.943&3.408&6.469&1.758&$-1.691$&0.489&$-6.803$&$-2.012$\\ 
         &  0.008 & 0.008  &0.123&0.217&0.307&0.161&0.229&0.369&0.197&$ 0.251$&0.167&$ 0.261$&$0.156$ \\
\hline
\end{tabular}
\caption{Lick/IDS line-strength indices within an aperture of
1~r$_{\rm eff}$ in our sample of galaxies.
The wavelength coverage allowed us to measure 13 of the 25 indices.
The second row for each galaxy lists the associated uncertainties
in the respective indices.\label{line-strength}}
\end{table*}

\addtocounter{table}{-1}
\begin{table*}
\begin{tabular}{lrrrrrrrrrrrrr}
\hline\hline
\small
1-2874   &  0.045 & 0.079  &0.988&5.301&4.422&1.161&2.993&6.584& --- &$-1.013$&0.491&$-5.730$&$-1.201$\\
         &  0.006 & 0.007  &0.107&0.190&0.269&0.142&0.210&0.336& --- &$ 0.212$&0.144&$ 0.223$&$ 0.134$\\
11-1278  & 0.107 & 0.149  &1.081&5.481&5.221&1.050&3.117&7.778& 2.108&$-1.042$&0.208&$-6.090$&$-1.241$ \\
         &  0.006 & 0.007  &0.101&0.166&0.224&0.121&0.179&0.267&0.119&$ 0.220$&0.148&$ 0.195$&$ 0.118$\\
11-1732  &$-0.019$& 0.005  &0.523&3.159&2.856&0.344&2.617&4.986&2.890&$ 2.422$&2.277&$-1.707$&$ 0.974$ \\
         &  0.003 & 0.004  &0.064&0.116&0.170&0.093&0.136&0.217&0.105&$ 0.115$&0.078&$ 0.124$&$ 0.076$\\
14-1401  &  0.046 & 0.078  &1.255&5.565&4.210&1.062&3.046&5.048&1.810&$-1.587$&0.747&$-5.361$&$-1.381$  \\
         &  0.010 & 0.012  &0.160&0.269&0.359&0.183&0.262&0.385&0.167&$ 0.375$&0.251&$ 0.307$&$ 0.186$\\
17-596   &  0.081 & 0.113  &1.081&5.304&4.604&0.840&2.948&6.786&1.882&$-1.621$&0.540&$-6.133$&$-1.621$  \\
         &  0.005 & 0.006  &0.084&0.143&0.197&0.106&0.154&0.232&0.101&$ 0.178$&0.118&$ 0.167$&$ 0.101$\\
17-681   &  0.068 & 0.100  &1.242&5.376&4.338&0.943&3.020&7.016&1.656&$-1.141$&0.582&$-6.180$&$-1.838$ \\
         &  0.007 & 0.008  &0.123&0.207&0.276&0.146&0.208&0.306&0.134&$ 0.259$&0.175&$ 0.240$&$ 0.144$\\
19-2242  &  0.033 & 0.068  &0.998&5.042&3.929&0.945&2.973&5.565&2.036&$-0.987$&0.855&$-4.997$&$-1.104$ \\
         &  0.003 & 0.004  &0.059&0.107&0.156&0.084&0.123&0.195&0.093&$ 0.120$&0.080&$ 0.126$&$ 0.076$\\
19-2206  &  0.060 & 0.094  &1.281&5.325&4.305&1.040&3.308&5.775&1.924&$-1.329$&0.574&$-6.315$&$-1.760$  \\
         &  0.006 & 0.007  &0.105&0.192&0.271&0.145&0.206&0.331&0.159&$ 0.223$&0.148&$ 0.228$&$ 0.137$\\
2-3070   &  0.019 & 0.046  &1.298&5.279&4.229&0.910&3.327&4.357&1.542&$-1.997$&0.299&$-5.280$&$-1.294$ \\
         &  0.008 & 0.010  &0.136&0.242&0.333&0.175&0.254&0.395&0.181&$ 0.305$&0.207&$ 0.281$&$ 0.170$\\
2-3102   &  0.058 & 0.096  &1.279&5.631&4.132&1.221&3.373&5.539&1.697&$-2.000$&0.993&$-5.557$&$-1.380$ \\
         &  0.011 & 0.013  &0.180&0.315&0.447&0.237&0.348&0.543&0.253&$ 0.404$&0.269&$ 0.376$&$ 0.228$\\
\hline
\end{tabular}
\caption{Continue}
\end{table*}

\section{rotation curves}
\label{rotationcurves}

\begin{figure*}
\resizebox{0.25\textwidth}{!}{\includegraphics[angle=-90]{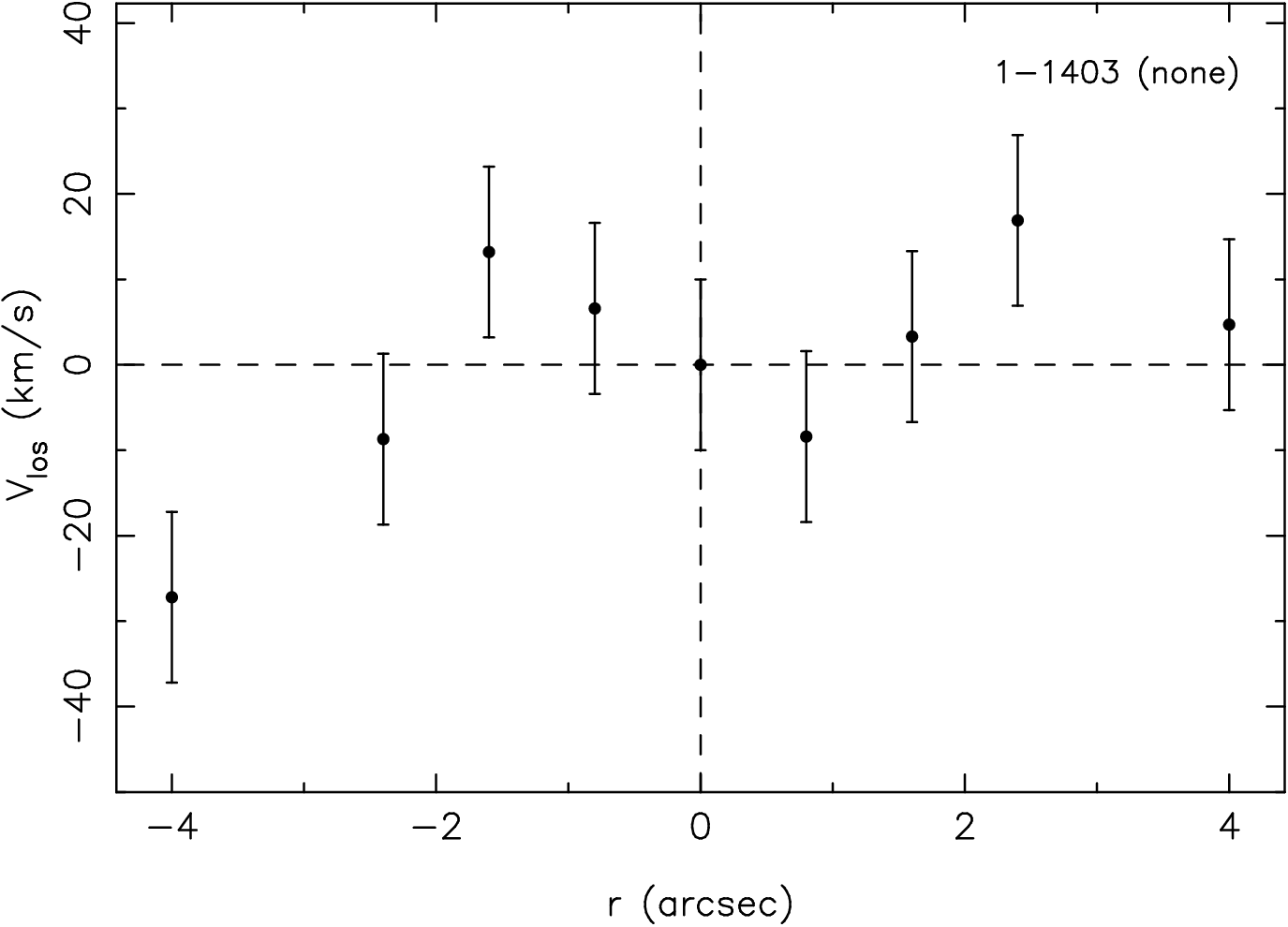}}
\resizebox{0.25\textwidth}{!}{\includegraphics[angle=-90]{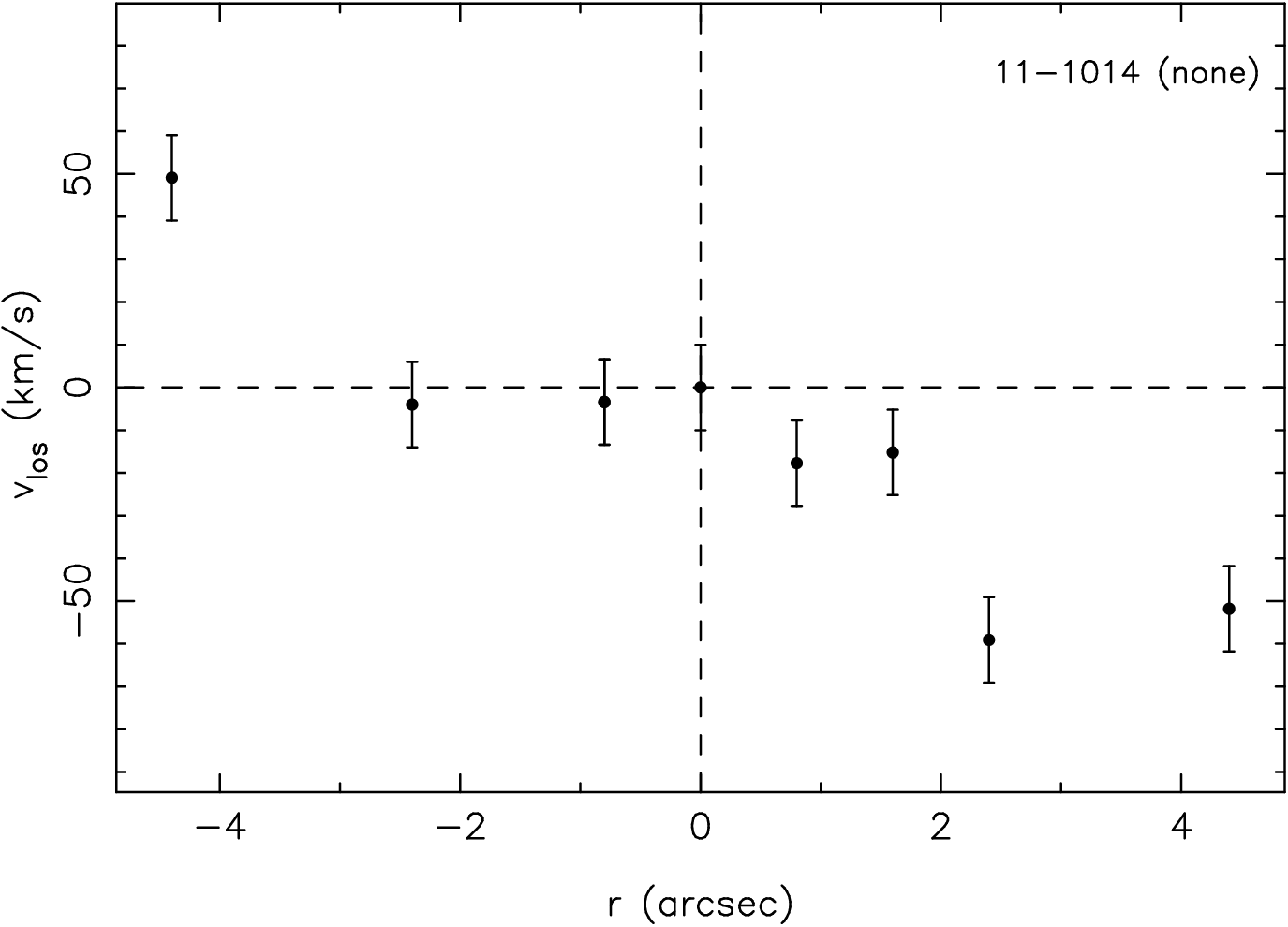}}
\resizebox{0.25\textwidth}{!}{\includegraphics[angle=-90]{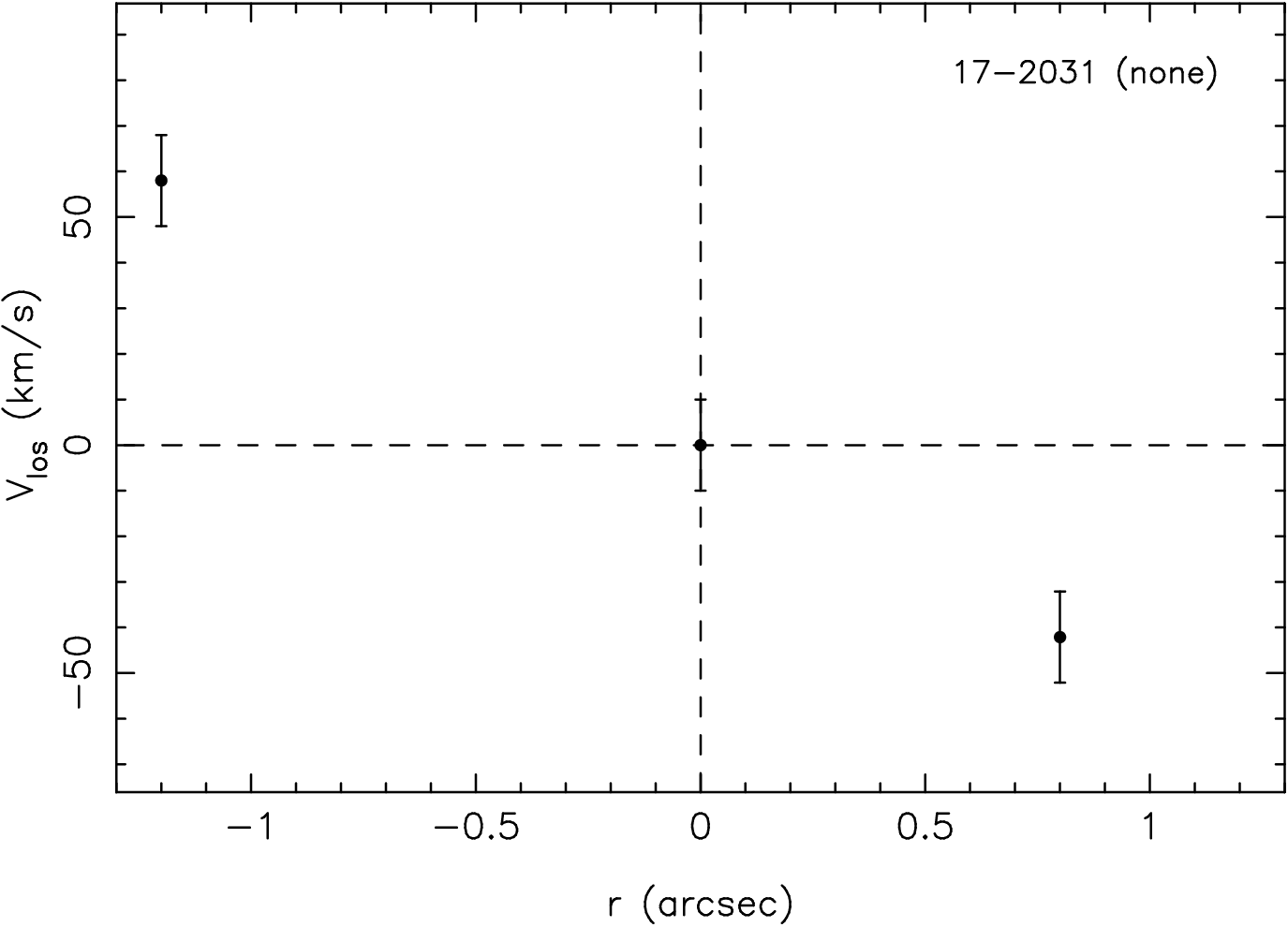}}
\resizebox{0.25\textwidth}{!}{\includegraphics[angle=-90]{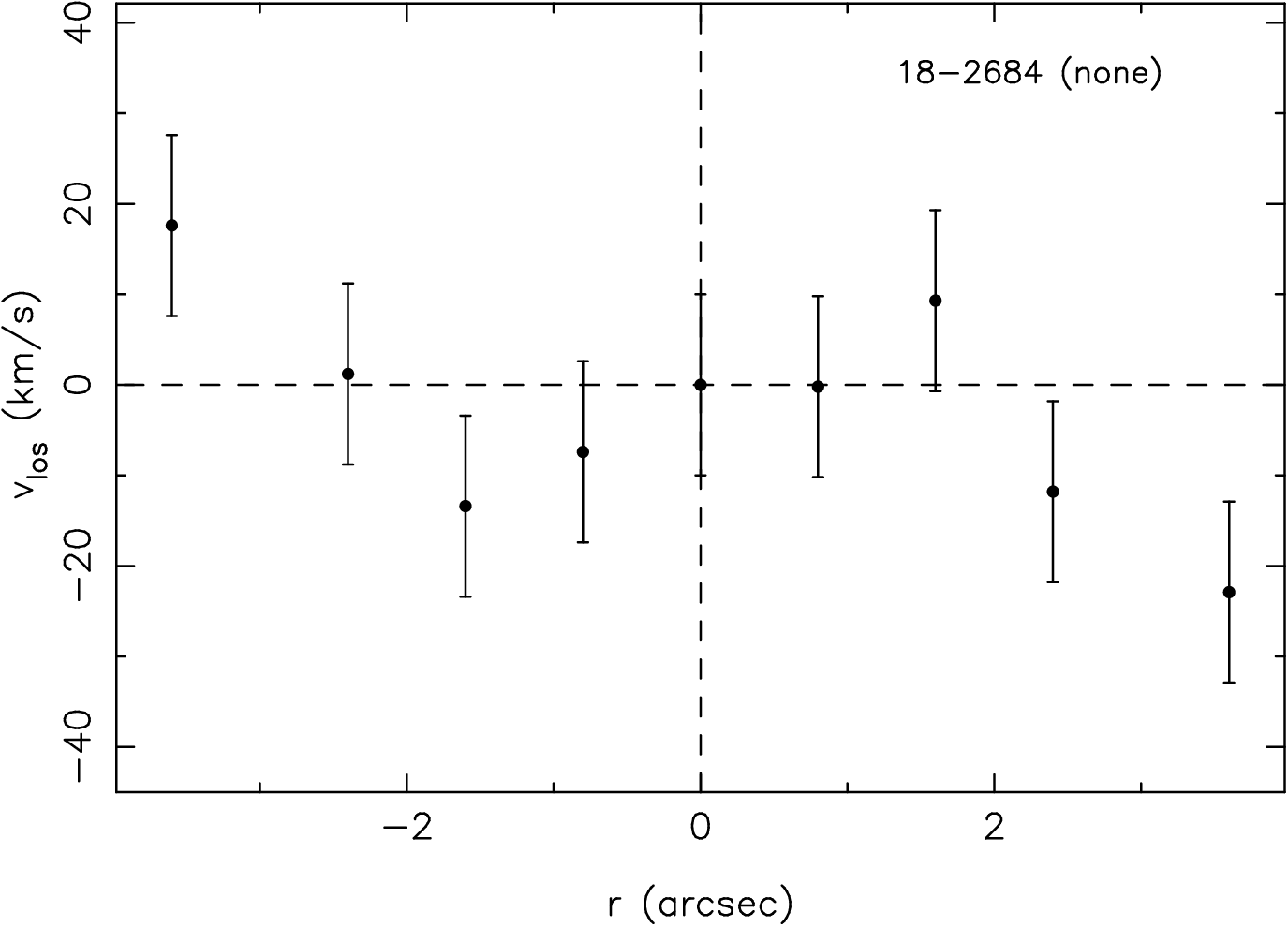}}
\resizebox{0.25\textwidth}{!}{\includegraphics[angle=-90]{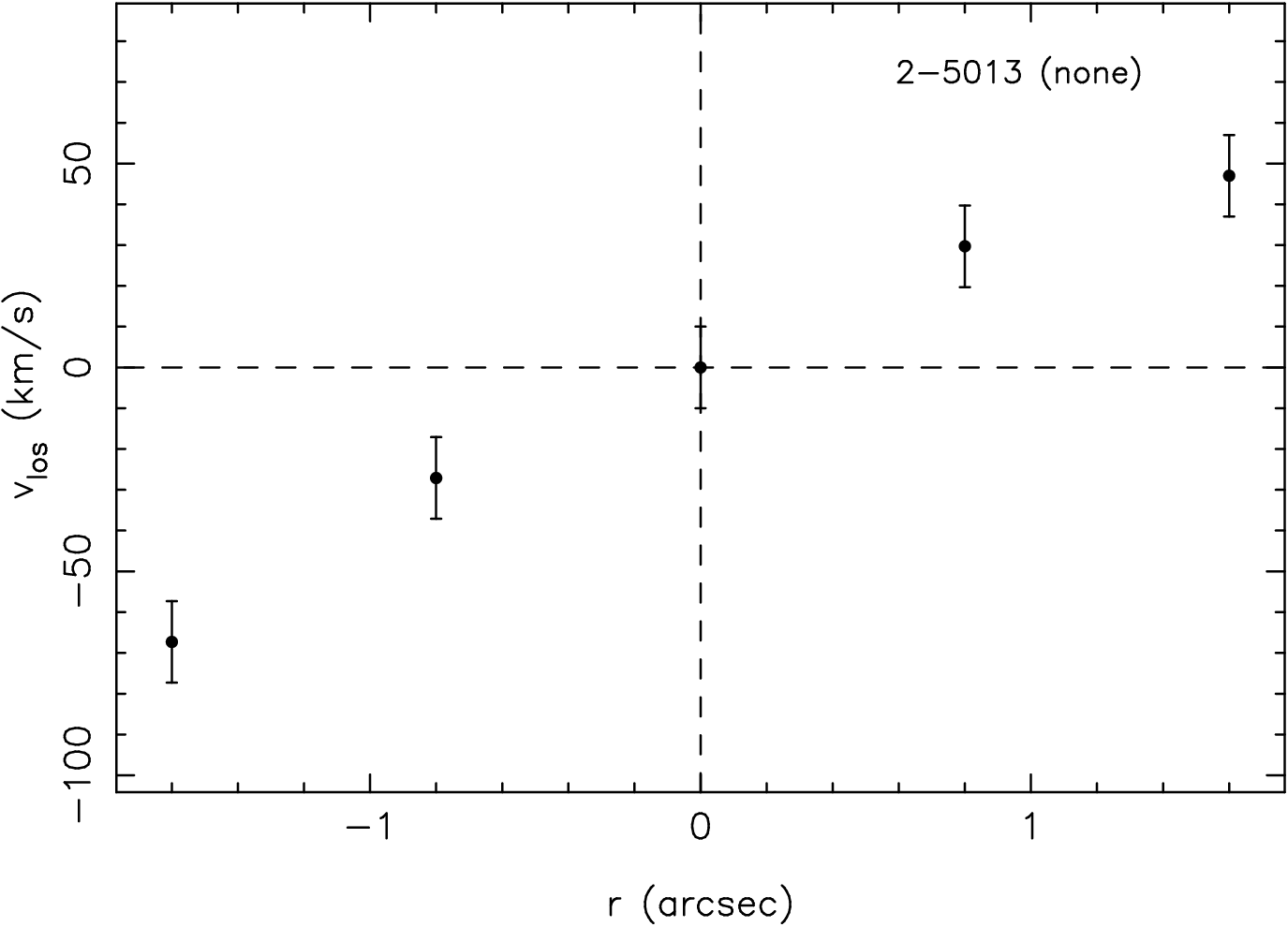}}
\resizebox{0.25\textwidth}{!}{\includegraphics[angle=-90]{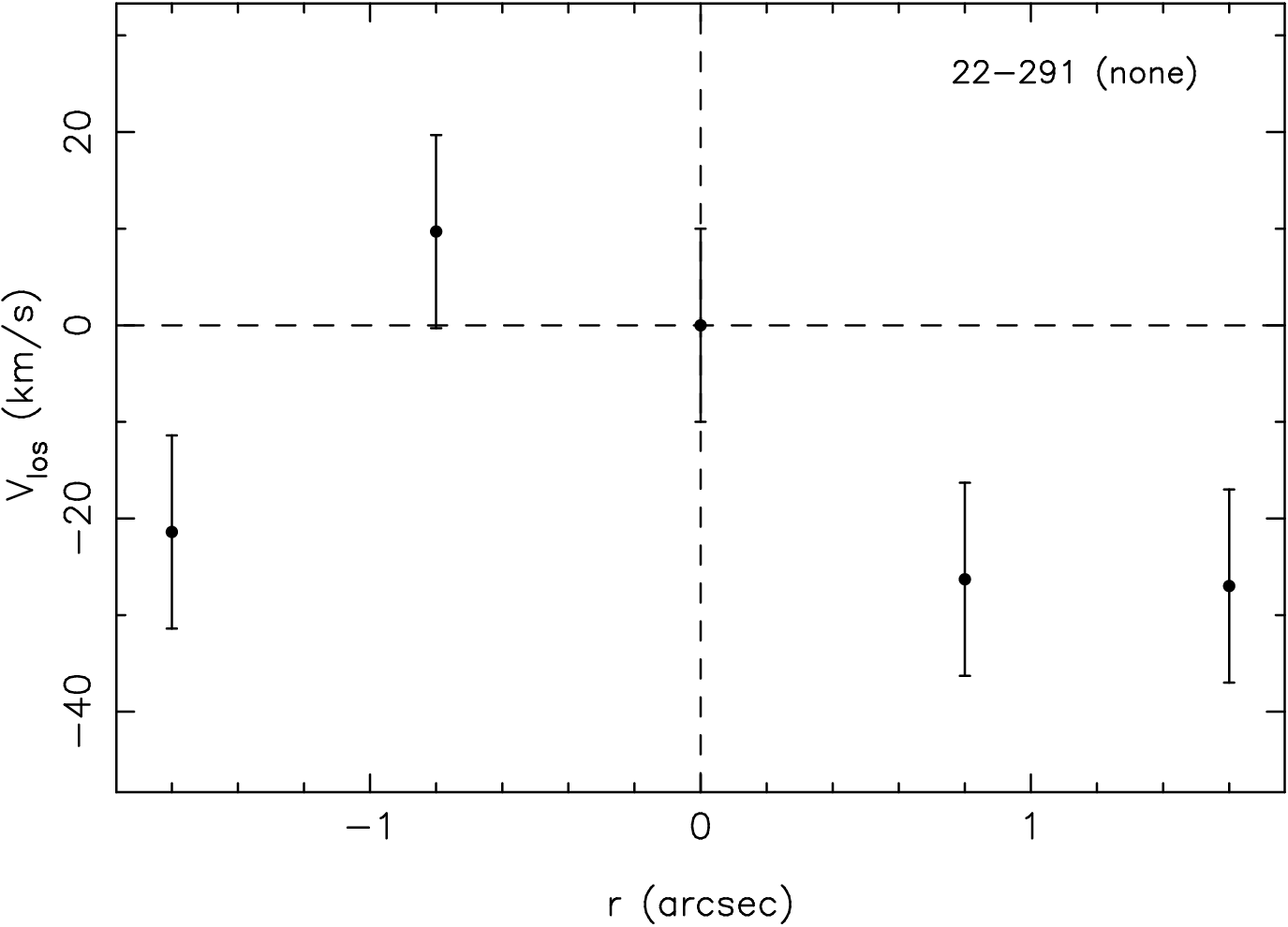}}
\caption{Rotation curves for undisturbed galaxies\label{fig:rotationcurves}}
\end{figure*}

Figure~\ref{fig:rotationcurves} shows the observed line-of-sight velocity 
as a function of the projected radius for the galaxies of our sample.

\addtocounter{figure}{-1}

\begin{figure*}
\resizebox{0.25\textwidth}{!}{\includegraphics[angle=-90]{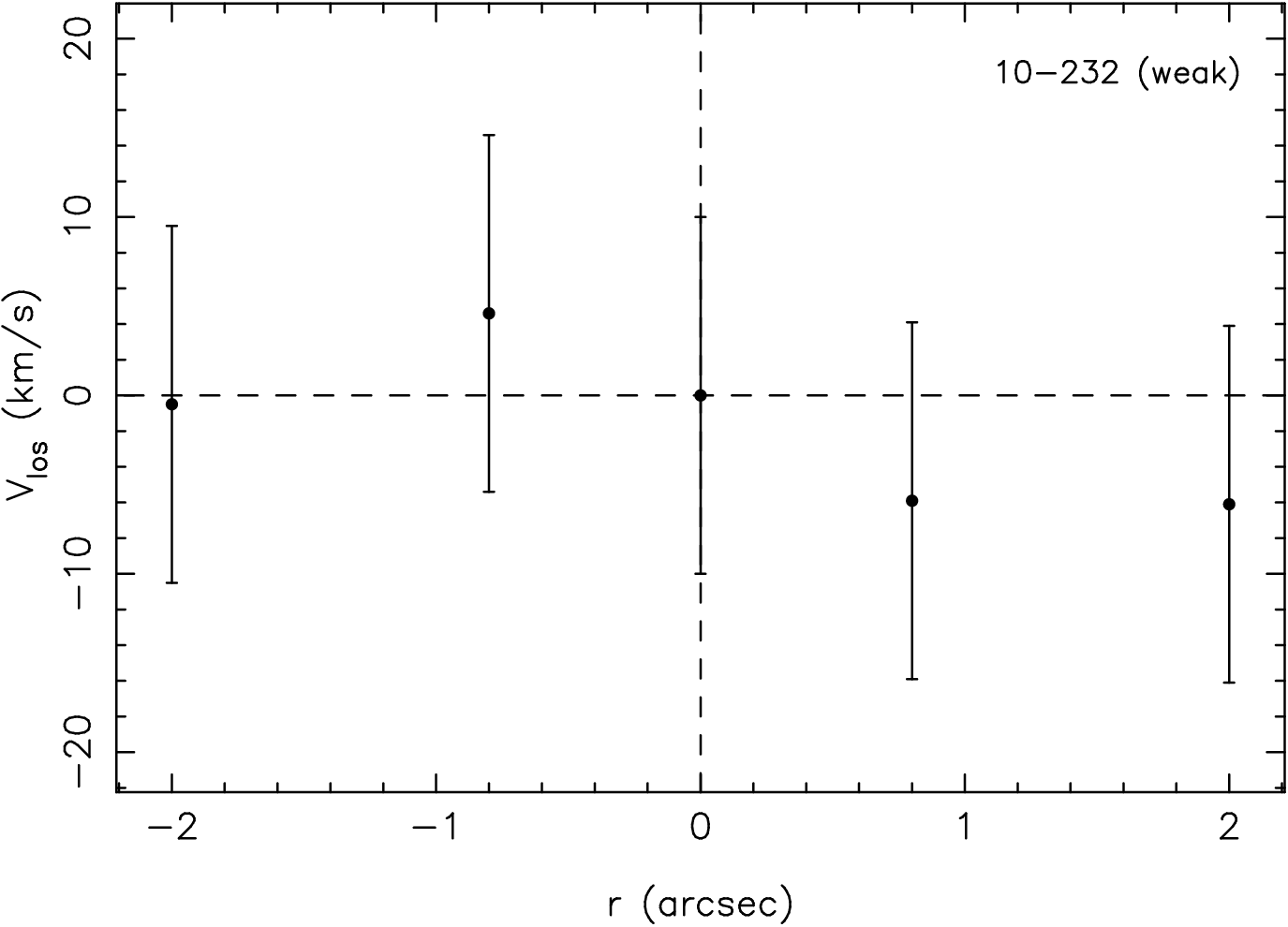}}
\resizebox{0.25\textwidth}{!}{\includegraphics[angle=-90]{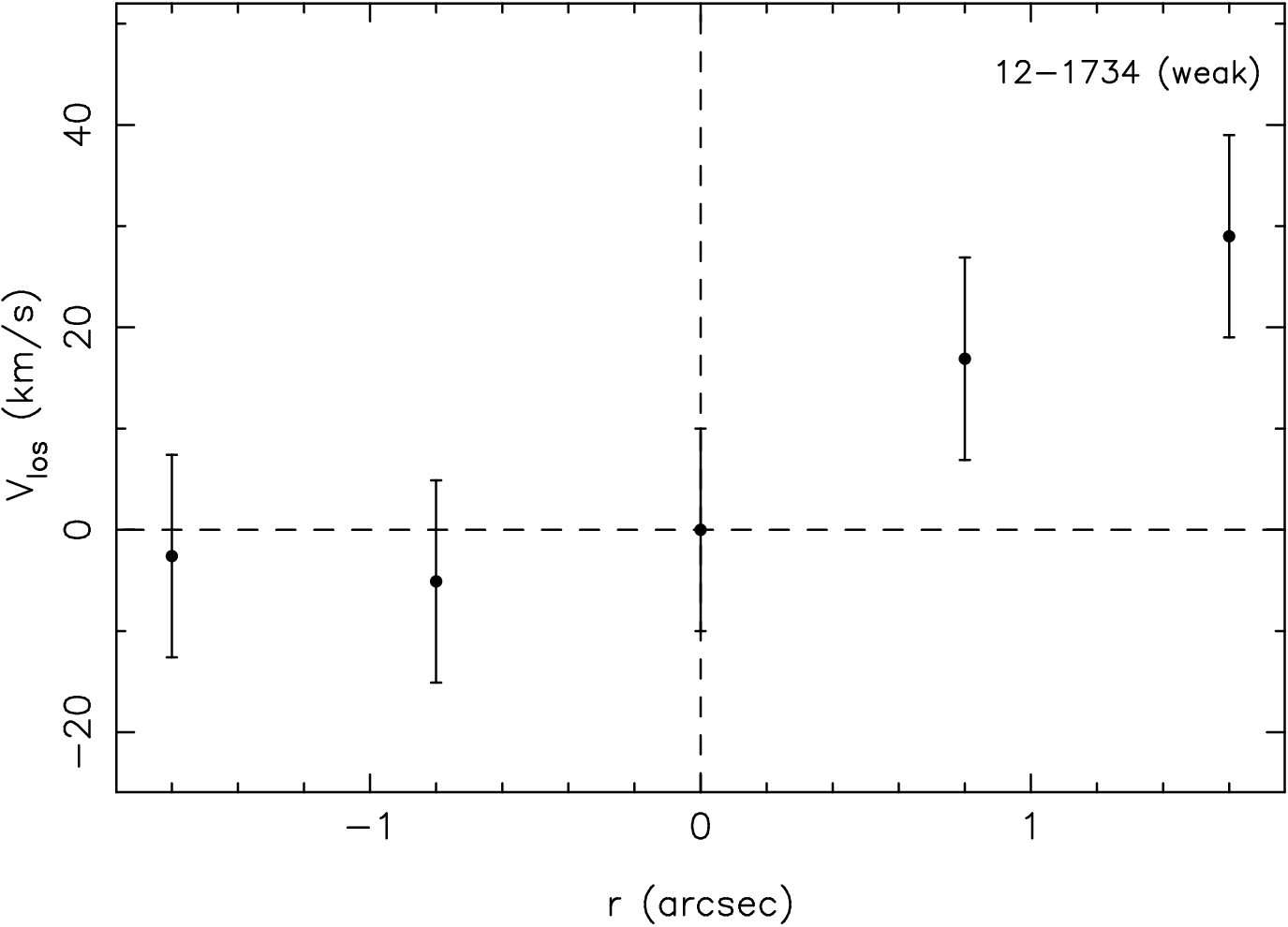}}
\resizebox{0.25\textwidth}{!}{\includegraphics[angle=-90]{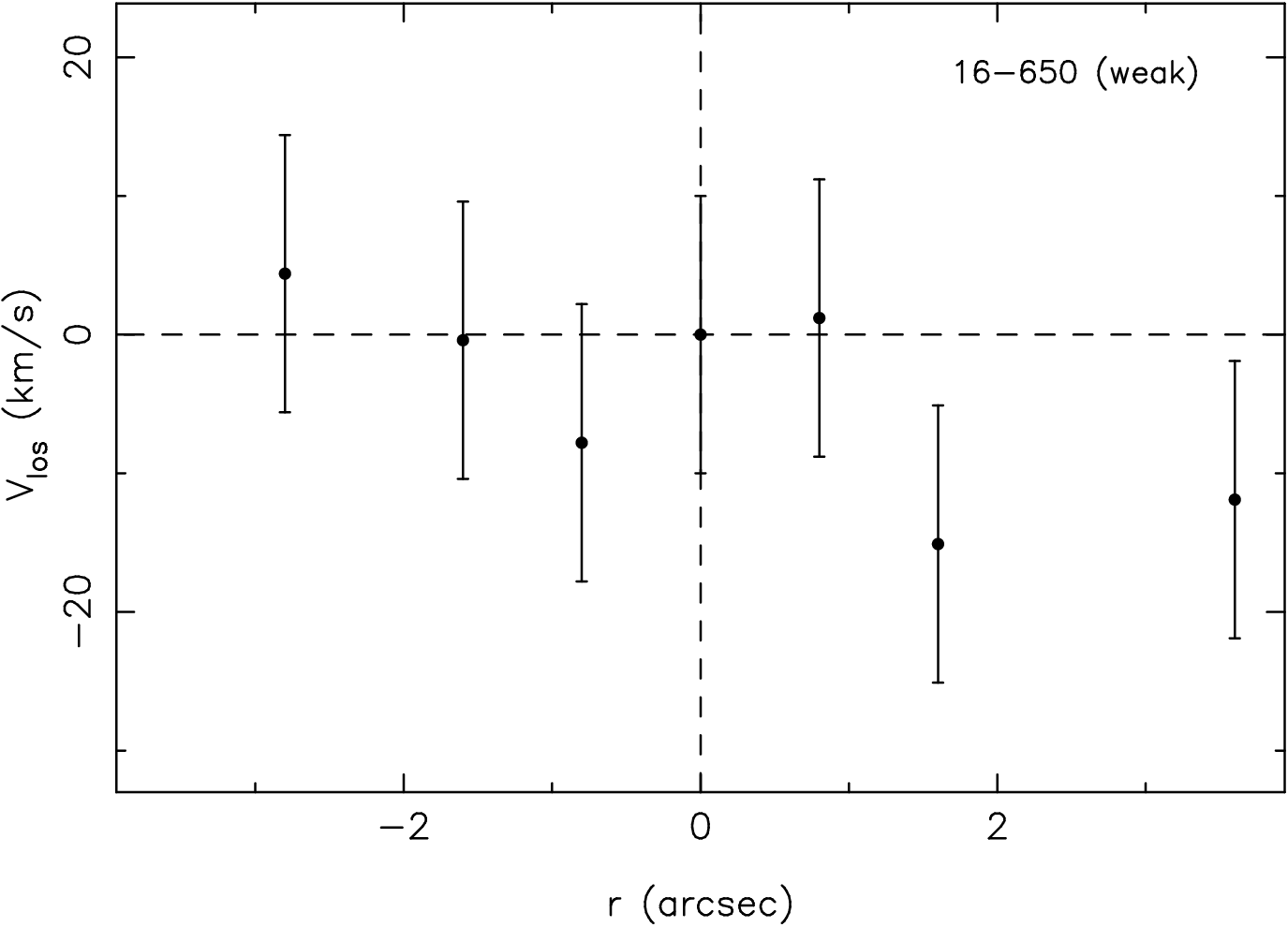}}
\resizebox{0.25\textwidth}{!}{\includegraphics[angle=-90]{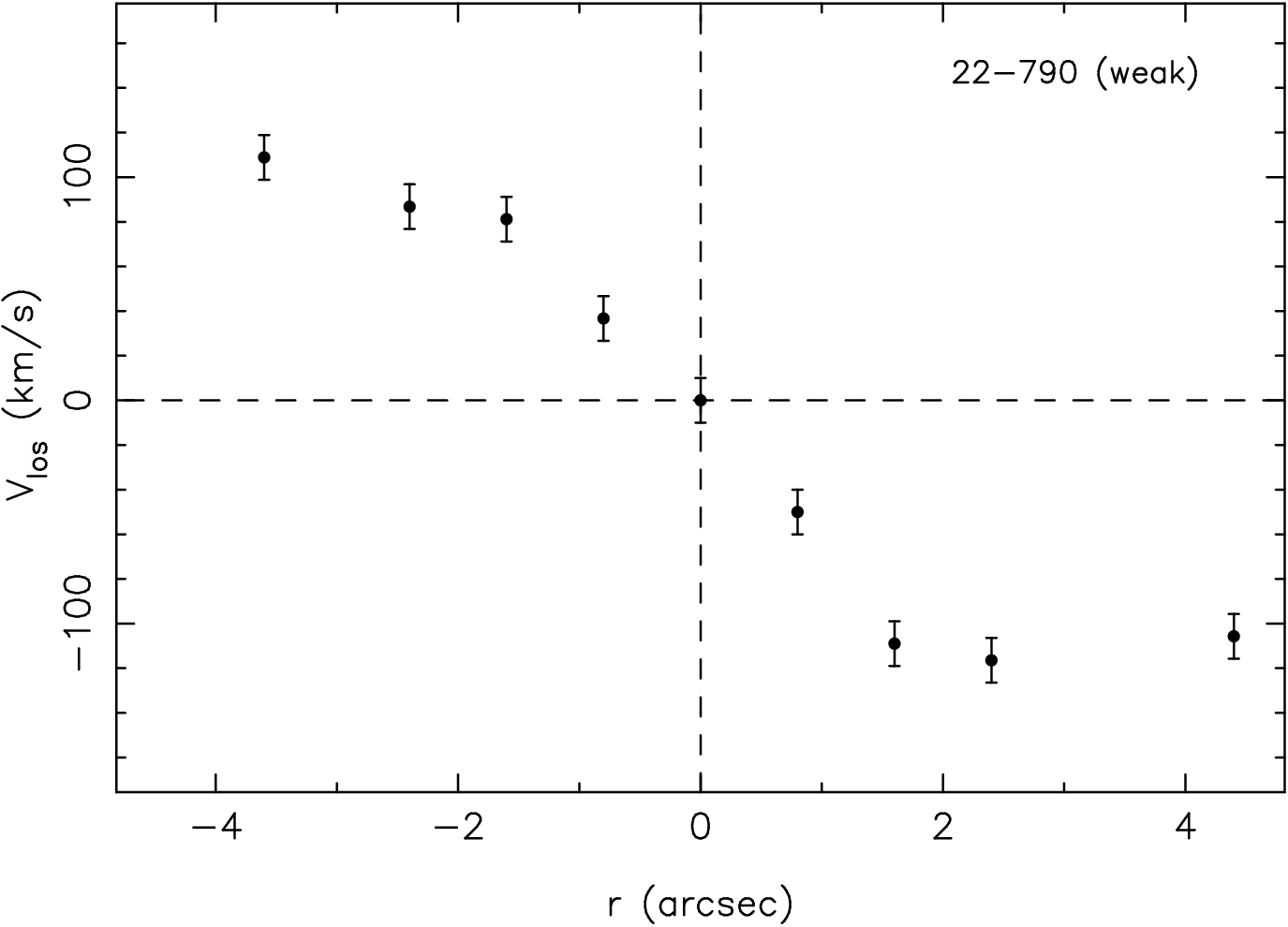}}
\resizebox{0.25\textwidth}{!}{\includegraphics[angle=-90]{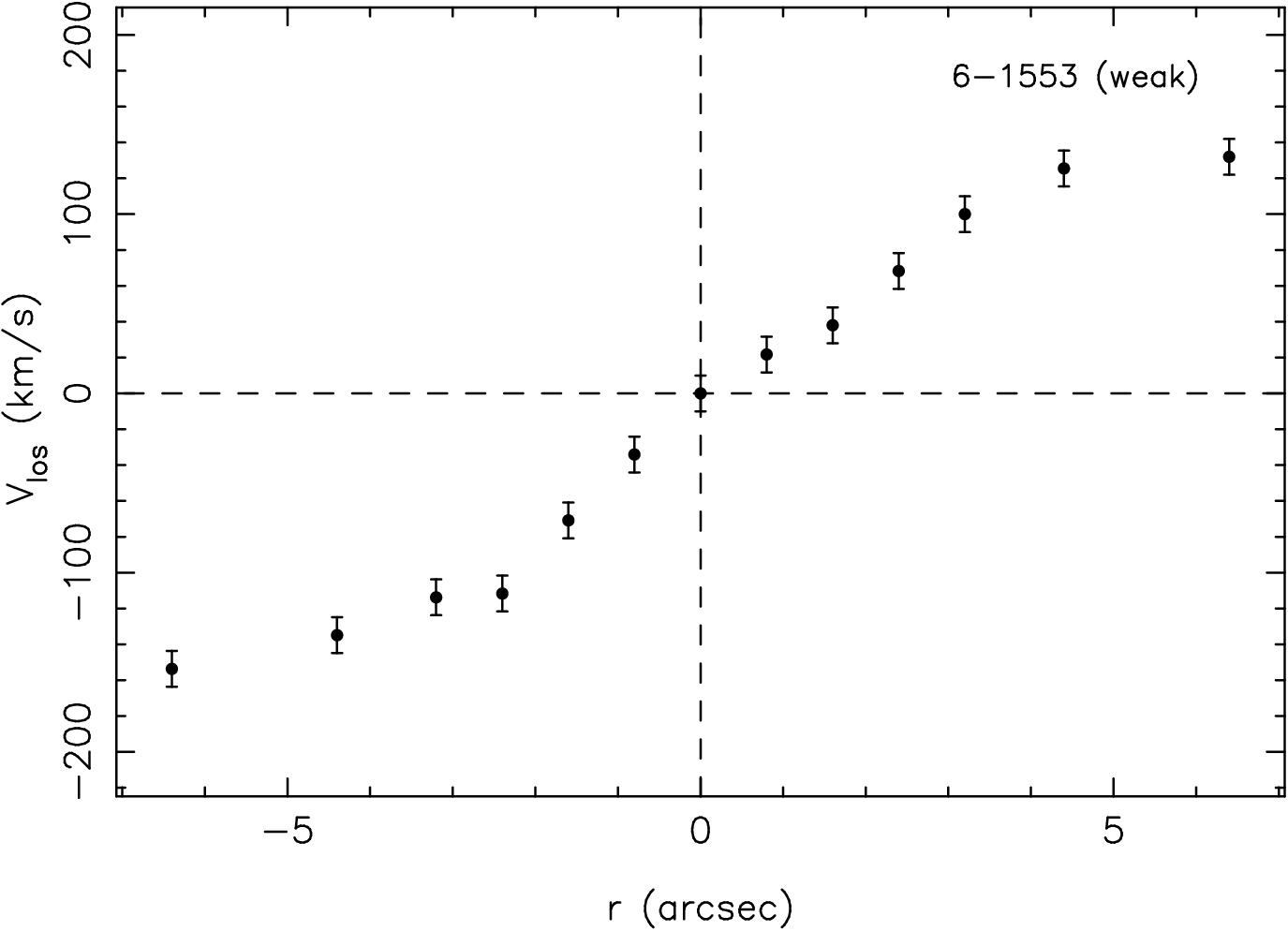}}
\resizebox{0.25\textwidth}{!}{\includegraphics[angle=-90]{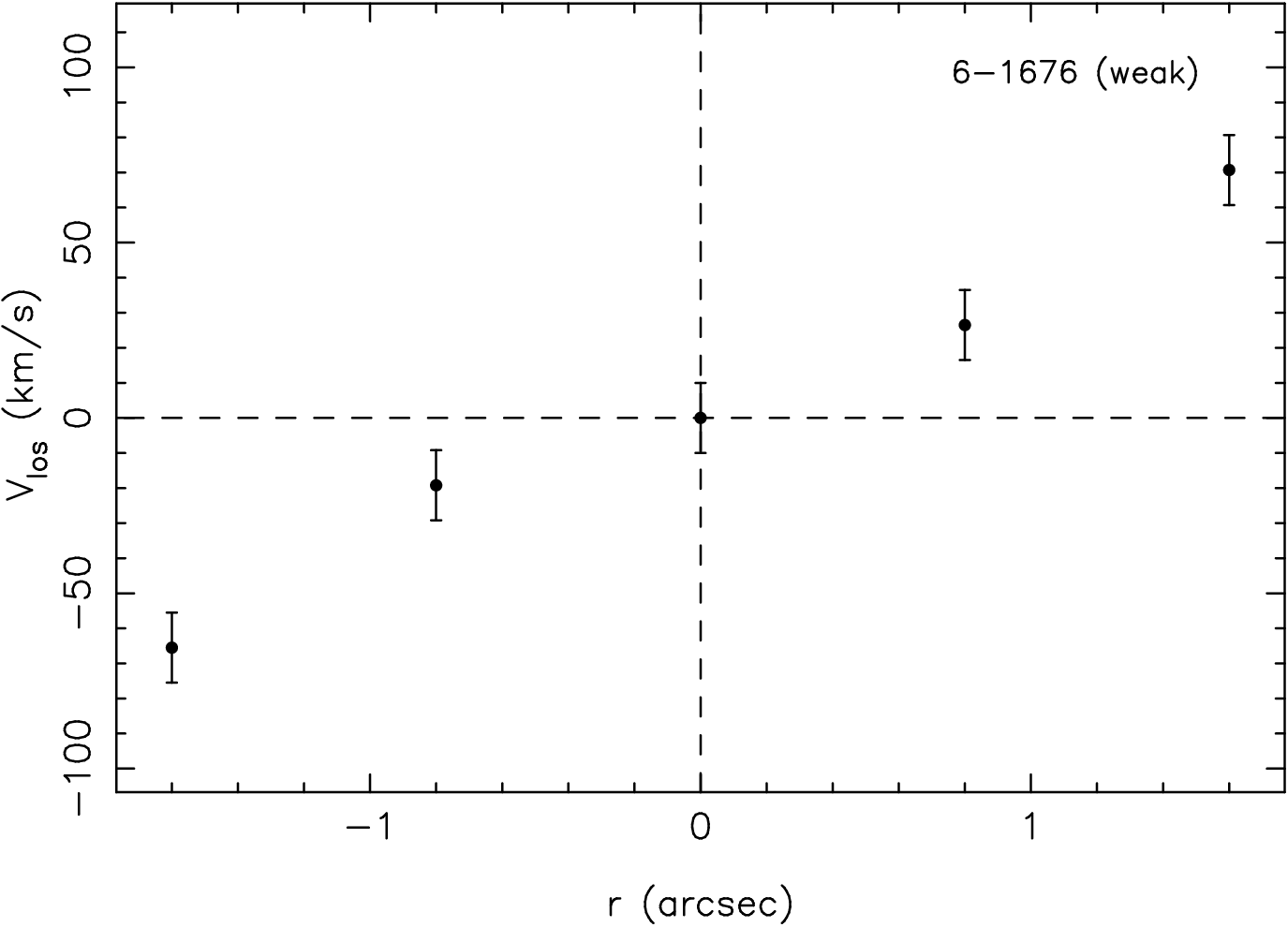}}
\resizebox{0.25\textwidth}{!}{\includegraphics[angle=-90]{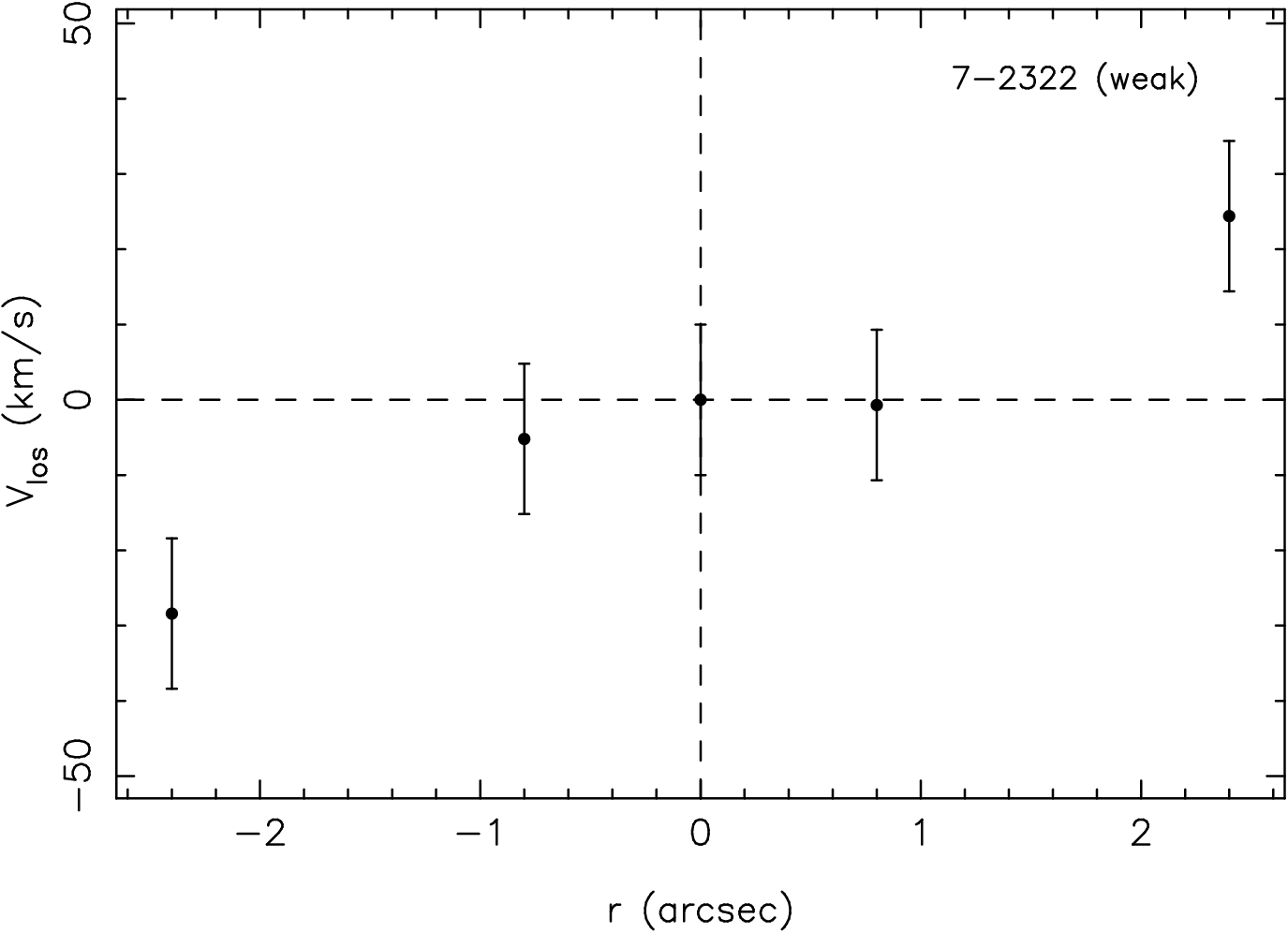}}
\resizebox{0.25\textwidth}{!}{\includegraphics[angle=-90]{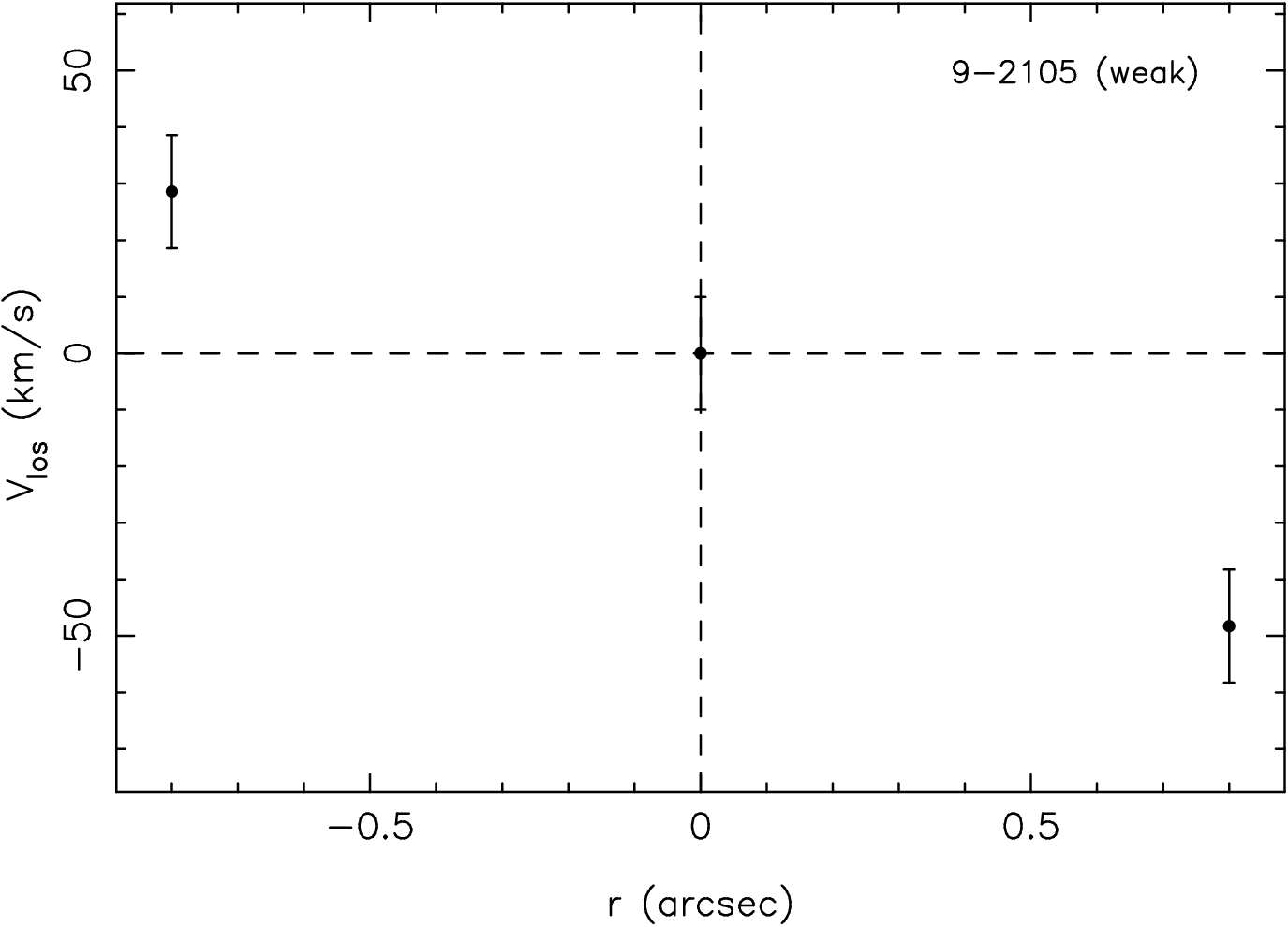}}
\caption{Rotation curve for galaxies showing weak morphological disturbances.}
\end{figure*}
\addtocounter{figure}{-1}

\begin{figure*}
\resizebox{0.25\textwidth}{!}{\includegraphics[angle=-90]{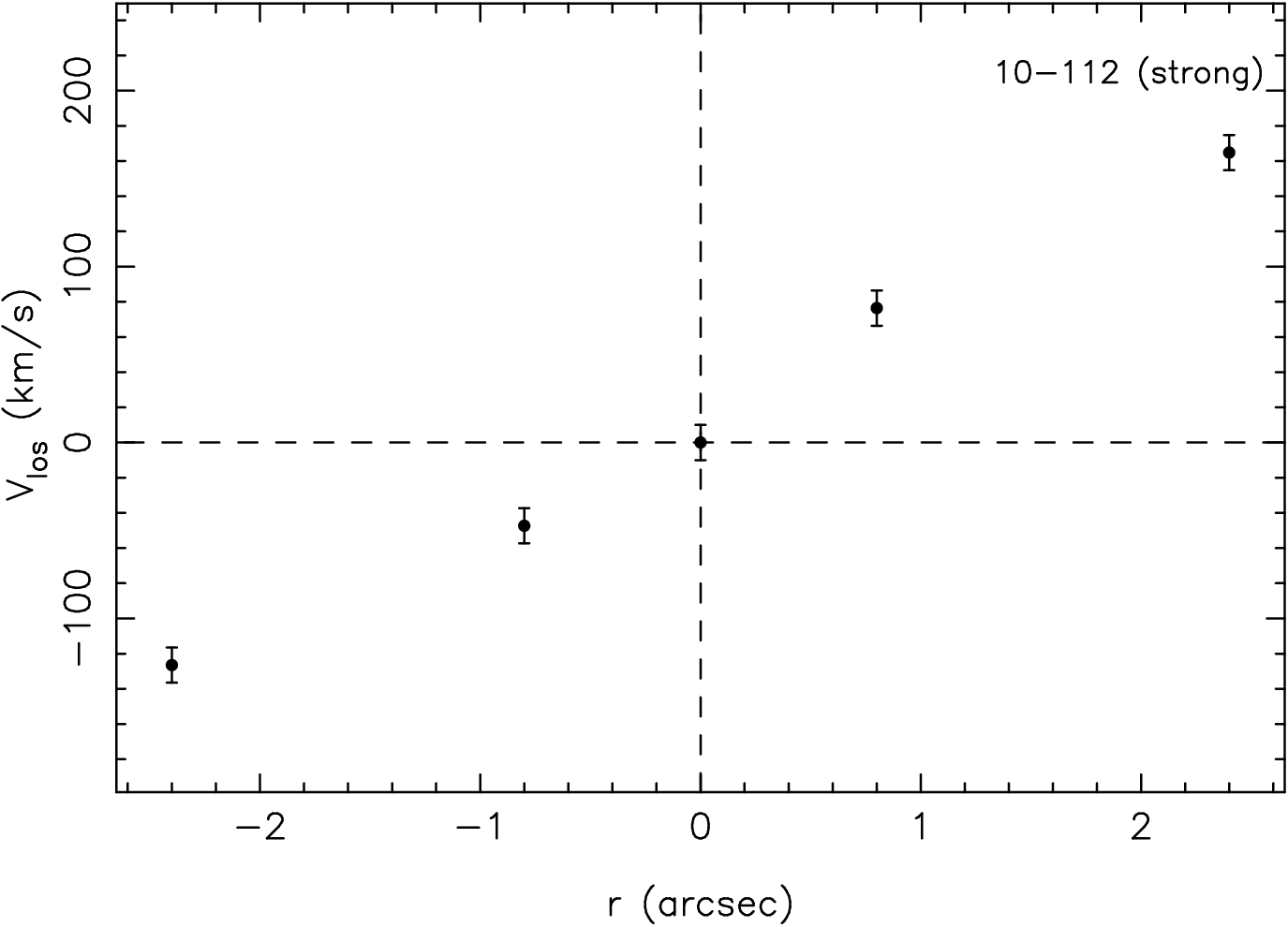}}
\resizebox{0.25\textwidth}{!}{\includegraphics[angle=-90]{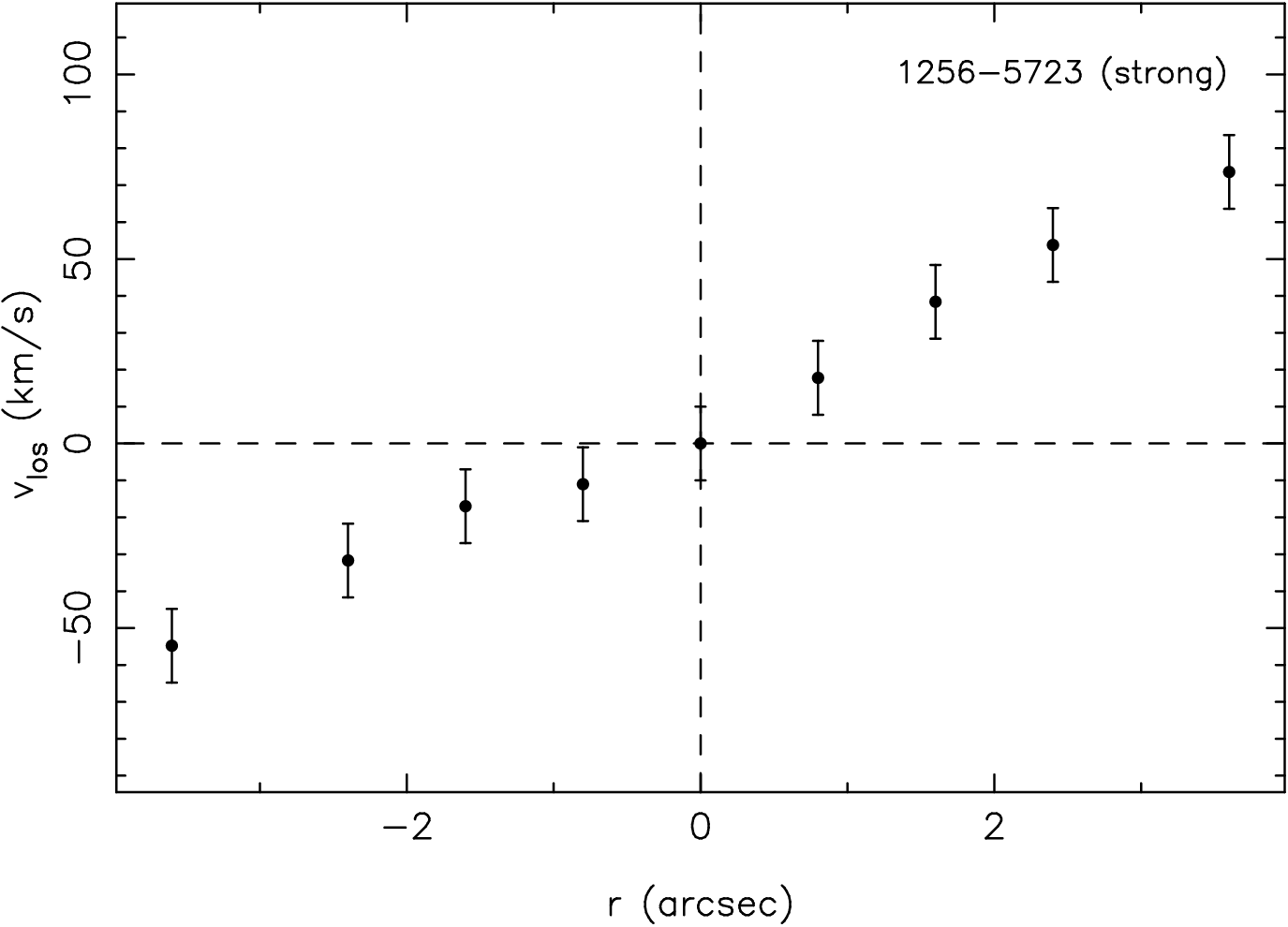}}
\resizebox{0.25\textwidth}{!}{\includegraphics[angle=-90]{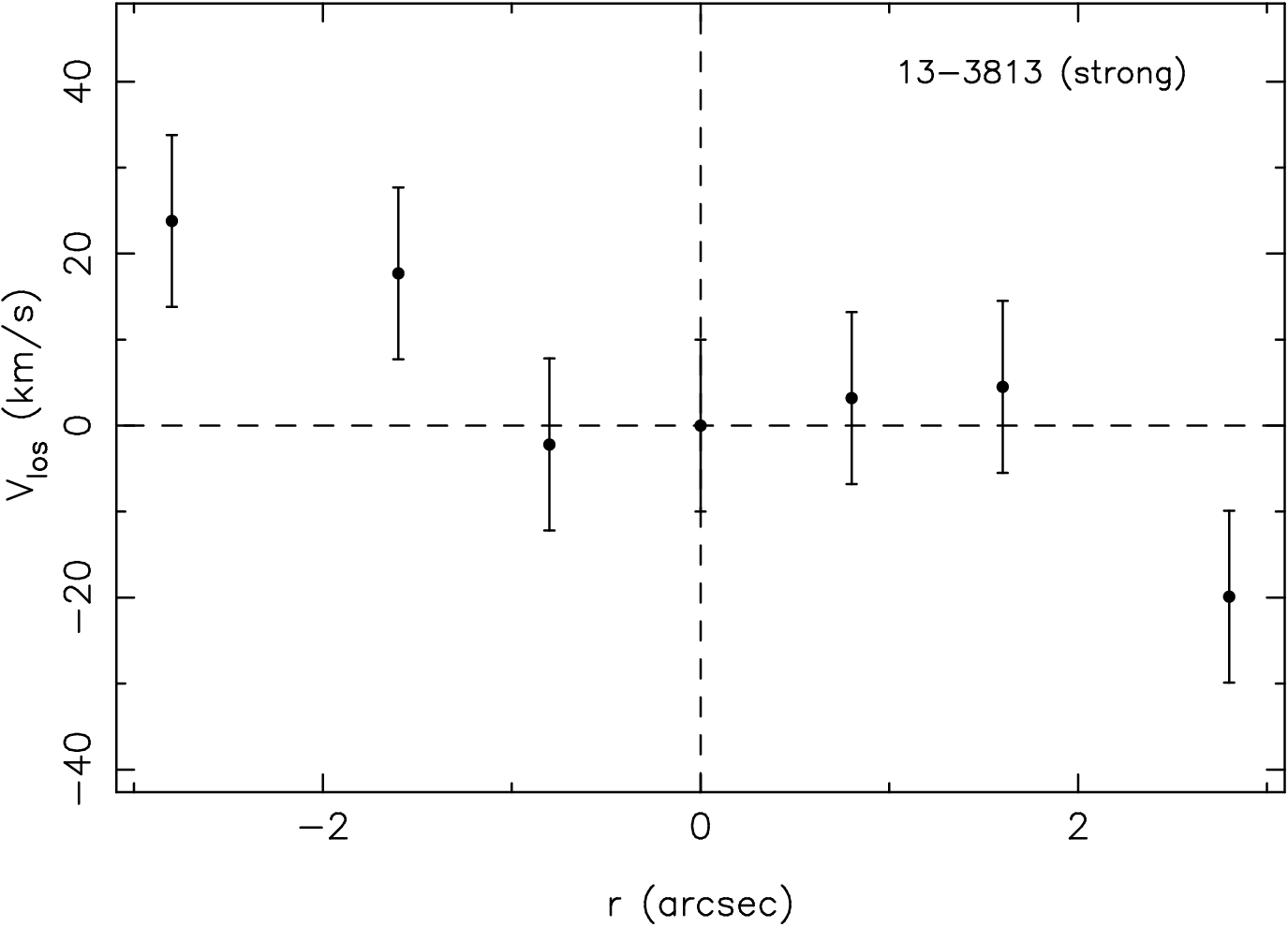}}
\resizebox{0.25\textwidth}{!}{\includegraphics[angle=-90]{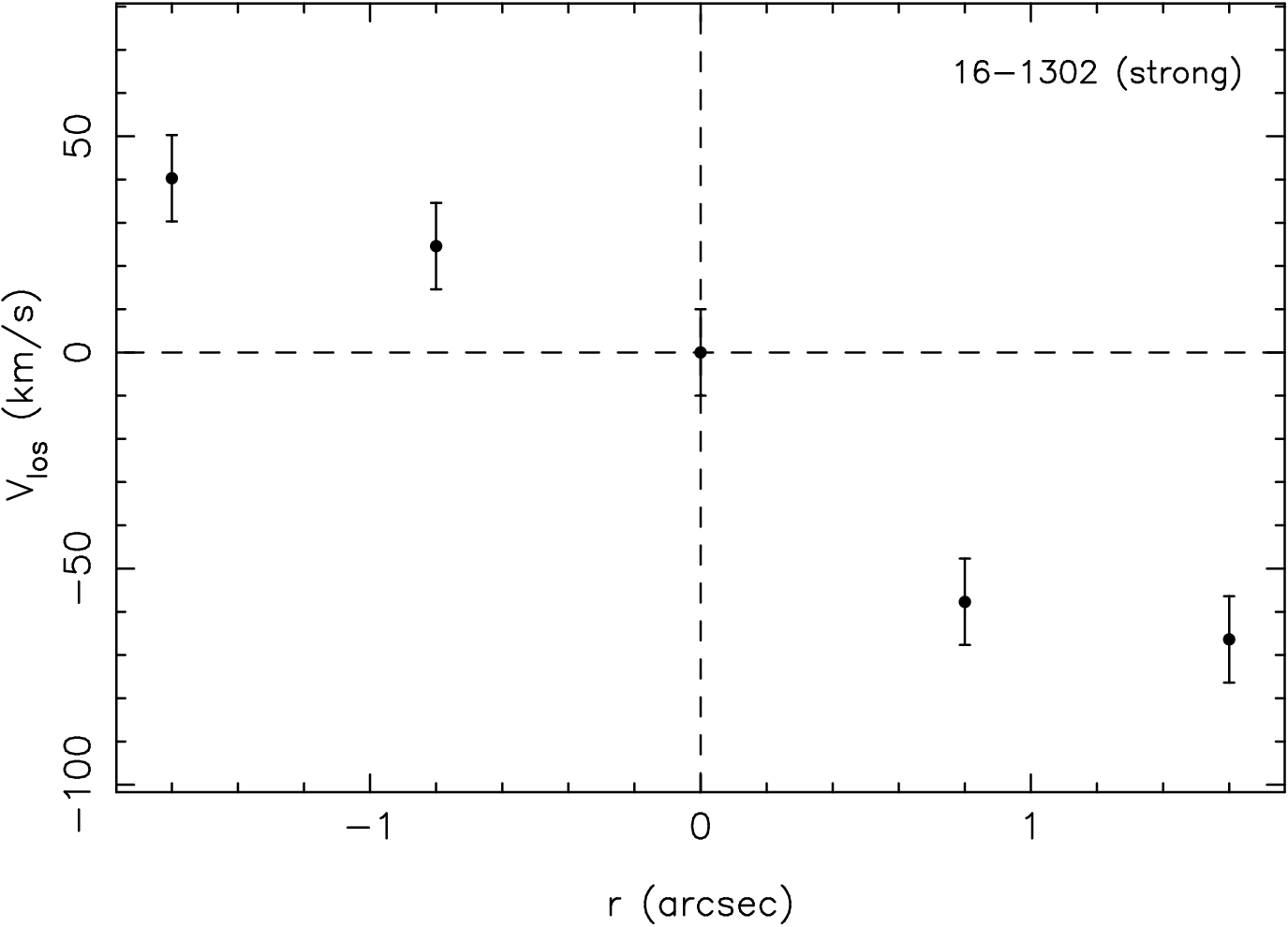}}
\resizebox{0.25\textwidth}{!}{\includegraphics[angle=-90]{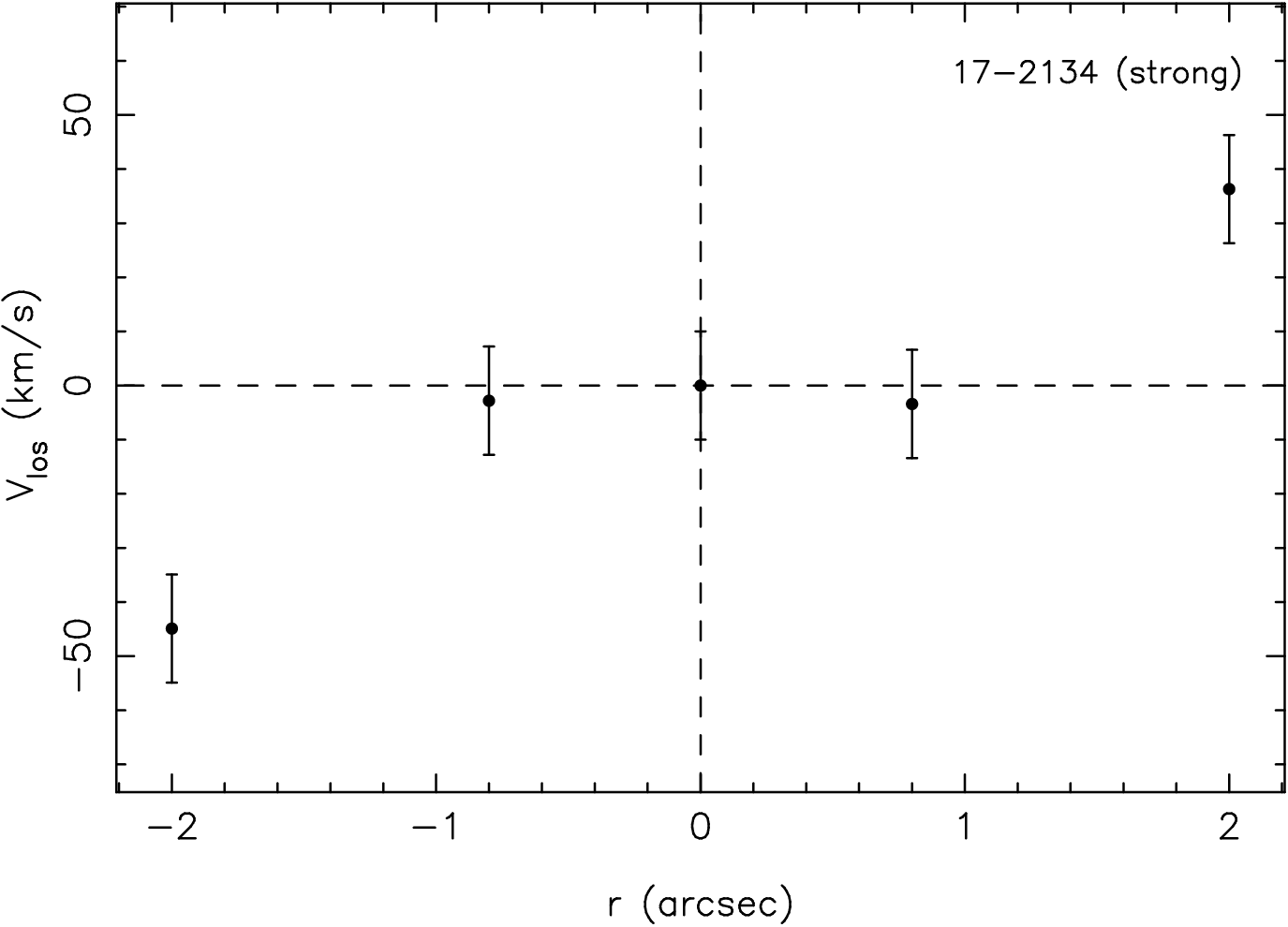}}
\resizebox{0.25\textwidth}{!}{\includegraphics[angle=-90]{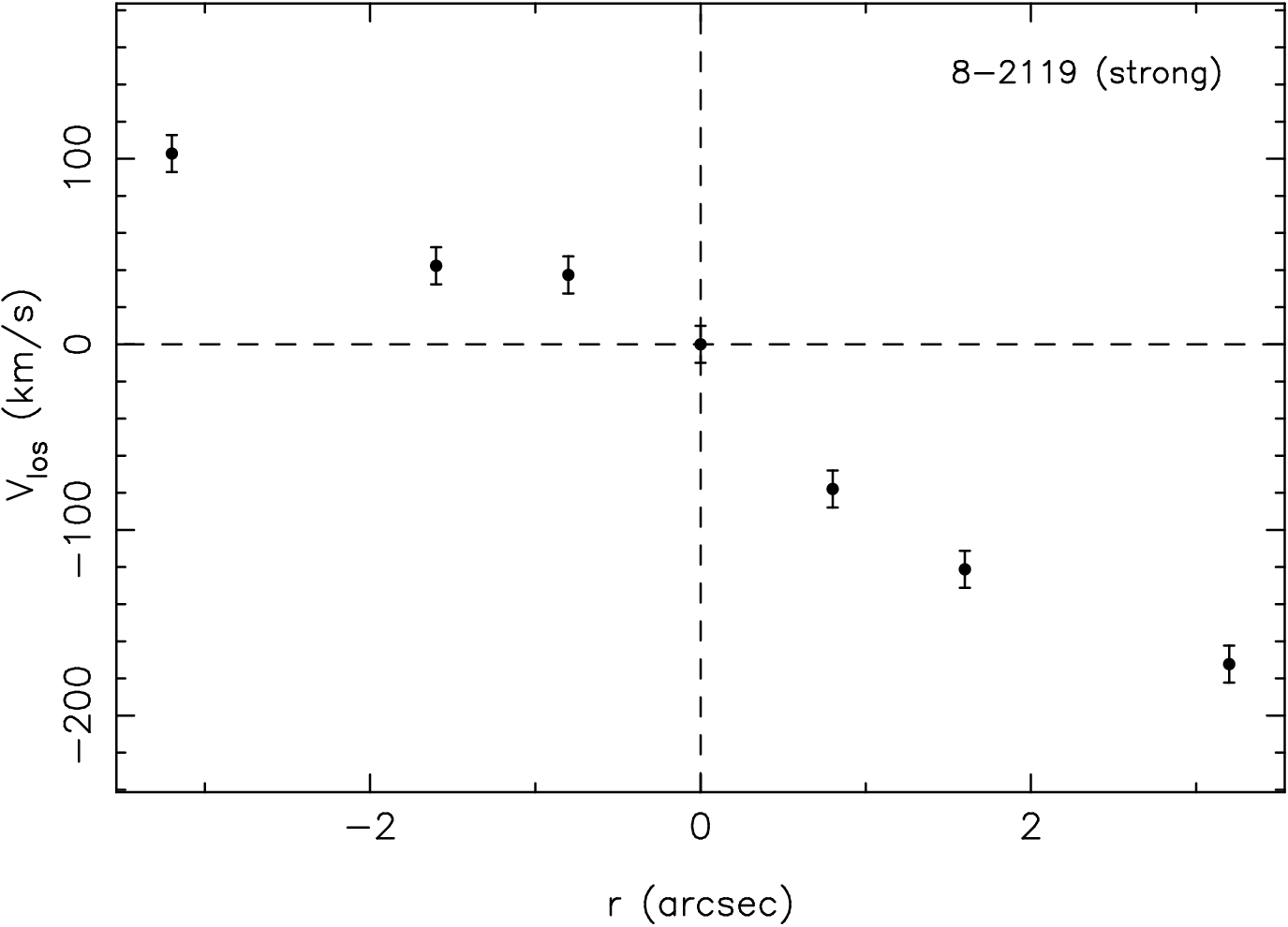}}
\resizebox{0.25\textwidth}{!}{\includegraphics[angle=-90]{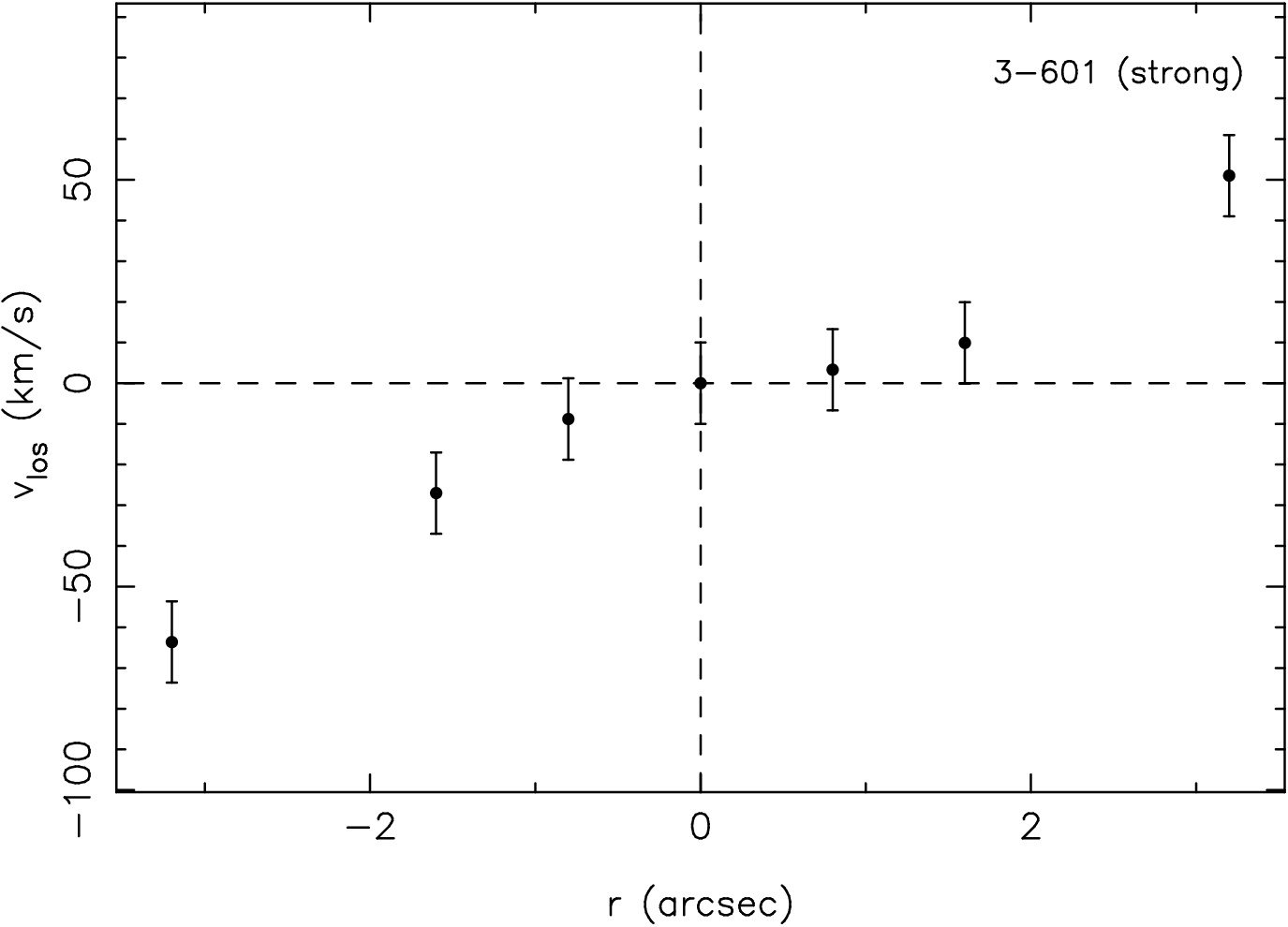}}
\resizebox{0.25\textwidth}{!}{\includegraphics[angle=-90]{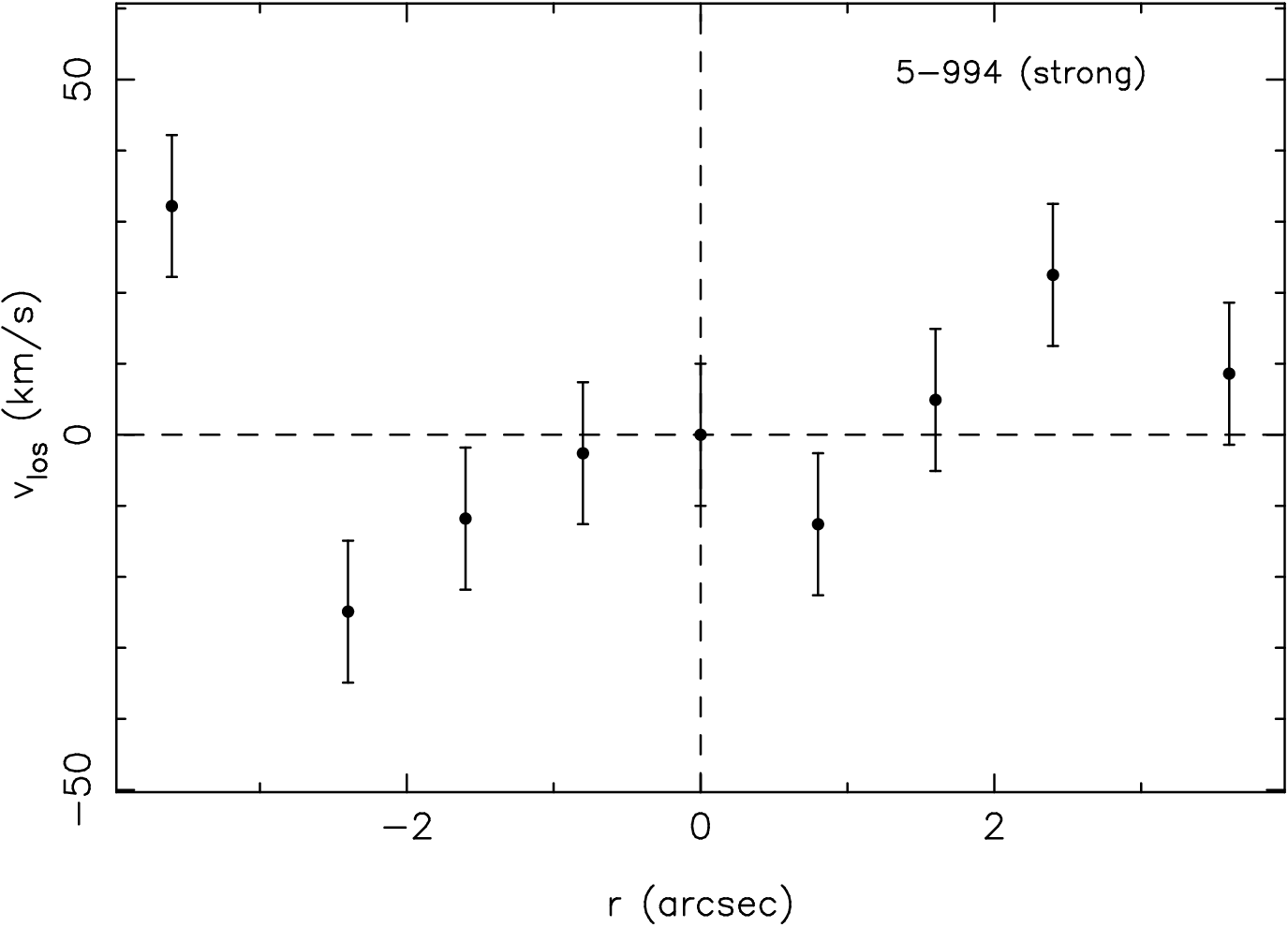}}
\resizebox{0.25\textwidth}{!}{\includegraphics[angle=-90]{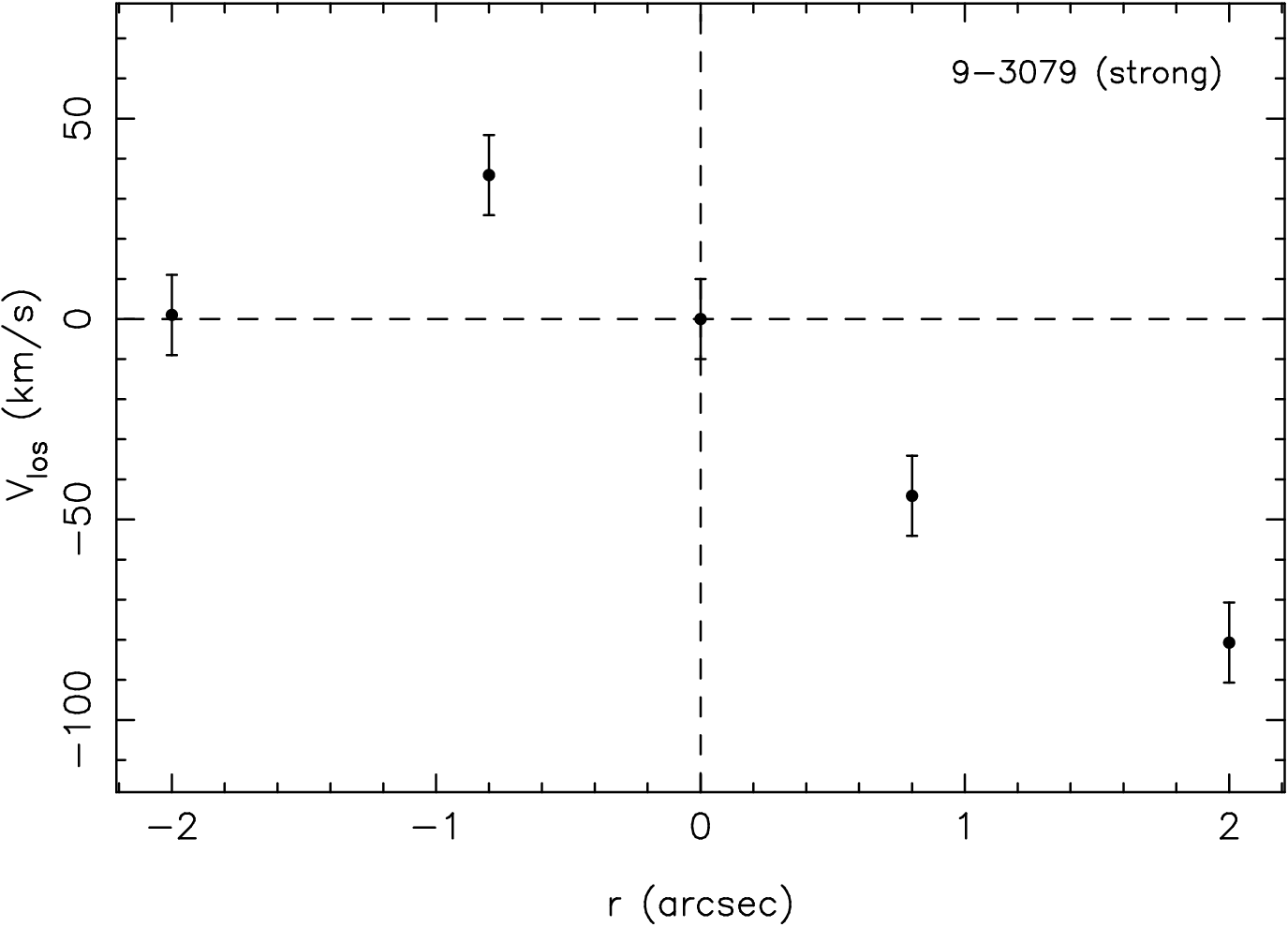}} 
\caption{Rotation curves for galaxies showing strong morphological disturbances.}
\end{figure*}
\addtocounter{figure}{-1}

\begin{figure*}
\resizebox{0.25\textwidth}{!}{\includegraphics[angle=-90]{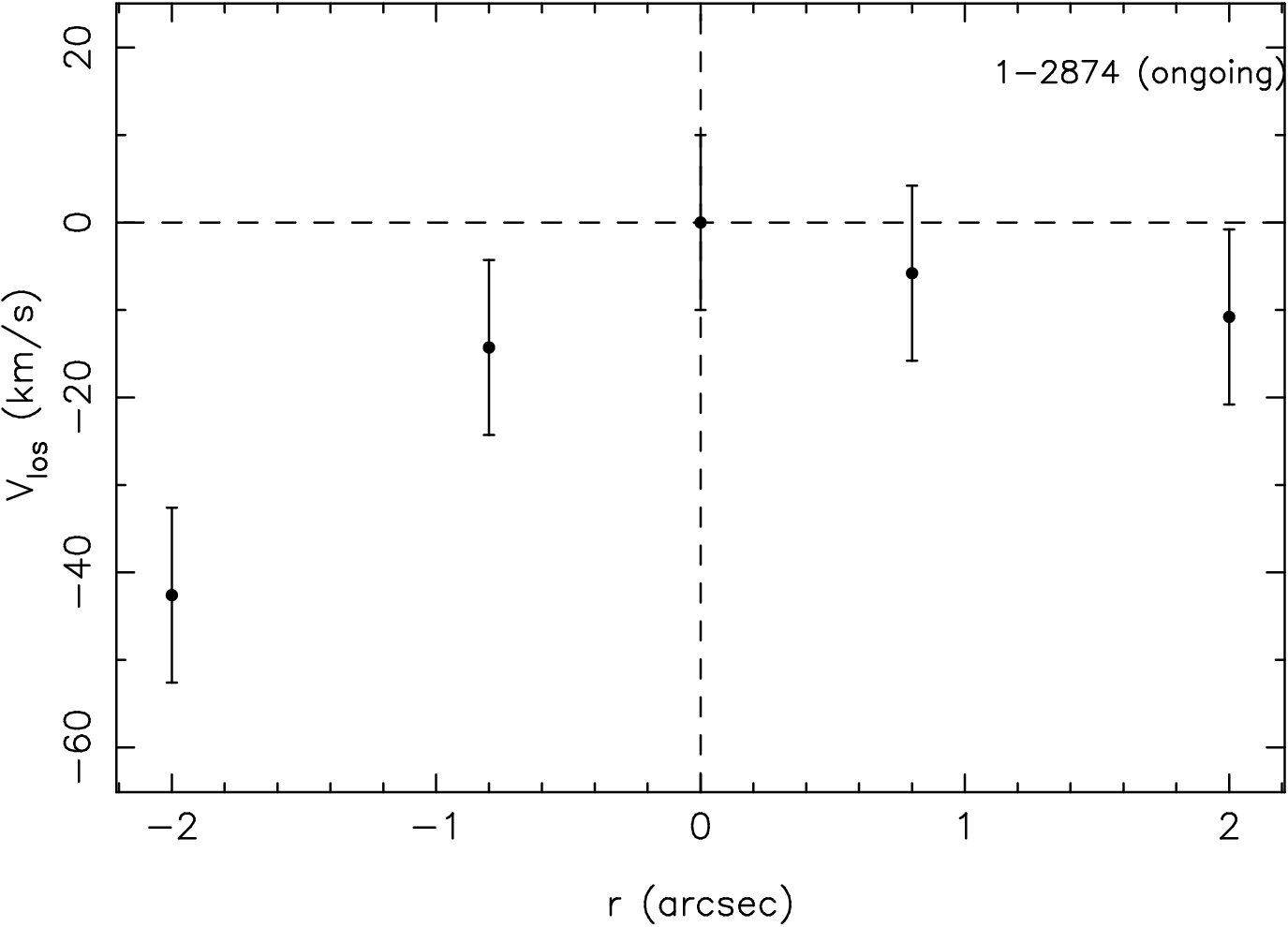}}
\resizebox{0.25\textwidth}{!}{\includegraphics[angle=-90]{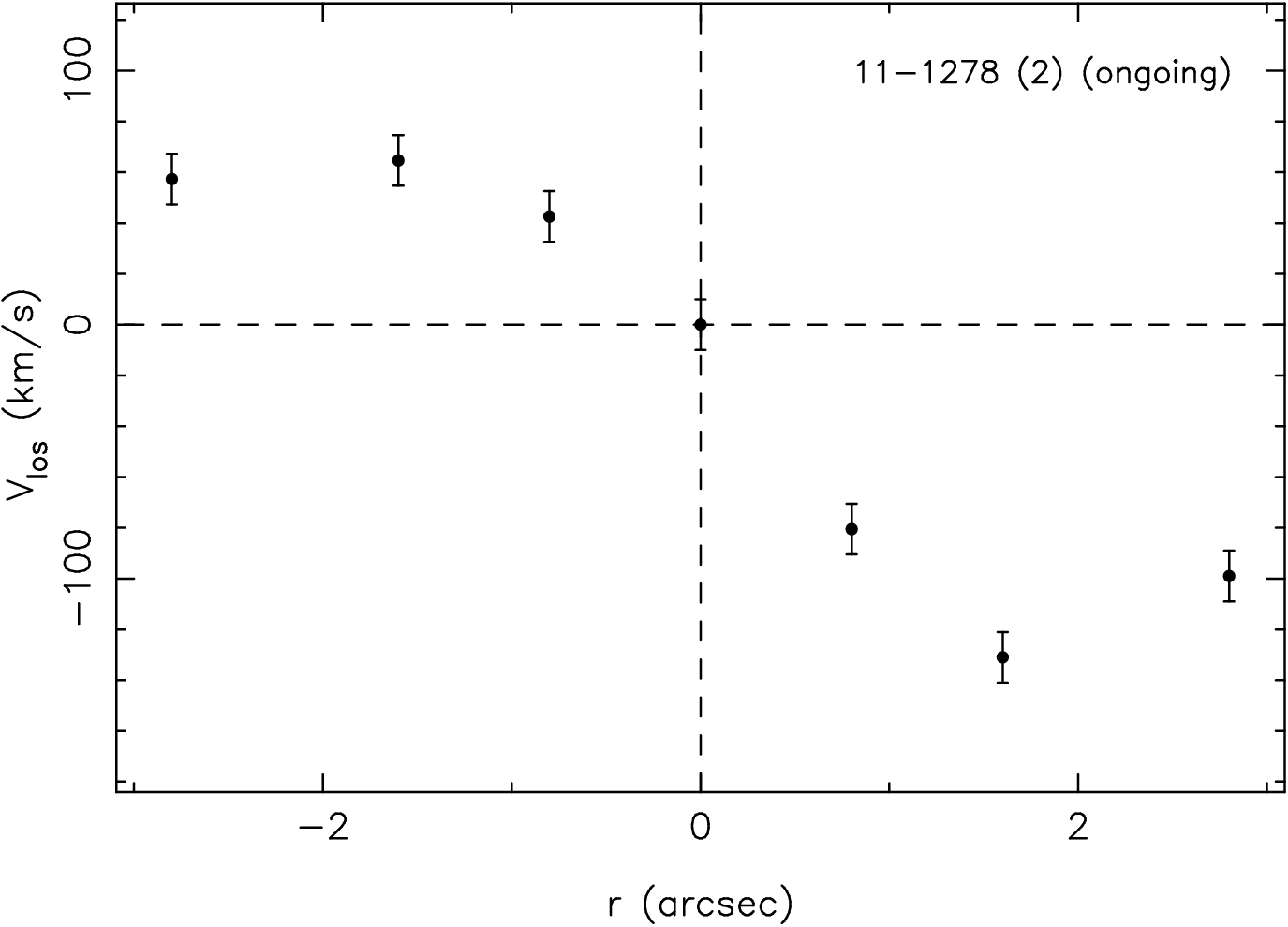}}
\resizebox{0.25\textwidth}{!}{\includegraphics[angle=-90]{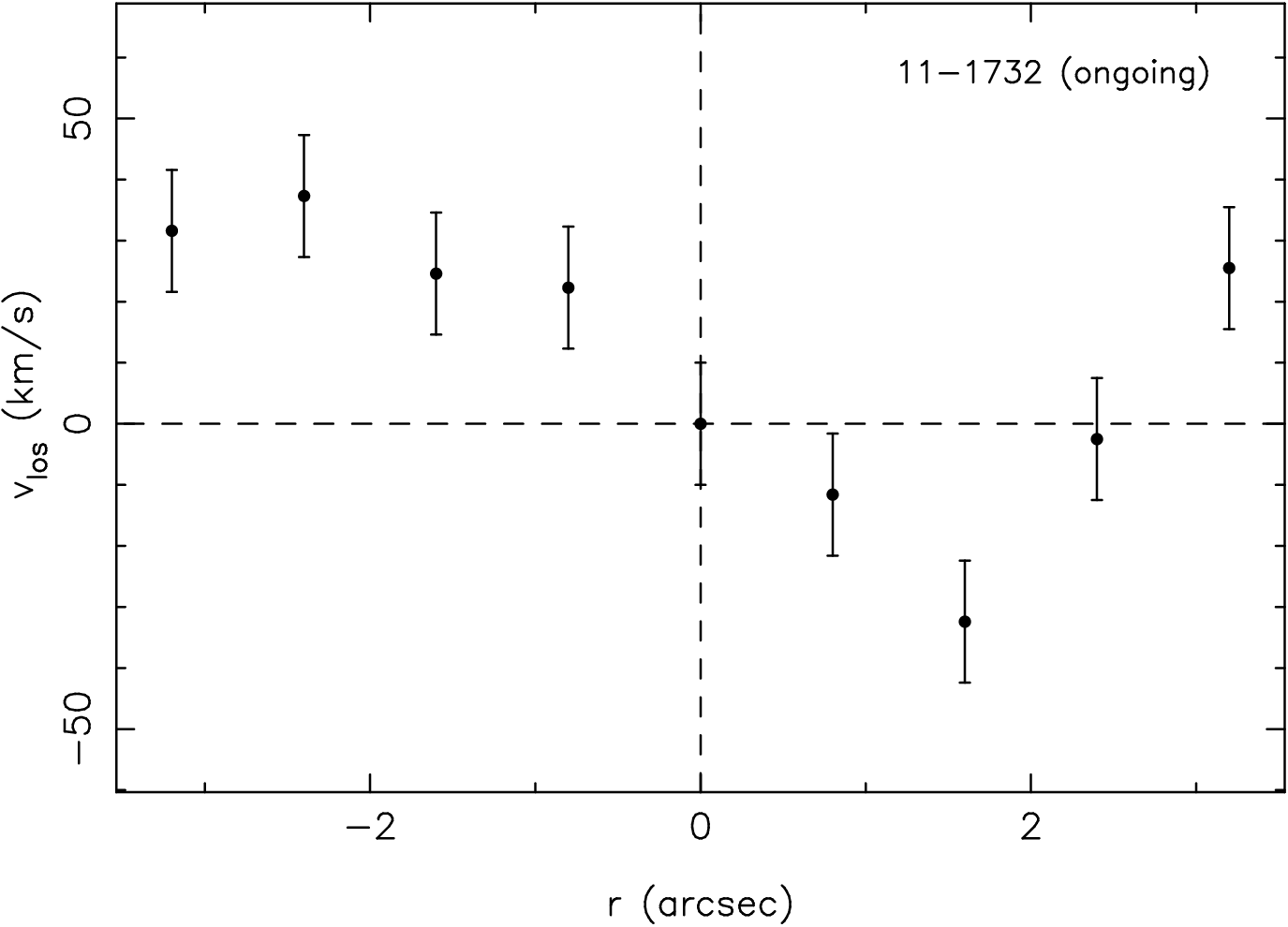}}
\resizebox{0.25\textwidth}{!}{\includegraphics[angle=-90]{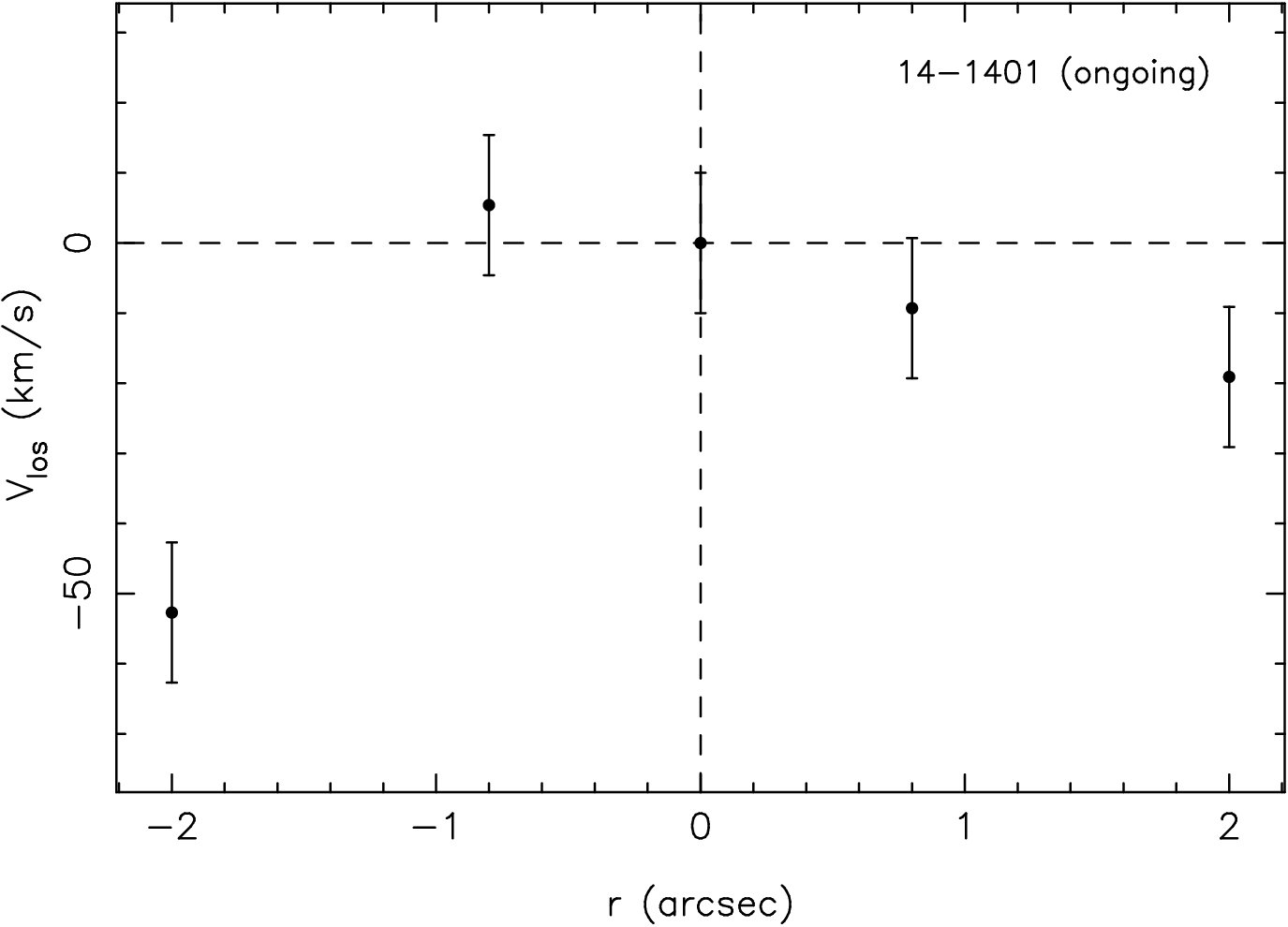}}
\resizebox{0.25\textwidth}{!}{\includegraphics[angle=-90]{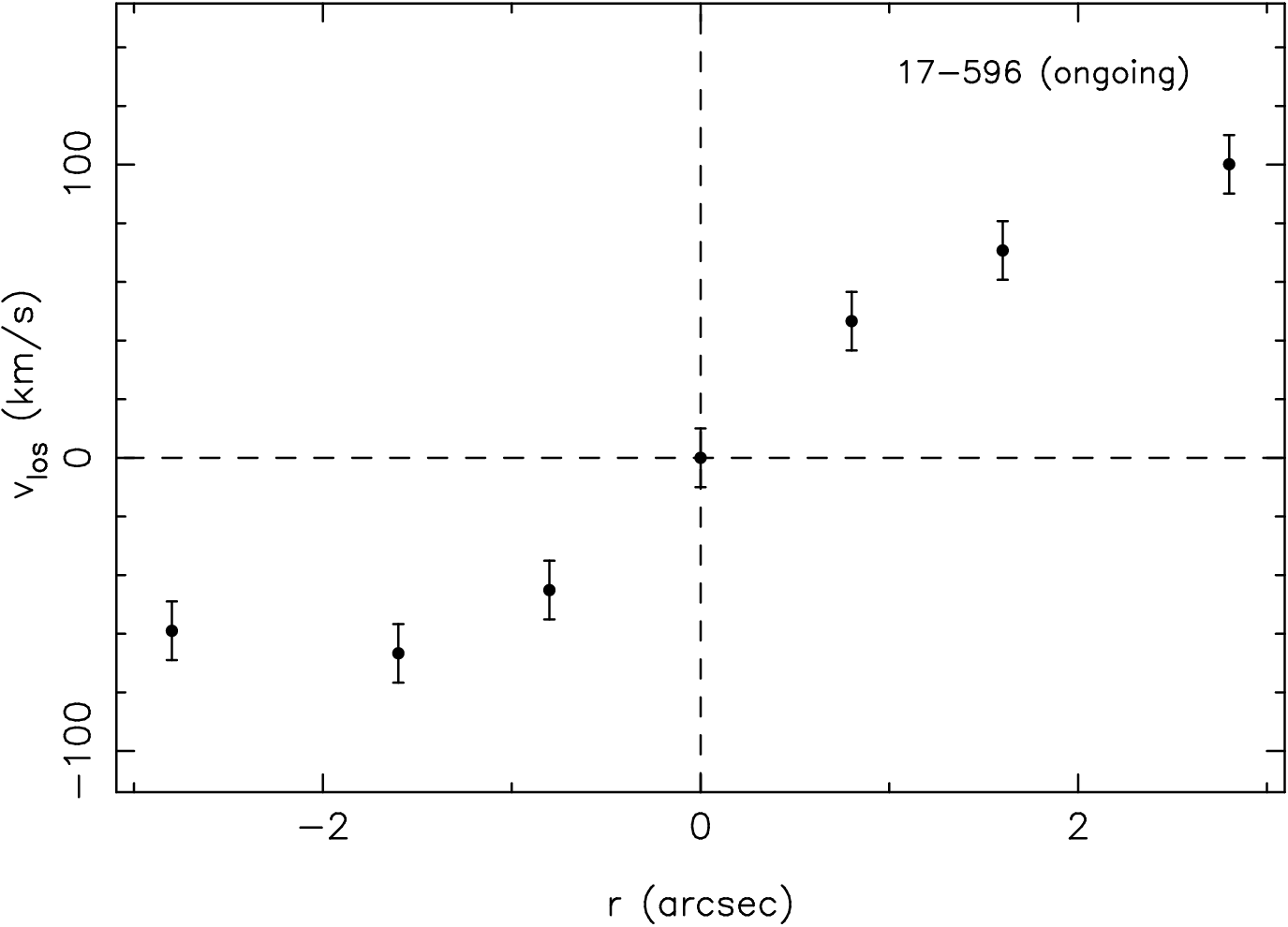}}
\resizebox{0.25\textwidth}{!}{\includegraphics[angle=-90]{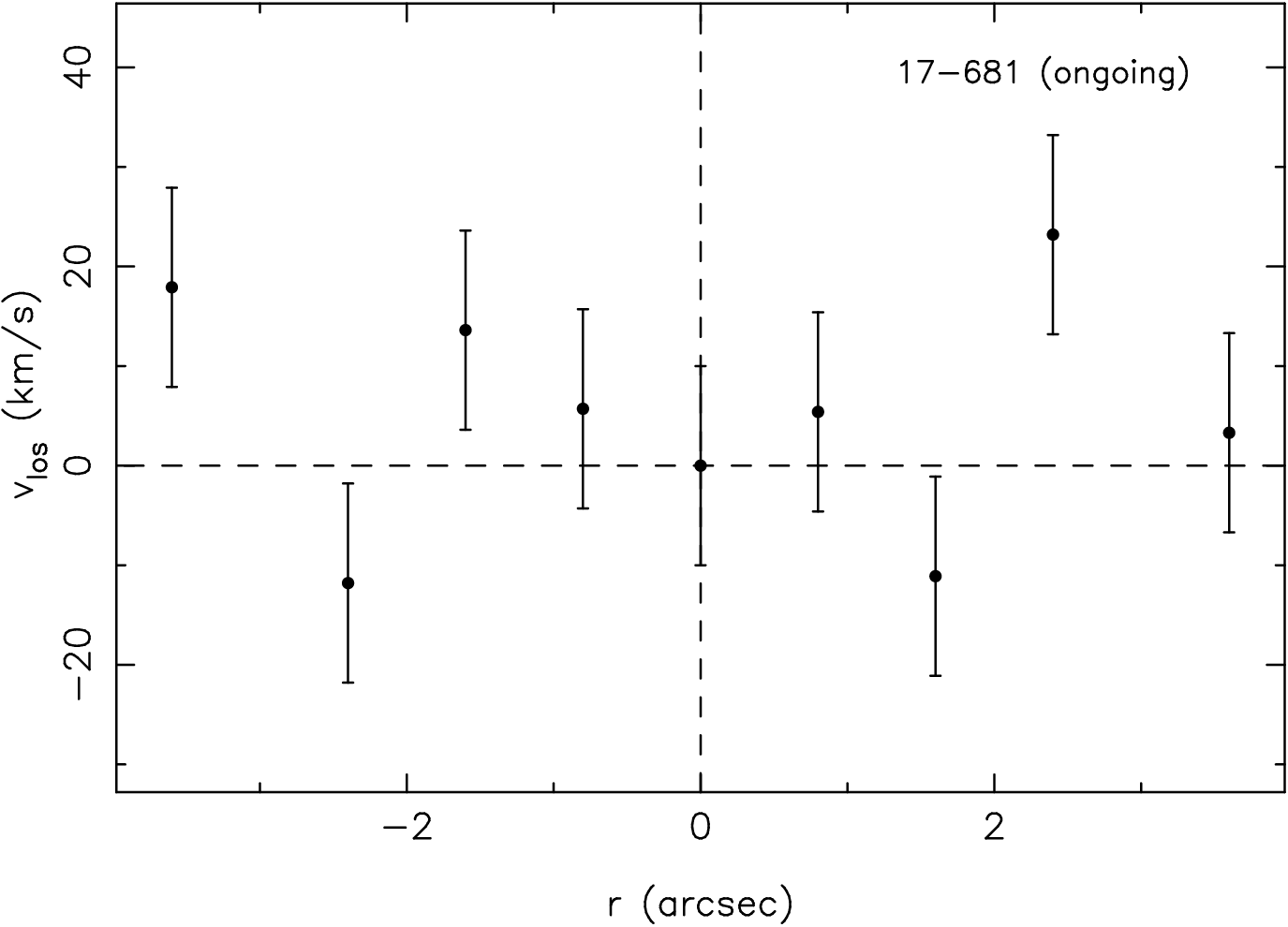}}
\resizebox{0.25\textwidth}{!}{\includegraphics[angle=-90]{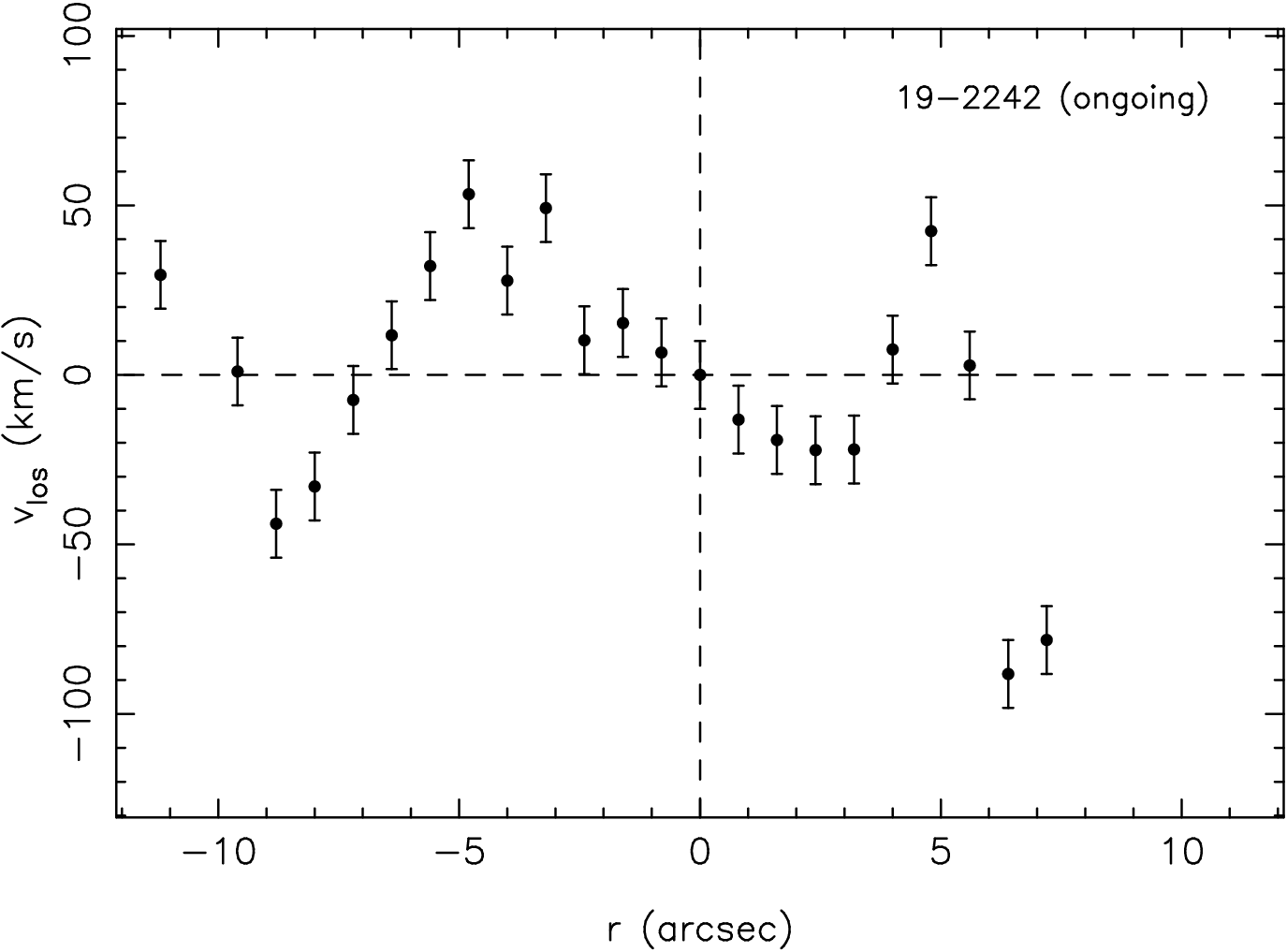}}
\resizebox{0.25\textwidth}{!}{\includegraphics[angle=-90]{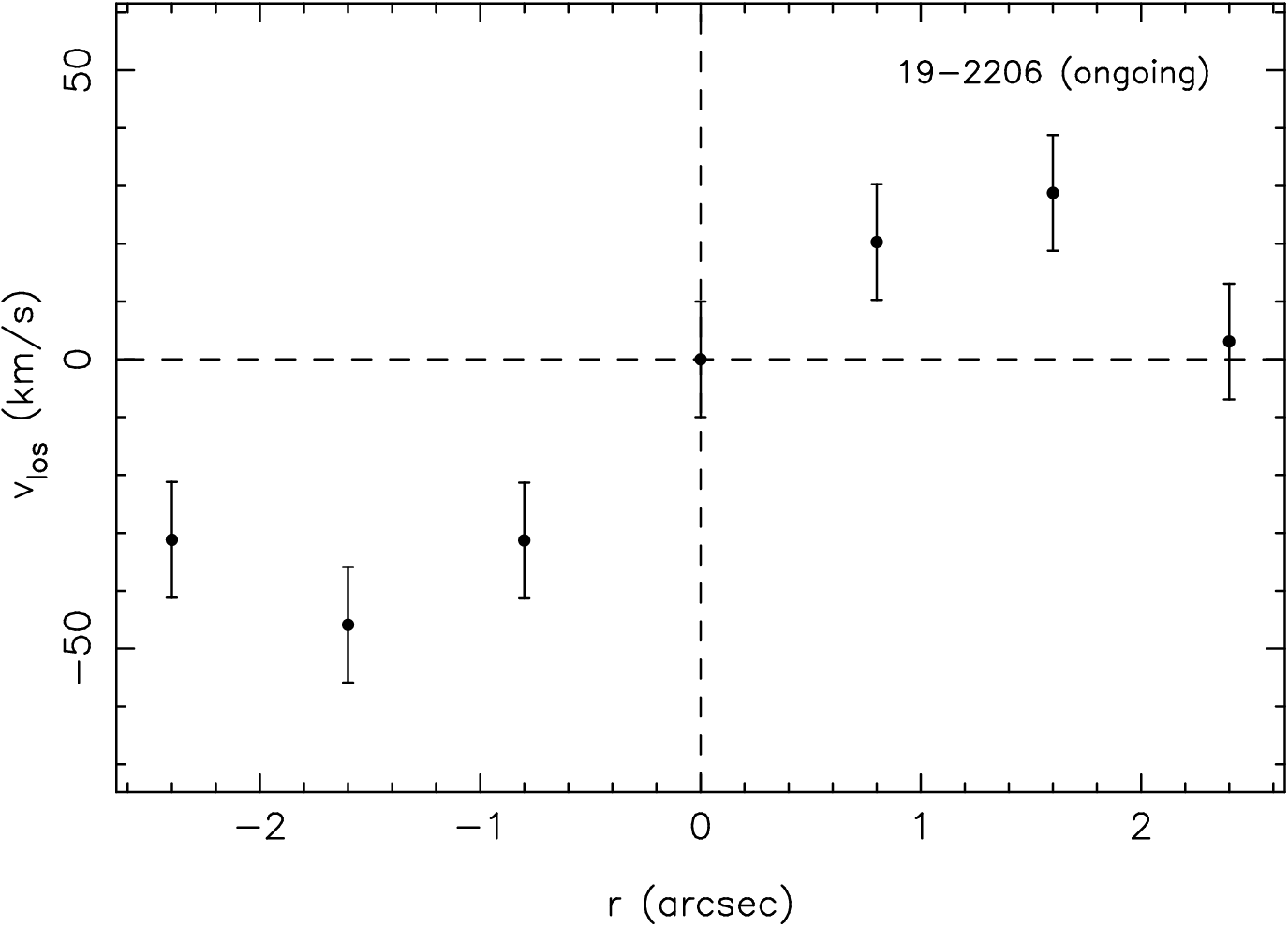}}
\resizebox{0.25\textwidth}{!}{\includegraphics[angle=-90]{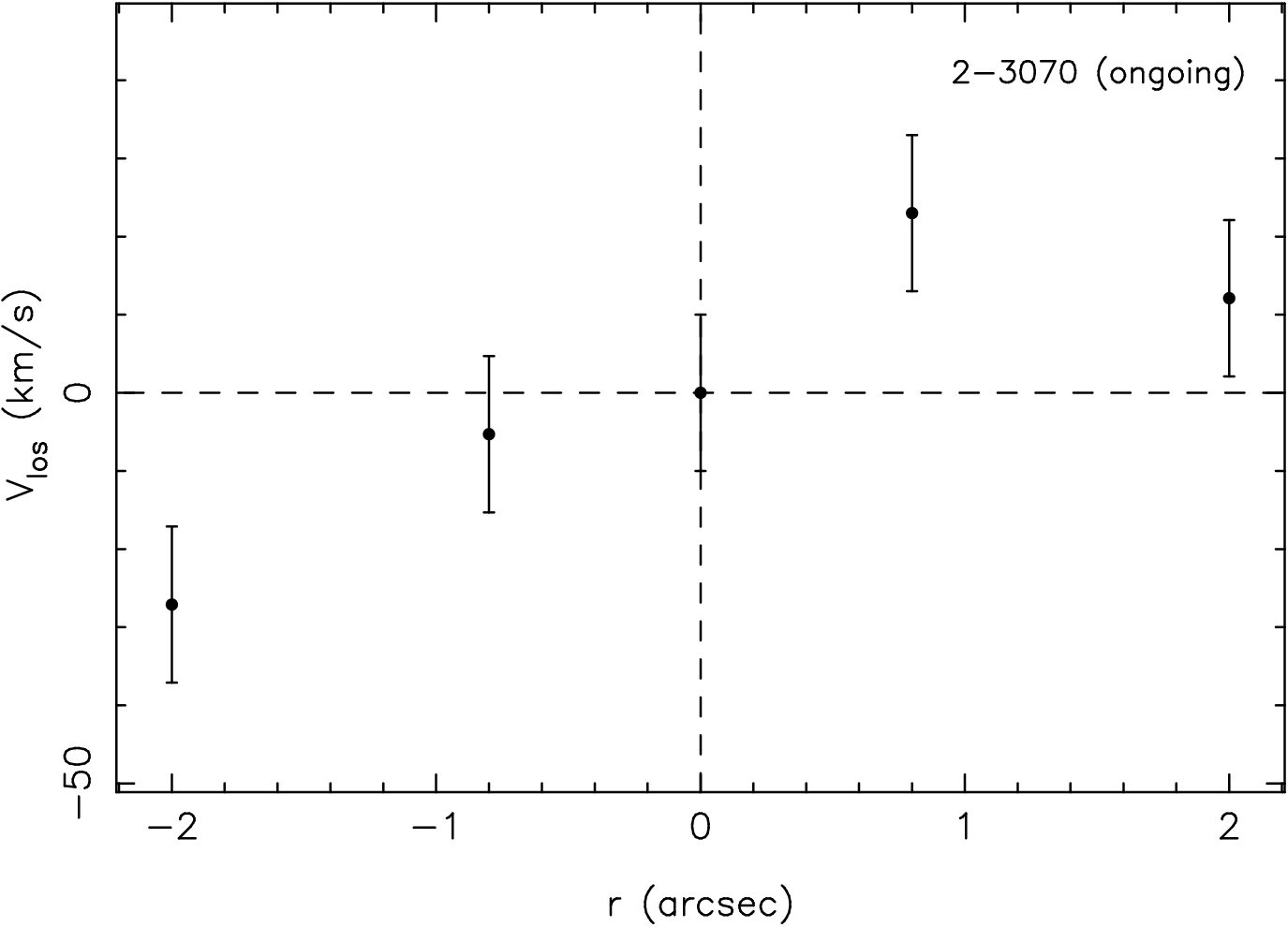}}
\resizebox{0.25\textwidth}{!}{\includegraphics[angle=-90]{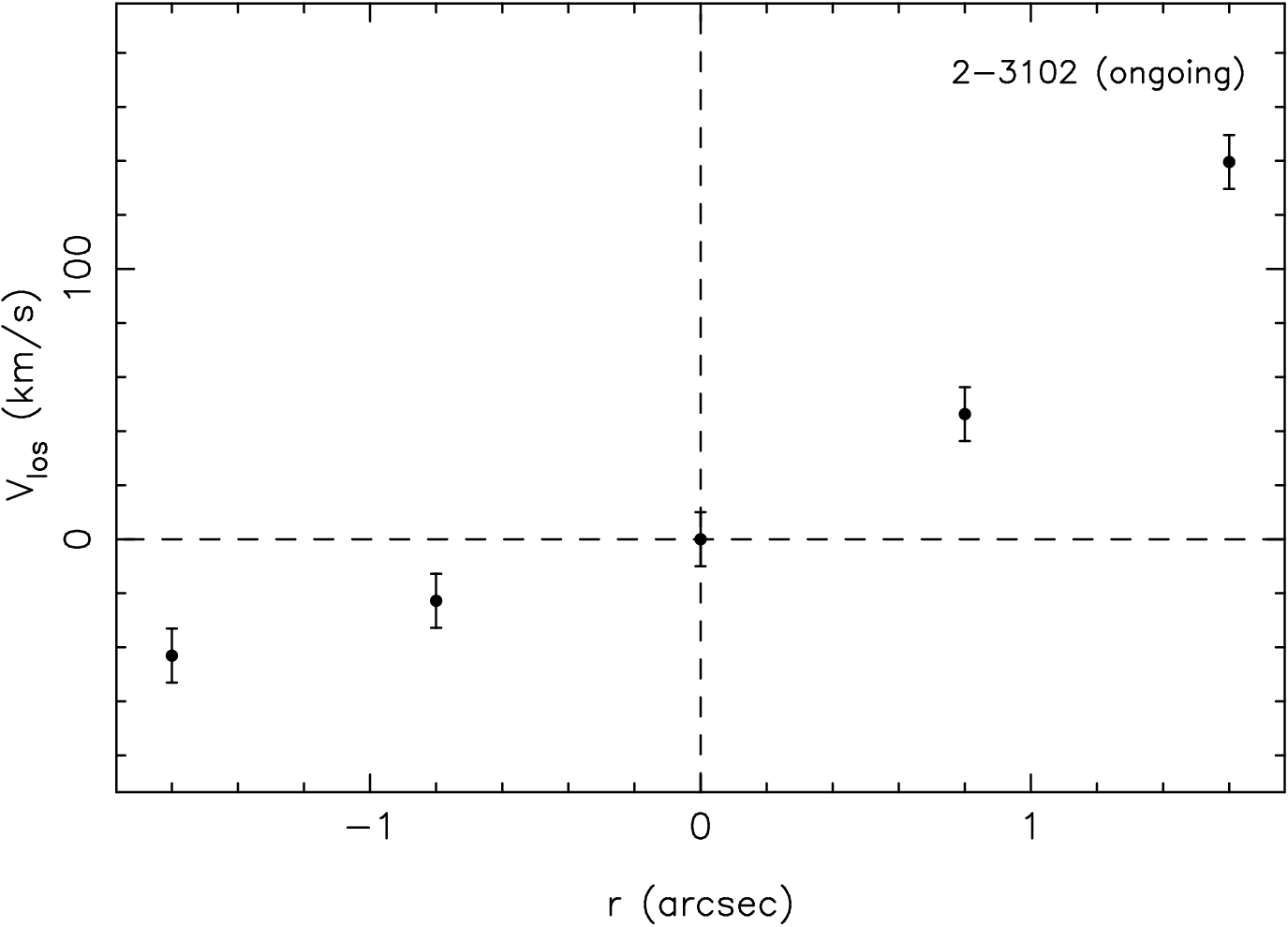}}
\caption{Rotation curves for galaxies in ongoing interactions.}
\end{figure*}
\end{document}